\documentclass{article}

\usepackage{amsmath, amsthm, amsfonts, amssymb}
\usepackage{bm}
\usepackage{a4wide}
\usepackage{graphicx} % for graphics
\usepackage{subfigure} % for putting different caption for lined figures
\usepackage{color}
\usepackage{euscript}
\usepackage{fancybox}

\newcommand{\cO}{\mathcal{O}}
\newcommand{\cA}{\mathcal{A}}
\newcommand{\cB}{\mathcal{B}}
\newcommand{\cC}{\mathcal{C}}
\newcommand{\cD}{\mathcal{D}}	
\newcommand{\cF}{\mathcal{F}}
\newcommand{\sP}{\mathsf{P}}
\newcommand{\sB}{\mathsf{B}}
\newcommand{\sY}{\mathsf{Y}}
\newcommand{\sy}{\mathsf{y}}
\newcommand{\sG}{\mathsf{G}}
\newcommand{\sm}{\mathsf{m}}
\newcommand{\sg}{\mathsf{g}}
\newcommand{\cT}{\mathcal{T}}
\newcommand{\cU}{\mathcal{U}}
\newcommand{\cH}{\mathcal{H}}
\newcommand{\cL}{\mathcal{L}}
\newcommand{\bC}{\mathbb{C}}
\newcommand{\txi}{\tilde{\xi}}
\newcommand{\tlambda}{\tilde{\lambda}}
\newcommand{\cI}{\mathcal{I}}
\newcommand{\cR}{\mathcal{R}}
\newcommand{\cS}{\mathcal{S}}
\newcommand{\cN}{\mathcal{N}}
\newcommand{\cM}{\mathcal{M}}
\newcommand{\cG}{\mathcal{G}}
\newcommand{\cK}{\mathcal{K}}
\newcommand{\bZ}{\mathbb{Z}}
\newcommand{\bR}{\mathbb{R}}
\newcommand{\ve}{\varepsilon}
\newcommand{\eH}{\EuScript{H}}
\newcommand{\cP}{\mathcal{P}}
\newcommand{\hT}{\hat{T}}
\newcommand{\htheta}{\hat{\theta}}
\newcommand{\sh}{\,\mathrm{sh}}
\newcommand{\ch}{\,\mathrm{ch}}
\newcommand{\fA}{\mathfrak{A}}
\newcommand{\fa}{\mathfrak{a}}
\newcommand{\fC}{\mathfrak{C}}
\newcommand{\fc}{\mathfrak{c}}
\newcommand{\fD}{\mathfrak{D}}
\newcommand{\fd}{\mathfrak{d}}
\newcommand{\fS}{\mathfrak{S}}
\newcommand{\ba}{\bar{a}}
\newcommand{\bb}{\bar{b}}
\newcommand{\dpsi}{\psi^\dag}
\newcommand{\pa}{a^{(+)}}
\newcommand{\dpa}{a^{(+)\dag}}
\newcommand{\ma}{a^{(-)}}
\newcommand{\dma}{a^{(-)\dag}}
\newcommand{\da}{a^\dag}
\newcommand{\ppsi}{\psi_+}
\newcommand{\dppsi}{\psi_+^\dag}
\newcommand{\mpsi}{\psi_-}
\newcommand{\dmpsi}{\psi_-^\dag}
\newcommand{\vphi}{\varphi}
\newcommand{\pS}{S^+}
\newcommand{\mS}{S^-}
\newcommand{\zS}{S^z}
\title{Boundary effects on the supersymmetric sine-Gordon model through light-cone lattice approach}
\author{Chihiro Matsui \\[3ex]
{\it Department of Mathematical Informatics, The University of Tokyo} \\
{\it 7-3-1 Hongo, Bunkyo-ku, Tokyo 113-8656, Japan}
}
\date{April 15, 2014}

\begin{document}
\maketitle

\begin{center}
{\bf Abstract}
\end{center}
\bigskip

{\small

We discussed subspaces of the $\mathcal{N}=1$ supersymmetric sine-Gordon model with Dirichlet boundaries through light-cone lattice regularization. In this paper, we showed, unlike the periodic boundary case, both of Neveu-Schwarz (NS) and Ramond (R) sectors of a superconformal field theory were obtained. 
Using a method of nonlinear integral equations for auxiliary functions defined by eigenvalues of transfer matrices, we found that an excitation state with an odd number of particles is allowed for a certain value of a boundary parameter even on a system consisting of an even number of sites. 
In a small-volume limit where conformal invariance shows up in the theory, we derived conformal dimensions of states constructed through the lattice-regularized theory. The result shows existence of the R sector, which cannot be obtained from the periodic system, while a winding number is restricted to an integer or a half-integer depending on boundary parameters. 

\ifx10
We studied the boundary effects on the $\cN = 1$ supersymmetric sine-Gordon model and the spin-$1$ XXZ model with the Dirichlet boundaries in the repulsive regime. The spin-$1$ inhomogeneous XXZ is known as a light-cone regularized model of the supersymmteric sine-Gordon model and inversely, the supersymmetric sine-Gordon model is obtained by taking the scaling limit of the spin-$1$ inhomogeneous XXZ model. 

The analysis of the supersymmetric sine-Gordon model via the light-cone regularization has been done for the periodic system in \cite{bib:HRS07} and then for the Dirichlet boundary conditions in \cite{bib:ANS07}. In the latter paper, they have discussed the infrared limit and the ultraviolet limit of the nonlinear integral equations of the light-cone regularized model on the ground state and the first-excited state in the regime where no boundary bound state exists.

%\bigskip
In this paper, we derived the nonlinear integral equations of the spin-$1$ inhomogeneous XXZ model for an arbitrary excited state. 
Our results are valid for all values of boundary parameters. From the asymptotic analysis of the nonlinear integral equations, the sum rules for the numbers of roots and holes were also derived. 

We obtained three different forms of boundary-dependent terms in the nonlinear integral equations depending on the values of boundary parameters. Such dependence on boundary parameters results from different analyticity structure of the auxiliary functions. We also showed that the boundary-dependent expressions are consistent with the boundary bootstrap relations through the analysis of the nonlinear integral equations in the large-volume limit. 

The ground state of the supersymmetric sine-Gordon model with the Dirichelt boundaries was studied by computing the eigenenergy via the light-cone regularized model. It was found that the ground state is not characterized only by the two-string roots, as for the periodic system, but also by a boundary bound state consists of imaginary roots for a certain value of the  boundary parameter. 
\fi

}

%%%%%%%%%%%%%%%%%%%%%%%%%%%%%%%%%%%%%%%%%%%%%%%%%%%%%%%%%%%%%%%%%%%%%%%%%%%%%%%%%%%
%%%%%%%%%%%%%%%%%%%%%%%%%%%%%%%%%%%%%%%%%%%%%%%%%%%%%%%%%%%%%%%%%%%%%%%%%%%%%%%%%%%
\section{Introduction} \label{sec:introduction}

Physical systems on finite volume show interesting features such as edge states and boundary critical exponents and their importance has been noticed for years. It is also important, as any real materials are finite-size systems, to know boundary effects on physical quantities. Nevertheless, existence of boundaries often destroys good symmetry obtained for periodic systems, which makes it more difficult to study a system with boundaries. 

For this reason, it would be nice to work on systems with good symmetries, even after adding non-trivial boundary conditions, which somehow allow us exact calculation of physical quantities. 
Although adding boundaries breaks symmetry of an integrable system at boundaries, whose integrability is ensured by the Yang-Baxter equation, there exist such boundary conditions that preserve integrability of the system satisfying the reflection relation \cite{bib:C84, bib:S88} at boundaries. 
Due to the Yang-Baxter equation and the reflection relation, a many-body scattering process can be decomposed into a sequence of two-body scatterings which allows us to find exact scattering and reflection matrices. 

An example which holds these symmetries is the spin-$\frac{1}{2}$ XXZ spin chain with boundary magnetic fields, whose $R$ and $K$-matrices can be obtained as solutions of the Yang-Baxter equation and the reflection relation. Another example is the sine-Gordon (SG) model with Dirichlet boundary conditions, which is obtained through bosonization of the spin-$\frac{1}{2}$ XXZ spin chain with boundary magnetic field. Both models has characterizing $R$-matrices associated with the $U_q(sl_2)$-algebra \cite{bib:KRS81, bib:J85}. 
%Although the same symmetry underlies these two models, they are not completely equivalent to each other; For instance, it is known that a whole Hilbert space of the SG model cannot be realized from the spin-$\frac{1}{2}$ XXZ chain \cite{}. 

\bigskip
Different methods have been developed for spin chains and quantum field theories, since the former model is a discrete system, while the latter a continuum one. For spin chains, a transfer matrix method is often used to solve a system by regarding a two-dimensional lattice with time sequences of a transfer matrix.
%\bigskip
The Bethe-ansatz method is one of the most successful method to diagonalize a transfer matrix \cite{bib:B31}. This method can be also applied to a system with non-trivial boundaries, as long as they satisfy the reflection relation. For instance, the XXZ model with boundary fields was first solved by the coordinate Bethe-ansatz method \cite{bib:ABBBQ87} and the method was algebraically formulated for the diagonal boundary case by introducing the double-row transfer matrix \cite{bib:S88}. 

%Exact solvability of the models makes it possible to compute physical quantities analytically. Such models that possess exactly solvability are called integrable systems and it is known that some of boundary conditions do not break integrability of the systems. 
%One of the most famous examples is the spin-$1/2$ Heisenberg chain with boundary magnetic fields. This model was first solved by the coordinate Bethe ansatz in \cite{bib:ABBBQ87}. The algebraic formulation to solve the integral open systems was accomplished in \cite{bib:S88}. As a result, any boundary matrices of integrable boundary conditions have been obtained as solutions of the reflection matrices. Especially, those which have diagonal boundary matrices were solved by the algebraic Bethe ansatz by introducing the double-row transfer matrix \cite{bib:S88}. 
In a presence of magnetic boundary fields, existence of boundary bound states have been found through a $q$-deformed vertex operator \cite{bib:JKKKM95} and later also by the Bethe-ansatz method \cite{bib:SS95,bib:KS96}. In a realm of the Bethe-ansatz method, boundary bound states are obtained as imaginary solutions of the Bethe-ansatz equations \cite{bib:SS95,bib:KS96}. 
One needs exact distribution of Bethe roots for computation of physical quantities by the Bethe-ansatz method. Existence of imaginary roots slightly deforms root density for the bulk, and as a result deforms root distribution for the ground state as well.  
% slightly deform (at the order $1/N$) the density of ground-state roots. 
This fact leads us to a question whether boundary bound states are to be included in the ground state or not. 
%If imaginary roots are included in the ground state, even the ground state is different for the open system from the periodic case. Therefore, studying the ground state of the open-boundary system is a good starting point of the analysis of boundary effects.  
The answer to this question was given for the repulsive regime \cite{bib:SS95} and for the attractive regime \cite{bib:KS96} by calculating a energy shift coming from emergence of imaginary roots themselves and a shift of root density driven by imaginary roots.

%For the spin-$1/2$ Heisenberg chain with the boundary magnetic fields, it was found that imaginary roots, which characterize boundary bound states, appear when the boundary fields are greater than the threshold through the $q$-deformed vertex operator method \cite{bib:JKKKM95} and the Bethe ansatz method \cite{bib:SS95,bib:KS96}.  
%Accordingly, it was found that the ground state is encoded by an imaginary root besides real Bethe roots unlike the periodic system. 
%\bigskip
On the other hand, analytical discussion of a continuum theory has been achieved by the bootstrap approach \cite{bib:ZZ79}. This method allows us to compute a scattering matrix between any particles subsequently from a soliton-soliton $S$-matrix obtained as a solution of the Yang-Baxter equation. Similarly, the boundary bootstrap principle was also developed which subsequently gives a reflection amplitude on a boundary with excitation particles. 

%the sine-Gordon model is a famous integrable quantum field theory and known that the Dirichlet boundary conditions conserve integrability of the model. 
%The scattering process of the sine-Gordon model has been well-studied. 
%As an integrable system, the $S$-matrix of the sine-Gordon model satisfies relation called the Yang-Baxter equation, which allows us exact computation of scattering amplitudes. Similarly, the reflection relation is known for integrable open-boundary systems, from which the exact reflection amplitudes are obtained. 
%There are some particles which do not scatter off each other but form bound states. For those cases, the procedure called the bootstrap principle has been developed to calculate the scattering amplitudes between bound states and their mass starting from the soliton mass, i.e. the mass of the lightest particle. 

In the context of a quantum field theory, boundary bound states are obtained as poles in a reflection matrix. Existence of boundary bound states in the SG model with Dirichlet boundary conditions was discussed in \cite{bib:GZ94, bib:G94} together with explicit forms of reflection matrices. Then spectrum of boundary states has been calculated in \cite{bib:MD00, bib:BPT01, bib:BPTT02, bib:BPT02}. 
However, it is hard to know whether boundary bound states are included in the ground state or not, since in a quantum field theory realm, the ground state is always considered as a vacuum.

\bigskip
If one correctly knows a corresponding lattice model to a quantum field theory, one can use a method valid only to discretized systems in analysis of a system which is originally continuum. 
Therefore, our main aim in this paper is to know correct correspondence between a lattice system and a quantum field theory. 
%On a lattice system, calculation of eigenenergy is accomplished by the Bethe-ansatz method using the commuting transfer matrix. One might be encouraged to map the quantum field theory on the lattice model to apply this method to the continuum theory. 
The notion to discretize an integrable quantum field theory was first introduced in \cite{bib:STF80}. Among various types of discretization, we employ the light-cone regularization \cite{bib:DV87, bib:V89, bib:V90}. 
The light-cone regularization is achieved by discretization of a light cone, at the same time with fixing a mass parameter \cite{bib:DV88}. This treatment is called ``scaling'' and we call a continuum limit to reproduce an original theory the scaling limit. 

A discretized light cone looks like a two-dimensional lattice system rotated by $45$-degrees. Each line is a trajectory of each particle and a right-mover runs over a line from left-bottom to right-top, while a left-mover runs over a line from right-bottom to left-top. A scattering occurs only at a vertex with a corresponding scattering amplitude to an original theory. 
This scattering matrix coincides with the $R$-matrix of the spin-$\frac{1}{2}$ XXZ with alternating inhomogeneity, which algebraically connects these two models. 
%The discretized quantum field theory has a cut-off parameter of light velocity $\Lambda$ and a lattice spacing $a$. With a proper choice of the scaling, which finally leads $\Lambda$ to infinity and $a$ to $0$, one recovers the original quantum field theory \cite{bib:RS94}. 
%The sine-Gordon model with the Dirichlet boundary conditions are related to the XXZ model with boundary magnetic fields via the lattice regularization \cite{}. 
%Roughly speaking, inhomogeneity of the spin chain and the lattice spacing turn into the mass parameter in the scaling limit \cite{bib:K79, bib:BT79}. The anisotropy of the XXZ model corresponds to the coupling constant of the SG model. The boundary parameters of the spin chain are related to the fixed values of bosonic fields at the boundaries \cite{bib:FS94}. 
%These relations were derived by comparing the $S$-matrix of the sine-Gordon model with that of the spin chain. 

%The $S$-matrix of the SG model was derived as a solution of the Yang-Baxter equation in \cite{bib:ZZ79}. 
In order to derive characteristic quantities in quantum field theories, such as $S$-matrices and conformal dimensions, from a light-cone regularized model, two different approaches have been developed to describe only excitation particles. The first one is based on the physical Bethe-ansatz equations calculated by assuming string solutions and deriving equations for density of those strings on an infinite system \cite{bib:K79, bib:DL82, bib:JNW83, bib:AD84, bib:R91}. The second is derived from the nonlinear integral equations (NLIEs) for counting functions or auxiliary functions defined from eigenvalues of transfer matrices \cite{bib:DV92, bib:FMQR97, bib:DV97, bib:KBP91}. This method, which allows us to deal with a finite-size system, is more algebraic in the sense that the equations are obtained based on $T$-systems and $Y$-systems, whose concept was first introduced in \cite{bib:Z91} and link with Dynkin diagram was explored in \cite{bib:RVT93}. 
%We use the second method throughout this paper for a reason advocated below. 

%The explicit relations between these parameters were obtained by comparing the known $S$-matrix ({\it resp.} reflection matrix) of the sine-Gordon model and that of the XXZ model \cite{}. In the reference \cite{}, they derived the physical Bethe ansatz equations by considering densities of any species of particles ({\it resp.} holes). 
%On the other hand, the reference \cite{} provides the alternative approach to derive the $S$-matrix of the XXZ model through the auxiliary functions defined from the $T$-$Q$ relations of the model. 
%Although the former method is available only for the (semi-)infinite system, the latter can be used also for the finite system. Consequently, the counting function of the finite number of real roots and the sum rules, which are conventionally derived from the Bethe ansatz equations, are reproduced from the nonlinear integral equations \cite{}. 

%UV limit

\bigskip
Correspondence between the SG model and the spin-$\frac{1}{2}$ XXZ model has been closely discussed through light-cone lattice approach. NLIEs of only excitation states with an even number of particles have been accessed under a periodic boundary, as a corresponding spin chain consists of an even number of sites. Consequently, in the ultraviolet (UV) limit, obtained conformal dimensions have an even winding number. Later in \cite{bib:FRT98}, it has been suggested that a subspace characterized by odd winding numbers is obtained from a spin chain consisting of an odd number of sites, although it has not been found yet how to define a scaling limit on an odd-site system. 
On the other hand, correspondence of these two models under Dirichlet boundaries has been discussed in \cite{bib:ABR04, bib:ABPR08} and found that a subsector consisting of odd winding numbers is also obtained for certain values of boundary parameters. 
%This is understood that a boundary bound state emerges for a strong enough boundary field, resulting in the outermost site occupied by a particle which makes a system effectively consisting of an odd number of sites. 

Our interest is, if we consider more complicated case with supersymmetry, how boundary fields affect on continuum-discrete correspondence. For this aim, we discuss the supersymmetric sine-Gordon (SSG) model \cite{bib:FGS78} and a corresponding spin chain, the Zamolodchikov-Fateev spin-$1$ XXZ chain \cite{bib:ZF80}. Correspondence of these two models has been discussed in \cite{bib:IO92, bib:A94} and under Dirichlet boundary conditions in \cite{bib:ANS07}. 
The periodic case was discussed from a light-cone point of view in \cite{bib:D03, bib:BDPTW04, bib:HRS07} and only the Neveu-Schwarz (NS) sector, {\it i.e.} one of two sectors in which the SSG model results in the UV limit, was obtained \cite{bib:HRS07}. In analysis of the SSG model from a light-cone regularization approach, they used NLIEs instead of a method based on string hypothesis, since Bethe roots of a higher-spin system are subjected to deviations of $\mathcal{O}(N^{-1})$ from string solutions. 

Higher-spin extension of a spin chain was first advocated in \cite{bib:KRS81}. Using a good property of the $U_q(sl_2)$ $R$-matrix, the fusion method has been developed. Applying a projection operator, the $R$-matrix of the spin-$1$ XXZ model is constructed. The $R$-matrix constructed in this way again satisfies the Yang-Baxter equation, which ensures integrability of a system associated with this $R$-matrix. The diagonal solution of the reflection relation for the spin-$1$ $R$-matrix results in Dirichlet boundaries on the SSG model \cite{bib:FLU02}. 

The BSSG model was first introduced in \cite{bib:IOZ95}. Boundary bound states and mass spectra have been discussed by a boundary bootstrap approach in \cite{bib:BPT02*}. Then light-cone regularization was also applied in \cite{bib:ANS07} where correspondence of a spin chain to the original theory has been intensively discussed in relation with a renormalization flow from the infrared (IR) limit to the UV limit. 
%in which they derived eigenenergies in the regime where no boundary bound state exists. They have derived the NLIEs of the corresponding spin chain to the BSSG model on the ground state and the first-excited state and studied the soliton reflection at the ground-state boundary. By taking the large-volume limit of the NLIEs, which corresponding to the infrared (IR) limit of the BSSG model, they have derived the relations between parameters of the $S$-matrix theory and the lattice system. 
%Also by considering the ultraviolet (UV) limit of the BSSG model, in which the model is reduced to the theory of free bosons and free Majorana fermions, and thus the conformal invariance shows up in the theory, they compared the parameters in the action of the BSSG model and the corresponding spin chain from the Casimir energy. 
Although they limited their discussion to a regime where no boundary bound state is obtained, we are more interested in how physics changes according to boundary parameters. Indeed, $T$-functions change their analytical structure in accordance with boundary parameters, which is physically interpreted as appearance of boundary bound states. Performing analytic continuation, we obtained different NLIEs for three regimes of boundary parameters. From each set of NLIEs, different counting equations were derived, which allows different types of excitations. We found, for certain values of boundary parameters, an odd number of particles is obtained in the system consisting of an even number of sites. Thus, we expect a similar sector separation obtained for the SG model under Dirichlet boundary conditions \cite{bib:ABPR08} but more complicated one due to supersymmetry.

\bigskip
Now we show the plan of this paper. 
%What we are going to discuss here is about boundary bound states of the spin-$1$ inhomogeneous XXZ model, which was left untouched in the paper \cite{bib:ANS07}. Through the derivation of the NLIEs for an arbitrary excited state with arbitrary values of boundary parameters, we show the ground state of the spin-$1$ inhomogeneous XXZ model (and the BSSG model) is described by either the two-strings or a boundary bound state besides two-string roots depending on boundary parameters. 
%
%The problem left is mainly about boundary bound states of the supersymmetric sine-Gordon model with the Dirichlet boundary conditions. Therefore, we investigate how the distribution of Bethe roots for the ground state varies according to the values of boundary parameters. In fact, as we will see in Section \ref{}, pure imaginary roots emerges for certain values of boundary parameters. In order to know the correct ground state, that is, whether these imaginary roots contribute to the ground state or not, we employ the lattice regularization of quantum field theory \cite{} on which the Bethe ansatz method becomes valid, as the method enables calculation of eigenenergies.  
Throughout this paper, we analyze the SSG model with Dirichlet boundaries on a finite volume. We focus on the repulsive regime where no breather {\it i.e.} a bound state of solitons exists in a system. Although two types of Dirichlet boundary conditions are allowed due to supersymmetry of Majorana fermions, we chose the condition referred by BSSG$^+$ in \cite{bib:BPT02}. 
In Section \ref{sec:ssg}, we first introduce the SSG model and review known results from a viewpoint of an integrable quantum field theory, including scattering and reflection matrices and a corresponding conformal field theory. A method of light-cone regularization is also explained in this section and properties obtained in a corresponding spin chain are referred. 
We use a method of NLIEs, since the spin-$1$ chain, a corresponding lattice model to the BSSG$^+$ model, is exposed to string deviations of $\mathcal{O}(N^{-1})$, which results in difficulty in calculation of physical quantities sensitive to a system size. This method also resolves a problem how to define a counting function of string solutions \cite{bib:ANS07}, which is also a fatal problem since a ground state of the SSG model is given by two-string roots. 
In Section \ref{sec:nlie}, we derive NLIEs of an arbitrary excitation state for a whole regime of boundary parameters. Derivation of NLIEs associated with a higher-spin representation of the $U_q(sl_2)$ algebra is based on 
%The method used is the boundary-extension of 
$T$-$Q$ relations \cite{bib:S88, bib:B82, bib:S99} together with analyticity structure of $T$-functions given by eigenvalues of transfer matrices. 
From asymptotic behaviors of NLIEs, counting equations are also derived. Counting equations relate the numbers of excitation particles by which we discuss allowed excitations in each regime of boundary parameters in connection with eigenenergy computed from NLIEs. 
%Moreover, scattering and reflection amplitudes are extracted from the sub-leading terms of the nonlinear integral equations in the infrared limit. By computing the eigenenergies in the infrared limit, we find the correct description of the ground state for any values of boundary parameters. 
In the next section, scattering and reflection amplitudes are discussed by taking the IR limit. Different NLIEs for three boundary regimes are connected via a boundary bootstrap method by interpreting a change of analyticity structure due to emergence of a boundary bound state. From symmetries obtained in reflection amplitudes, it is also referred how lattice symmetries survive in the scaling limit.
Then in Section \ref{sec:UVlim}, the UV limit is considered. Conformal dimensions are computed for a state obtained from a light-cone regularized BSSG$^+$ model and we show that both the NS and Ramond (R) sectors are obtained. A similar restriction on a winding number to the Dirichlet SG case is also obtained, which strongly motivate us to construct a corresponding spin chain to a subspace of the BSSG$^+$ model which cannot be obtained from a conventional light-cone regularization. 
The last section is devoted to concluding remarks and future works.

\ifx10
\documentclass{article}

%#############
%packages 
%#############
%\usepackage[utf8x]{inputenc}
\usepackage{amsmath, amsthm, amsfonts, amssymb}
\usepackage{bm}
\usepackage{a4wide}
\usepackage{graphicx} % for graphics
\usepackage{color}
\usepackage{euscript}
\usepackage{fancybox}

%#############
%newtheorems
%############
%\renewcommand{\thefootnote}{\fnsymbol{footnote}}
%\theoremstyle{plain}
%\newtheorem{thm}{Theorem}
%\newtheorem{claim}{Claim}
%\newtheorem{propn}{\bfseries Proposition}
%\newtheorem{propnn}{Proposition}
%\renewcommand{\thepropnn}{\arabic{propn}\alph{propnn}}
%\newtheorem{lem}{\bfseries Lemma}
%\newtheorem{cor}{Corollary}
%\newtheorem{conj}{\bfseries Conjecture}
%\newtheorem{defn}{Definition}
%\theoremstyle{remark}
%\newtheorem{rem}{Remark}
%\newtheorem{df}{Definition}
%\newtheorem{th1}[df]{Theorem}
%\newtheorem{lem}[df]{Lemma}
%\newtheorem{conj}[df]{Conjecture}
%\newtheorem{prop}[df]{Proposition}
%\newtheorem{cor}[df]{Corollary}
%\newtheorem{lem2}{Lemma}[section]
%\newtheorem{prop2}{Proposition}[section]
%\newtheorem{cor2}{Corollary}[section]
%\newtheorem{df2}{Definition}[section]

%##############
%newcommands
%#############
\newcommand{\cO}{\mathcal{O}}
\newcommand{\cA}{\mathcal{A}}
\newcommand{\cB}{\mathcal{B}}
\newcommand{\cC}{\mathcal{C}}
\newcommand{\cD}{\mathcal{D}}	
\newcommand{\cF}{\mathcal{F}}
\newcommand{\sP}{\mathsf{P}}
\newcommand{\sB}{\mathsf{B}}
\newcommand{\sY}{\mathsf{Y}}
\newcommand{\sy}{\mathsf{y}}
\newcommand{\sG}{\mathsf{G}}
\newcommand{\sm}{\mathsf{m}}
\newcommand{\sg}{\mathsf{g}}
\newcommand{\cT}{\mathcal{T}}
\newcommand{\cU}{\mathcal{U}}
\newcommand{\cH}{\mathcal{H}}
\newcommand{\cL}{\mathcal{L}}
\newcommand{\bC}{\mathbb{C}}
\newcommand{\txi}{\tilde{\xi}}
\newcommand{\tlambda}{\tilde{\lambda}}
\newcommand{\cI}{\mathcal{I}}
\newcommand{\cR}{\mathcal{R}}
\newcommand{\cS}{\mathcal{S}}
\newcommand{\cN}{\mathcal{N}}
\newcommand{\sM}{\mathsf{M}}
\newcommand{\cG}{\mathcal{G}}
\newcommand{\cK}{\mathcal{K}}
\newcommand{\bZ}{\mathbb{Z}}
\newcommand{\bR}{\mathbb{R}}
\newcommand{\ve}{\varepsilon}
\newcommand{\eH}{\EuScript{H}}
\newcommand{\cP}{\mathcal{P}}
\newcommand{\hT}{\hat{T}}
\newcommand{\htheta}{\hat{\theta}}
\newcommand{\sh}{\,\mathrm{sh}}
\newcommand{\ch}{\,\mathrm{ch}}
\newcommand{\fA}{\mathfrak{A}}
\newcommand{\fa}{\mathfrak{a}}
\newcommand{\fC}{\mathfrak{C}}
\newcommand{\fc}{\mathfrak{c}}
\newcommand{\fD}{\mathfrak{D}}
\newcommand{\fd}{\mathfrak{d}}
\newcommand{\fS}{\mathfrak{S}}
\newcommand{\ba}{\bar{a}}
\newcommand{\bb}{\bar{b}}
\newcommand{\dpsi}{\psi^\dag}
\newcommand{\pa}{a^{(+)}}
\newcommand{\dpa}{a^{(+)\dag}}
\newcommand{\ma}{a^{(-)}}
\newcommand{\dma}{a^{(-)\dag}}
\newcommand{\da}{a^\dag}
\newcommand{\ppsi}{\psi_+}
\newcommand{\dppsi}{\psi_+^\dag}
\newcommand{\mpsi}{\psi_-}
\newcommand{\dmpsi}{\psi_-^\dag}
\newcommand{\vphi}{\varphi}
\newcommand{\bpsi}{\bar{\psi}}
\newcommand{\pS}{S^+}
\newcommand{\mS}{S^-}
\newcommand{\zS}{S^z}
%%%
%\newcommand{\bm}[1]{\mbox{\boldmath$#1$}}
%\newcommand{\be}{\begin{equation}}
%\newcommand{\ee}{\end{equation}}
%\newcommand{\bea}{\begin{eqnarray}}
%\newcommand{\eea}{\end{eqnarray}}
%\newcommand{\non}{\nonumber}
%\newcommand{\ra}{\rangle}
%\newcommand{\la}{\langle}
%\newcommand{\lam}{\lambda} 
%\newcommand{\Lam}{\Lambda} 
%\newcommand{\tht}{\theta} 
%\newcommand{\al}{\alpha} 
%\newcommand{\bt}{\beta} 
%\newcommand{\gm}{\gamma} 
%\newcommand{\dt}{\delta} 

%#############
%document
%#############
\begin{document}
\fi

\section{SSG model with Dirichlet boundary conditions}\label{sec:ssg}
The SSG model is an integrable one-dimensional quantum field theory consisting of a real scalar field $\Phi$ and a Majorana fermion $\Psi$. On a finite system size $L$, the action of the SSG model is given by 
\begin{equation}\label{SSG_action}
\begin{split}
  &\cA_{\rm SSG} = \int_{-\infty}^{\infty} dt \int_0^L dx\; \cL_{\rm SSG}(x;t),
  \\
  &\cL_{\rm SSG} = \frac{1}{2} \partial_{\mu} \Phi \partial^{\mu} \Phi + \frac{i}{2} \bar{\Psi} \gamma^{\mu} \partial_{\mu} \Psi - \frac{m_0}{2} \cos(\beta \Phi) \bar{\Psi} \Psi + \frac{m_0^2}{2\beta^2} \cos^2(\beta \Phi), 
\end{split}
\end{equation}
where
\begin{equation}
 \Psi = \begin{pmatrix} \psi \\ \bar{\psi} \end{pmatrix}, 
 \quad
  \gamma^0 = \begin{pmatrix} 0 & i \\ -i & 0 \end{pmatrix}, 
  \quad
   \gamma^1 = \begin{pmatrix} 0 & i \\ i & 0 \end{pmatrix}. 
%   \quad \gamma^3 = \gamma ^0 \gamma ^1. 
\end{equation}
A mass parameter $m_0$ determined in such a way that realizes a proper scaling limit \cite{bib:DV88} is related to the physical soliton mass via the relation found in \cite{bib:BF98}. 

The theory behaves differently depending on a value of the coupling constant $\beta$; In the attractive regime ($0<\beta^2<\frac{4\pi}{3}$), solitons form bound states called breathers, while the repulsive regime ($\frac{4\pi}{3}<\beta^2<4\pi$) does not admit breathers. Throughout this paper, we concentrate on the repulsive regime. 

Besides, we impose the Dirichlet boundary conditions \cite{bib:IOZ95}: 
\begin{equation} \label{dirichlet}
\begin{split}
 &\Phi (0;t) = \Phi_-, \hspace{8mm}
 \Psi(0;t) \mp \bar{\Psi}(0;t) = 0, 
 \\
 &\Phi (L;t) = \Phi_+, \qquad
 \Psi(L;t) \mp \bar{\Psi}(L;t) = 0. 
\end{split}
\end{equation}
By following the notations used in \cite{bib:BPT02*}, we call the conditions given by (\ref{dirichlet}) the BSSG$\pm$ model, respectively.

%%%%%%%%%%%%%%%%%%%%%%%%%%%%%%%%%%%%%%%%%%%%%
\subsection{SSG model as a perturbed CFT}
From a viewpoint of a renormalization group theory, the SSG model is considered as a perturbation from a $\mathcal{N}=1$ superconformal field theory consisting of free bosons and free fermions compactified on a cylinder with radius $R = \frac{4\sqrt{\pi}}{\beta}$. The third term in Lagrangian (\ref{SSG_action}) is an irrelevant perturbation in the UV limit given as a small-volume limit ($L \rightarrow 0$). 
The bosonic part is also obtained from the SG model, while the fermionic part from the tricritical Ising model \cite{bib:AK96, bib:NA02}. 
%Thus the Hilbert space of the $c = \frac{3}{2}$ CFT consists of a tensor product of a bosonic part with $c=1$ and a fermionic part with $c=\frac{1}{2}$. 

\subsubsection{Free boson}
A free boson theory compactified on a radius $R$ is defined by the following action: 
\begin{equation}
\cA_{\text{FB}} = \frac{1}{8\pi} \int_{-\infty}^{\infty}d\tau \int_0^{2\pi} d\sigma\; \partial_{\mu} \varphi \partial^{\mu} \varphi, 
\end{equation}
which is identified with the first term of the SSG action (\ref{SSG_action}) by the relation $\Phi = \frac{1}{\sqrt{4\pi}}\varphi$. 

\begin{figure}
\begin{center}
 \includegraphics[scale=0.65]{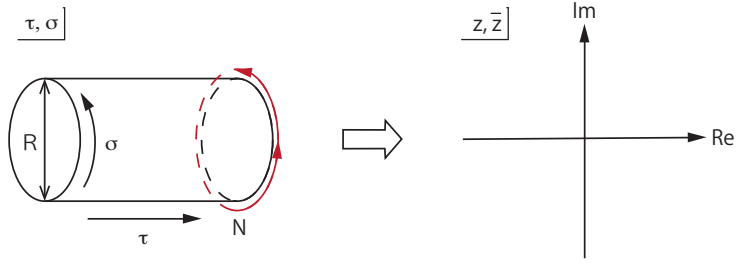}
 \caption{A conformal map from a cylinder onto a complex plain.} 
 \label{fig:CFT_map}
\end{center}
\end{figure}

A conformal boson has a $\widehat{U}(1) \times \widehat{U}(1)$ symmetry. 
Applying a conformal map from a cylinder onto a complex plane (Figure \ref{fig:CFT_map}): 
\begin{equation}
 \sigma = \frac{1}{2i} (\ln z - \ln \bar{z}),
  \qquad
  \tau = \frac{1}{2i} (\ln z + \ln \bar{z}), 
\end{equation}
a boson field is decomposed into a holomorphic part and an anti-holomorphic part: 
\begin{equation}
 \varphi(z, \bar{z}) = \frac{1}{2}(\phi(z) + \bar{\phi}(\bar{z})). 
\end{equation}
% figure
Subsequently, mode expansion of a boson field is obtained as 
\begin{equation}
% J(z) = i\partial \phi(z) = \sum_n a_n z^{-n-1}, 
%  \qquad
%  \bar{J}(\bar{z}) = i\bar{\partial}\bar{\phi}(\bar{z}) = \sum_n \bar{a}_n \bar{z}^{-n-1}, 
 \phi(z) = Q - ia_0 \ln z + i \sum_{n\neq 0} \frac{1}{n} z^{-n} a_n, 
 \qquad
 \bar{\phi}(\bar{z}) = Q - i\bar{a}_0 \ln \bar{z} + i \sum_{n\neq 0} \frac{1}{n} \bar{z}^{-n} \bar{a}_n, 
\end{equation}
where $Q$ is a zero mode of $\varphi(z,\bar{z})$. Bosonic modes satisfy commutation relations given by 
\begin{equation}
 [a_k,\, a_l] = k \delta_{k+l}, 
  \qquad
  [a_k,\, \bar{a}_l] = 0,
  \qquad
  [\bar{a}_k,\, \bar{a}_l] = k \delta_{k+l}. 
\end{equation}

Space of states of a free boson theory is spanned by highest weight vectors and their descendants created by bosonic modes with negative labels: 
\begin{equation}
\begin{split}
 \underset{m,n \in \bZ}{\oplus} \underset{p_i q_i > 0}{\oplus}
 \prod_i \bar{a}_{-p_i} \prod_j a_{-q_j} 
 |m, n \rangle, 
\end{split}
\end{equation}
where a highest weight vector $|m, n \rangle$ is created from the vacuum state $|0, 0 \rangle$ by applying a vertex operator: 
\begin{align}
 &|m, n \rangle = V_{(m, n)}(z, \bar{z}) |0, 0 \rangle, \\
 &V_{(m, n)} = :e^{i(mR + \frac{n}{R}) \phi(z) + i(mR - \frac{n}{R}) \bar{\phi}(\bar{z})}:. 
\end{align}
Here we used a notation $:*:$ for the normal order. 
Therefore, infinitely many highest weight vectors are obtained in a free boson theory. 
One of characteristic quantities of a conformal state $|m, n \rangle$ is a conformal dimension which shows up in energy as a function of system size: 
\begin{align}
 &E(L) = -\tfrac{\pi}{6L} (1- 12(\Delta^+_{\rm FB} + \Delta^-_{\rm FB})) + \cO(L^{-2}), \\
 &\Delta_{\text{FB}}^{\pm} = \tfrac{1}{2} \left(mR \pm \tfrac{n}{R}\right)^2. 
\end{align}

If one imposes Dirichlet boundary conditions (\ref{dirichlet}), a boson field should satisfy the following conditions: 
\begin{equation}
 \tfrac{1}{\sqrt{4\pi}} \varphi(z,\bar{z})|_{\sigma=0} = \Phi_-, 
 \qquad
 \tfrac{1}{\sqrt{4\pi}} \varphi(z,\bar{z})|_{\sigma=\frac{L}{R}} = \Phi_+, 
\end{equation}
which lead us to obtain $\bar{\alpha}_n = -\alpha_n$. As a result, the theory is described only by a holomorphic part and a momentum part of a conformal dimension vanishes: 
\begin{equation}
 \Delta_{\rm BFB} = \tfrac{1}{2}
  \left(\tfrac{1}{\sqrt{\pi}}(\Phi_+ - \Phi_-) + mR\right)^2, 
\end{equation}
Consequently, energy is obtained as 
\begin{equation} \label{energy_BFB}
 E(L) = -\tfrac{\pi}{24L} (1 - 24\Delta_{\rm BFB}) + \mathcal{O}(L^{-2}). 
\end{equation}

%%%%%%%%%%%%%%%%%%%%%%%%%%%%%%%%%%%%%%%%%%
\subsubsection{Free fermion}
A free fermion theory appears with a bosonic coupling in the SSG theory in the second term (\ref{SSG_action}) whose action is given by
\begin{equation}
\cA_{\text{FF}} = \iint\frac{dz d\bar{z}}{2\pi}\Big(\psi \frac{\partial}{\partial\bar{z}} \psi + \bar{\psi} \frac{\partial}{\partial z} \bar{\psi}\Big). 
\end{equation}
Mode expansion of a fermion field is given by
\begin{equation}
 \psi(z) = \sum_{n \in \bZ + r} b_n z^{-n -1/2}, 
  \qquad
  \bar{\psi}(\bar{z}) = \sum_{n \in \bZ + r} \bar{b}_n \bar{z}^{-n -1/2}, 
\end{equation}
where $r$ is a free parameter, in principle, but takes only $0$ or $\frac{1}{2}$ under compactification with an arbitrary radius. In the case of $r=\frac{1}{2}$, the theory results in the NS sector, {\it i.e.} an periodic boundary condition for a fermion part of the superconformal field theory, while $r=0$ leads to the R sector, {\it i.e.} an anti-periodic boundary condition. 

Fermionic modes satisfy anti-commutation relations given by 
\begin{equation}
 \{b_s,\, b_t\} = \delta_{s+t}, 
  \qquad
  \{b_s,\, \bar{b}_t\} = 0, 
  \qquad
  \{\bar{b}_s,\, \bar{b}_t\} = \delta_{s+t}. 
\end{equation}
Space of states of a free fermion theory is spanned by 
\begin{equation}
\begin{split}
 \underset{\hat{f} \in \mathcal{V}}{\oplus} \underset{p_i, q_j > 0}{\oplus}
  \prod_i \bar{b}_{-p_i} \prod_j b_{-q_j} \hat{f}(z, \bar{z}) |0,0 \rangle, 
\end{split}
\end{equation}
where highest vectors are constructed from the vacuum by applying an operator $\hat{f}(z,\bar{z})$ ($\mathcal{V} \in \{\mathbb{I}, \psi(z)\bar{\psi}(\bar{z}), \sigma(z,\bar{z})\}$). One may notice that, unlike the free boson theory, there are only three highest weight vectors whose conformal dimensions are given by
\begin{equation}
\begin{split}
 (\Delta_{\mathbb{I}}^+, \Delta_{\mathbb{I}}^-) = (0,0), 
  \qquad
 (\Delta_{\psi \bar{\psi}}^+, \Delta_{\psi \bar{\psi}}^-) = (\tfrac{1}{2}, \tfrac{1}{2}),
  \qquad
 (\Delta_{\sigma}^+, \Delta_{\sigma}^-) = (\tfrac{1}{16}, \tfrac{1}{16}), 
\end{split}
\end{equation}
for the first two belonging to the NS sector, while the last one belonging to the R sector. 
Free energy is then expressed in terms of conformal dimensions:  
\begin{equation}
 E(L) = -\tfrac{\pi}{6L}\left(\tfrac{1}{2} - 12(\Delta^+ + \Delta^-)\right) + \mathcal{O}(L^{-2}), 
\end{equation}
from which two sectors of a superconformal field theory are distinguished. 

If one imposes Dirichlet boundary conditions (\ref{dirichlet}), we obtain $b_n = \pm \bar{b}_n$. As a result, the free fermion theory is also written only by a holomorphic part and then energy is given by 
\begin{equation} \label{energy_BFF}
 E(L) = -\tfrac{\pi}{24L} 
  \left(\tfrac{1}{2} - 24 \Delta\right) + \mathcal{O}(L^{-2}), 
\end{equation}
where a conformal dimension $\Delta$ takes either $0$ or $\frac{1}{2}$ in the NS sector and $\frac{1}{16}$ in the R sector.

%%%%%%%%%%%%%%%%%%%%%%%%%%%%%%%%%%%%%%%
%%%%%%%%%%%%%%%%%%%%%%%%%%%%%%%%%%%%%%%
\subsection{Scattering theory of the SSG model}
Supersymmetric solitons are described by non-commuting symbols $A_{a_j a_{j+1}}^{\epsilon_j}$. A superscript represents a soliton charge $\epsilon_j \in \{\pm\}$, a set of subscripts represents RSOS indices $a_j, a_{j+1} \in \{0,\pm 1\}$ with an adjacency condition $|a_j - a_{j+1}| = 1$. 

\subsubsection{Bulk $S$-matrix}
Corresponding to soliton-soliton scattering, the following commutation relations are obtained: 
\begin{equation}
 A_{ab}^{\epsilon_1}(\theta_1) A_{bc}^{\epsilon_2}(\theta_2) 
  = \sum_{\epsilon_1', \epsilon_2'} \sum_{d} 
  S^{\epsilon_1 \epsilon_2}_{\epsilon_1' \epsilon_2'}|^{ac}_{bd} (\theta_1 - \theta_2) 
  A_{ad}^{\epsilon_2'}(\theta_2) A_{dc}^{\epsilon_1'}(\theta_1) 
\end{equation}
with a parameter $\theta_j$ as rapidity of a supersymmetric soliton. 

As a known fact, the $S$-matrix of the SSG model is decomposed into a tensor product of the SG part and the RSOS part: 
%\begin{equation}
% S_{\rm SSG}(\theta) = S_{\rm SG}(\theta) \otimes S_{\rm RSOS}(\theta),  
%\end{equation}
\begin{equation}
 S^{\epsilon_1 \epsilon_2}_{\epsilon_1' \epsilon_2'}|^{ac}_{bd} (\theta) 
 =
  S^{\epsilon_1 \epsilon_2}_{\epsilon_1' \epsilon_2'}(\theta)
  \times
  S^{ac}_{bd}(\theta). 
\end{equation}
As a result of integrability, each $S$-matrix satisfies the Yang-Baxter equation: 
\begin{align}
 &S^{\epsilon_1 \epsilon_2}_{\epsilon_1' \epsilon_2'}(\theta_1 - \theta_2)
 S^{\epsilon_1' \epsilon_3}_{\epsilon_1'' \epsilon_3'}(\theta_1 - \theta_3)
 S^{\epsilon_2' \epsilon_3'}_{\epsilon_2'' \epsilon_3''}(\theta_2 - \theta_3)
 =
 S^{\epsilon_2 \epsilon_3}_{\epsilon_2' \epsilon_3'}(\theta_2 - \theta_3)
 S^{\epsilon_1 \epsilon_3'}_{\epsilon_1' \epsilon_3''}(\theta_1 - \theta_3)
 S^{\epsilon_1' \epsilon_2'}_{\epsilon_1'' \epsilon_2''}(\theta_1 - \theta_2), 
 \\
 &S^{ac}_{bg}(\theta_1 - \theta_2)
 S^{gd}_{ce}(\theta_1 - \theta_3)
 S^{ae}_{gf}(\theta_2 - \theta_3)
 =
 S^{bd}_{cg'}(\theta_2 - \theta_3)
 S^{ag'}_{bf}(\theta_1 - \theta_3)
 S^{fd}_{g'e}(\theta_1 - \theta_2), 
\end{align}
from which the exact $S$-matrix is derived. 

A solution to the SG part has been obtained in \cite{bib:ZZ79, bib:Z77}: 
\begin{align}
%\begin{split}
 &S_{\epsilon \epsilon}^{\epsilon \epsilon}(\theta) = S(\theta), \\
 &S_{\epsilon -\epsilon}^{\epsilon -\epsilon}(\theta) = \frac{\sinh \lambda\theta}{\sinh \lambda(i\pi - \theta)} S(\theta), \quad
 S_{\epsilon -\epsilon}^{-\epsilon \epsilon}(\theta) = i \frac{\sin \pi\lambda}{\sinh \lambda(i\pi - \theta)} S(\theta), 
%S_{\rm SG} (\theta) = 
%\begin{pmatrix}
%S(\theta) & 0 & 0 & 0 \\
%0 & S_T(\theta) & S_R(\theta) & 0 \\
%0 & S_R(\theta) & S_T(\theta) & 0 \\
%0 & 0 & 0 & S(\theta)
%\end{pmatrix}, 
%\end{split}
\end{align}
where $\epsilon \in \{\pm\}$, 
and found that it is closely related to the $R$-matrix of the six-vertex model. The overall factor $S(\theta)$ is obtained by setting $u=i\theta$: 
\begin{align} 
 S(\theta) 
 &= -\prod_{l=1}^{\infty} 
%  \left[
 \frac{\Gamma(2(l-1)\lambda - \frac{\lambda u}{\pi}) \Gamma (2l\lambda +1 -\frac{\lambda u}{\pi})}
 {\Gamma ((2l-1)\lambda - \frac{\lambda u}{\pi}) \Gamma ((2l-1)\lambda + 1 - \frac{\lambda u}{\pi})}/(u\rightarrow -u) \label{SGamp}
 \\
 &= \exp\left[
       i \int_0^{\infty} \frac{dt}{t} \frac{\sin \frac{\theta t}{\pi} \sinh(\frac{1}{\lambda} - 1)\frac{t}{2}}{\cosh\frac{t}{2} \sinh\frac{t}{2\lambda}}
       \right]. \label{SGamp_int}
\end{align}
A parameter $\lambda$ is determined by a coupling constant $\beta$ via $\lambda = \frac{2\pi}{\beta^2} - \frac{1}{2}$ \cite{bib:ANS07}. 

A solution to the RSOS part is also obtained in \cite{bib:GZ94, bib:A91, bib:AK96, bib:NA02} as 
\begin{equation} \label{RSOSamp*}
S^{ac}_{bd}(\theta)
 = X^{ac}_{bd}(\theta) K(\theta), 
\end{equation}
where 
\begin{equation}
\begin{split}
 &X^{\sigma \sigma}_{0 0} (\theta) 
 = 2^{(i\pi - \theta)/2\pi i} \cos\left(\frac{\theta}{4i} - \frac{\pi}{4}\right), 
 \qquad
 X^{0 0}_{\sigma \sigma} (\theta) 
 = 2^{\theta/2\pi i} \cos\left(\frac{\theta}{4i}\right), 
 \\
 &X^{\sigma -\sigma}_{0 0} (\theta) 
 = 2^{(i\pi - \theta)/2\pi i} \cos\left(\frac{\theta}{4i} + \frac{\pi}{4}\right), 
 \qquad
 X^{0 0}_{\sigma -\sigma} (\theta) 
 = 2^{\theta/2\pi i} \cos\left(\frac{\theta}{4i} - \frac{\pi}{2}\right), 
\end{split}
\end{equation}
with $\sigma\in\{\pm 1\}$. 
The overall factor $K(\theta)$ is given by 
\begin{align} 
 K(\theta) 
 &= \frac{1}{\sqrt{\pi}}
  \prod_{k-1}^{\infty} \frac{\Gamma (k-\frac{1}{2}+\frac{\theta}{2\pi i}) \Gamma (k-\frac{\theta}{2\pi i})}
  {\Gamma(k+\frac{1}{2}-\frac{\theta}{2\pi i}) \Gamma (k+\frac{\theta}{2\pi i})} \label{RSOSamp}
 \\
 &= \frac{-i}{\sqrt{2} \sinh\frac{\theta - i\pi}{4}}
 \exp\left[
 i \int_{0}^{\infty} \frac{dt}{t}
  \frac{\sin \frac{\theta t}{\pi} \sinh \frac{3t}{2}}{\sinh 2t \cosh\frac{t}{2}} 
 \right]. \label{RSOSamp_int}
\end{align}
%This is realized, up to the twist coming from the first term, by taking $p = \lambda^{-1} = 4$ in the function $S(\theta)$ of the SG $S$-matrix (\ref{SGamp}). 

%Here let us remark that the $S$-matrix of the SG part has poles in the physical strip $\theta \in [0, \frac{\pi}{2})$ at $\theta = $, while the RSOS part does not (See (\ref{SGamp}) and (\ref{RSOSamp}). Noting that $\theta$ in the $S$-matrix denotes the difference of rapidities of two SSG solitons, those two particles whose rapidity difference is given by $\theta = $ form a bound state. 
%The bootstrap approach developed in \cite{bib:K80} provides a systematic way to compute the mass spectrum and $S$-matrices of bound states. 

%%%%%%%%%%%%%%%%%%%%%%%%%%%%%%%%%%%%%%%%%%%%%%%
\subsubsection{Boundary $S$-matrix}
In a finite and non-periodic system, a soliton is reflected at a boundary with a reflection amplitude obtained from the following algebraic relations: 
\begin{equation}
 A_{ab}^{\epsilon}(\theta) B 
  =
  \sum_c \sum_{\epsilon'} R_{\epsilon'}^{\epsilon}|_{ac}^b
  A_{bc}^{\epsilon'}(-\theta) B. 
\end{equation}
Here we used a boundary creation operator $B$. 

As in the case of the bulk $S$-matrix, the reflection matrix of the SSG model is also written by a tensor product of the SG part and the RSOS part: 
\begin{equation}
 R_{\epsilon'}^{\epsilon}|_{ab}^c(\theta) = R_{\epsilon'}^{\epsilon}(\theta)
  \times R_{ab}^c(\theta). 
\end{equation}
Each reflection matrix of integrable boundaries like the Dirichlet boundary conditions independently satisfies the reflection relation: 
\begin{align}
 &S^{\epsilon_1 \epsilon_2}_{\epsilon_2' \epsilon_1'} (\theta_1 - \theta_2)
 R^{\epsilon_2'}_{\epsilon_2''} (\theta_2)
 S^{\epsilon_2'' \epsilon_1'}_{\epsilon_1'' \epsilon_2'''} (\theta_1 + \theta_2)
 R^{\epsilon_1''}_{\epsilon_1'''} (\theta_1)
 =
 R^{\epsilon_1}_{\epsilon_1'} (\theta_1)
 S^{\epsilon_1' \epsilon_2}_{\epsilon_2'' \epsilon_1'} (-\theta_1 - \theta_2)
 R^{\epsilon_2'}_{\epsilon_2''} (\theta_2)
 S^{\epsilon_2' \epsilon_1''}_{\epsilon_1''' \epsilon_2'''} (-\theta_1 + \theta_2), \label{refrel_sg}
 \\
 &S^{ac}_{bf} (\theta_1 - \theta_2)
 R_{ag}^f (\theta_2)
 S^{gc}_{fd} (\theta_1 + \theta_2)
 R_{ge}^d (\theta_1)
 =
 R_{af'}^b (\theta_1)
 S^{f'c}_{bg'} (-\theta_1 - \theta_2)
 R_{f'e}^{g'} (\theta_2)
 S^{ec}_{g'd} (-\theta_1 + \theta_2). \label{refrel_susy}
\end{align}

A solution of the reflection relation has only diagonal elements under Dirichlet boundaries as obtained in \cite{bib:GZ94}: 
\begin{equation}
 R_{\pm}^{\pm}(\theta) = \cos(\xi \pm \lambda u) R_0(u) \frac{\sigma(\theta, \xi)}{\cos \xi}, 
%R_{\rm SG}(\theta)=
% \begin{pmatrix}
%  \cos(\xi+\lambda u) & 0 \\
%  0 & \cos(\xi-\lambda u)
% \end{pmatrix}
% R_0(u) \frac{\sigma(\theta, \xi)}{\cos \xi}, 
\end{equation}
where $R_0(u)$  is given by 
\begin{equation} 
 R_0(u)=
 \prod_{l=1}^{\infty} \left[
 \frac{\Gamma(4l\lambda - \frac{2\lambda u}{\pi}) \Gamma (4\lambda (l-1) + 1 -\frac{2\lambda u}{\pi})}
 {\Gamma((4l-3) \lambda - \frac{2\lambda u}{\pi}) \Gamma((4l-1)\lambda +1 -\frac{2\lambda u}{\pi})}
 / (u\rightarrow -u)
 \right]. 
\end{equation}
The overall factor $\sigma(\theta,\xi)$ is written by $\Gamma$-functions: 
\begin{equation}
 \sigma(\theta, \xi) = \frac{\cos\xi}{\cos(\xi+\lambda u)}
 \prod_{l=1}^{\infty}\left[
 \frac{\Gamma(\frac{1}{2} + \frac{\xi}{\pi} + (2l-1)\lambda -\frac{\lambda u}{\pi}) \Gamma (\frac{1}{2} - \frac{\xi}{\pi} + (2l-1)\lambda -\frac{\lambda u}{\pi})}
 {\Gamma(\frac{1}{2} - \frac{\xi}{\pi} + (2l-2)\lambda - \frac{\lambda u}{\pi}) \Gamma(\frac{1}{2} + \frac{\xi}{\pi} + 2l\lambda - \frac{\lambda u}{\pi})}
 /(u\rightarrow -u)
 \right]. 
\end{equation}
Treating $\Gamma$-functions with negative real parts separately, one can write a soliton reflection $R_{+}^{+}(\theta)$ by integral forms \cite{bib:SS95, bib:FS94}: 
\begin{equation}
\begin{split}
 &\frac{R_{+}^{+}(\theta)}{R_0(\theta)} = R^+_1(\theta) + R_2(\theta), \\
 &R^+_1(\theta)
  = \exp\left[ i \int_0^{\infty} \frac{dt}{t}
  \left(
   \frac{\sinh(1 - \frac{2\xi}{\pi\lambda})\frac{t}{2}}{2 \sinh\frac{t}{2\lambda} \cosh\frac{t}{2}}
   + \frac{\sinh(\frac{\xi}{\pi} - \lfloor \frac{\xi}{\pi} - \frac{1}{2} \rfloor - 1) \frac{t}{\lambda}}{\sinh\frac{t}{2\lambda}}
  \right) \sin \frac{\theta t}{\pi}
  \right], \\
 &R_2(\theta)
 = \exp\left[
 i \int_{0}^{\infty} \frac{dt}{t}
 \frac{\sinh\frac{3t}{4} \sinh(\frac{1}{\lambda} - 1)\frac{t}{4}}{\sinh t \sinh\frac{t}{4\lambda}} 
 \right]. 
\end{split}
\end{equation}
On the other hand, an anti-soliton reflection is given by \cite{bib:SS95} 
\begin{equation} \label{SG-antiref}
\begin{split}
 &\frac{R_{-}^{-}(\theta)}{R_0(\theta)} = R^-_1(\theta) + R_2(\theta), \\
 &R^-_1(\theta)
  = \exp\left[ i \int_0^{\infty} \frac{dt}{t}
   \frac{\sinh(1 - \frac{2\xi}{\pi\lambda})\frac{t}{2}}{2 \sinh\frac{t}{2\lambda} \cosh\frac{t}{2}}
   \sin \frac{\theta t}{\pi}
  \right]. 
\end{split}
\end{equation}
A boundary parameter $\xi$ is connected to field values at boundaries through $\xi_{\pm} = \frac{2\pi}{\beta} \Phi_{\pm}$ \cite{bib:ANS07}.

The RSOS part of the reflection relation has been solved. Different solutions were obtained for two sectors of the superconformal field theory \cite{bib:AK96, bib:NA02}. For the NS sector, a solution is given by 
\begin{align} 
%\begin{split}
 &R_{\sigma \sigma}^0(\theta; \xi)
 = P(\theta; \xi), \label{ramp_NS1}
 \\
 &R_{00}^{\pm 1}(\theta; \xi)
 = \left(\cos\frac{\xi}{2} \pm i \sinh\frac{\theta}{2}\right)
 2^{i\theta/\pi} K(\theta - i\xi) K(\theta + i\xi) P(\theta; \xi), \label{ramp_NS2}
%\end{split}
\end{align}
where 
\begin{align}
 &P(\theta, \xi) = \frac{\sin\xi - i\sinh\theta}{\sin\xi + i\sinh\theta}
 P_0(\theta), 
 \\
 &P_0(\theta) = \prod_{k=1}^{\infty} \left[
 \frac{\Gamma (k - \frac{\theta}{2\pi i}) \Gamma(k - \frac{\theta}{2\pi i})}
 {\Gamma(k-\frac{1}{4} - \frac{\theta}{2\pi i}) \Gamma(k+\frac{1}{4} - \frac{\theta}{2\pi i})}
 /(\theta \rightarrow -\theta)
 \right]
 \\
 &\hspace{8mm}= \exp\left(
 -\frac{\theta}{2\pi} \ln 2 + \frac{1}{8} \int_0^\infty \frac{dt}{t}
 \frac{\sin\frac{2\theta t}{\pi}}{\cosh^2 t \cosh^2 \frac{t}{2}} 
 \right). \label{ref_susy_int}
\end{align}
Thus only diagonal matrix elements are non-zero in the reflection matrix of the NS sector. 

On the other hand, a solution to the R sector is obtained as 
\begin{align} 
%\begin{split}
 &R_{\sigma \sigma}^0(\theta; \xi) = \cos\frac{\xi}{2} K(\theta - i\xi) K(\theta + i\xi) P(\theta; \xi), \label{ramp_R1}
 \\
 &R_{-\sigma \sigma}^0(\theta; \xi) = -ir^{\sigma} \sinh\frac{\theta}{2} K(\theta - i\xi) K(\theta + i\xi) P(\theta; \xi), \label{ramp_R2}
 \\
 &R_{00}^{\sigma}(\theta; \xi) = 2^{i\theta/\pi} P(\theta; \xi). \label{ramp_R3}
%\end{split}
\end{align}
Unlike the NS sector, the reflection matrix of the R sector has non-diagonal elements $R^0_{-\sigma \sigma}(\theta; \xi)$. The matrix (\ref{ramp_R1})-(\ref{ramp_R3}) is block diagonal whose non-diagonal subspace is diagonalized with eigenvalues $\cos\frac{\xi}{2} \pm i\sinh\frac{\theta}{2}$, which are clearly the same as (\ref{ramp_NS2}) up to a factor $2^{i\theta/\pi}$ which can be removed by a similarity transformation.

%The other RSOS reflection factors can be obtained from the boundary bootstrap principle \cite{bib:GZ94, bib:AK96} (we will show it in the next subsection), as we expect the ground state is non-degenerate and must have $\frac{1}{2}$ as its RSOS index. 
%Here we gave the expression of the RSOS reflection matrix only for $R_{\frac{1}{2} \frac{1}{2}}^a(\theta)$. 
%As the ground state is non-degenerate, we assume the RSOS index for the ground state is $\frac{1}{2}$. Therefore, the other RSOS reflection factors are expected to be obtained for the reflection on an excited boundary. For this reason, the expression for $R_{ab}^{\frac{1}{2}}(\theta)$ ($a,b=0,1$) will be given in the context of the boundary bootstrap approach in the next subsection. 

%%%%%%%%%%%%%%%%%%%%%%%%%5
\subsection{Light-cone regularization}
The light-cone regularization of a quantum field theory is achieved by discretizing a light-cone with a lattice spacing $a$ \cite{bib:DV87, bib:V89, bib:V90}. A trajectory of each particle then forms a two-dimensional lattice. Particle scattering occurs only at a vertex with an amplitude properly scaled from the original quantum field theory. If one works on an integrable quantum field theory in which an exact $S$-matrix can be derived, one may expect that an amplitude assigned on each vertex of a regularized light-cone can be identified with a Boltzmann weights of an integrable lattice model. Indeed, it was found that a light-cone of the lattice-regularized SG (LSG) model is obtained as a $90$-degree rotation of the six-vertex model. In the case of the SSG model, the light-cone regularization leads to the $19$-vertex model \cite{bib:ANS07, bib:HRS07}. 
% figure
%\begin{figure}
%\begin{center}
% \includegraphics[scale=0.5]{light-cone}
% \caption{Light-cone lattice with a lattice spacing $a$. Inhomogeneity in a corresponding spin chain is regarded as rapidity of a rigth-mover or a left-mover.}
% \label{fig:light-cone}
%\end{center}
%\end{figure}
%

This fact leads us to discuss the SSG model on an integrable lattice system, on which the transfer matrix method has been developed intensively. As a well-known fact, a transfer matrix of the spin-$1$ Zamolodchikov-Fateev model is defined on the $19$-vertex model \cite{bib:ZF80}. Since time development of a SSG state is also defined on the $19$-vertex model with inhomogeneities $\pm \Theta$ corresponding to rapidity of a right or left mover, a transfer matrix of the Zamolodchikov-Fateev spin chain with inhomogeneity describes time development of an SSG state. 

The spin-$1$ Zamolodchikov-Fateev model is defined by the following Hamiltonian: 
\begin{equation} \label{hamiltonian}
%\begin{split}
 \cH=\sum_{j=1}^{N-1} \left[
 T_j - (T_j)^2 - 2\sin^2\gamma \;( T_j^z + (S_j^z)^2 + (S_{j+1}^z)^2 - (T_j^z)^2 )
 + 4\sin^2\tfrac{\gamma}{2} \; (T_j^{\bot} T_j^z + T_j^z T_j^{\bot})
 \right]
 + \cH_{\rm B},
%\end{split}
\end{equation}
where 
\begin{equation}
 T_j = \vec{S}_j \cdot \vec{S}_{j+1}, \qquad
  T^{\bot}_j = S^x_j S^x_{j+1} + S^y_j S^y_{j+1}, \qquad
  T^z_j = S^z_j S^z_{j+1}. 
\end{equation}
Operators $S_j^{\alpha}$ ($\alpha \in \{x,y,z\}$) are three-dimensional $SU(2)$ spin operators which nontrivially act on the $j$th space of $N$-fold tensor product of three-dimensional vector space. 
A parameter $\gamma$ is an anisotropy parameter which determines a coupling constant of the SSG model in the scaling limit by $\beta^2 = 4(\pi - 2\gamma)$. Since $\beta^2$ in the SSG model takes a real value, an allowed value for $\gamma$ is less than $\frac{\pi}{2}$. In a spin chain realm, a parameter $\gamma$ in this condition makes the system gapless. 

Corresponding to Dirichlet boundaries, which do not change a soliton charge, the boundary Hamiltonian $\cH_{\rm B}$ is given by diagonal operators: 
\begin{equation} \label{bhamiltonian}
 \cH_{\rm B} = h_1(H_-) S_1^z + h_2(H_-)(S_1^z)^2
  + h_1(H_+) S_N^z +  h_2(H_+) (S_N^z)^2, 
\end{equation}
where two types of boundary fields are connected by a common parameter $H$ as 
\begin{align}
 &h_1(H) = \tfrac{1}{2} \sin 2\gamma \left(\cot\tfrac{\gamma H}{2} + \cot\tfrac{\gamma(H+2)}{2}\right), 
 \\
 &h_2(H) = \tfrac{1}{2} \sin 2\gamma \left(-\cot\tfrac{\gamma H}{2} + \cot\tfrac{\gamma (H+2)}{2}\right). 
\end{align}
% figure
\begin{figure}
\begin{center}
 \includegraphics[scale=0.75]{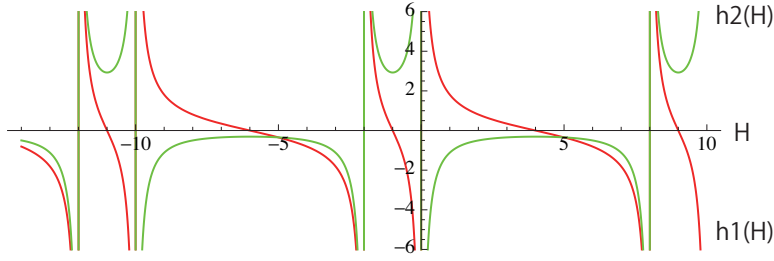}
 \caption{Boundary magnetic fields as functions of a boundary parameter $H$. Anisotropy is taken to be $\gamma = \frac{\pi}{5}$.}
 \label{fig:boundary_fields}
\end{center}
\end{figure}
These boundary fields are $\frac{2\pi}{\gamma}$-periodic functions with respect to $H$ (Figure \ref{fig:boundary_fields}). Each periodicity cell apparently consists of two domains $[-2+\frac{2\pi n}{\gamma}, \frac{2\pi n}{\gamma}]$ (domain NS) and $[\frac{2\pi (n-1)}{\gamma}, -2+\frac{2\pi n}{\gamma}]$ ($n \in \mathbb{Z}$) (domain R), and therefore, we expect different behaviors for the corresponding quantum field theory obtained after taking the scaling limit. 
For instance, this system has symmetries with respect to boundary magnetic fields and they have different meanings in each domain; We first obtain $\frac{2\pi}{\gamma}$-periodicity for both domains NS and R. In contrast, a symmetry $H \leftrightarrow -H-\frac{2\pi}{\gamma}-2$ is understood as a $S^z \leftrightarrow -S^z$-symmetry in domain NS, while $H \leftrightarrow -H-2$ gives the same symmetry but for domain $R$. 

The transfer matrix of the Zamolodchikov-Fateev spin chain is obtained from the $19$-vertex model by taking a trace over an auxiliary space. If one inserts inhomogeneities corresponding to rapidities of right and left movers, the transfer matrix of the LSSG model is given by a set of the following operators: 
\begin{equation} \label{T-for-SSG}
 T_{\rm R} = {\rm tr}_0 [K_+(\theta) T(\theta) K_-(\theta) \widehat{T}(\theta)]_{\theta = \Theta}, 
  \qquad
  T_{\rm L} = {\rm tr}_0 [K_+(\theta) T(\theta) K_-(\theta) \widehat{T}(\theta)]_{\theta = -\Theta}, 
\end{equation}
where 
\begin{equation}
\begin{split}
 &T(\theta) = R_{0,2N}(\tfrac{\gamma}{\pi}(\theta-\Theta)) R_{0,2N-1}(\tfrac{\gamma}{\pi}(\theta+\Theta)) \dots R_{02}(\tfrac{\gamma}{\pi}(\theta-\Theta)) R_{01}(\tfrac{\gamma}{\pi}(\theta+\Theta)), 
\\
 &\widehat{T}(\theta) = R_{10}(\tfrac{\gamma}{\pi}(\theta+i\pi+\Theta)) R_{20}(\tfrac{\gamma}{\pi}(\theta+i\pi-\Theta)) \dots R_{2N-1,0}(\tfrac{\gamma}{\pi}(\theta+i\pi+\Theta)) R_{2N,0}(\tfrac{\gamma}{\pi}(\theta+i\pi-\Theta))
\end{split}
\end{equation}
and $R_{ij}(\theta)$ is the $R$-matrix of the $19$-vertex model \cite{bib:ZF80} constructed from that of the six-vertex model through the fusion procedure \cite{bib:KRS81}. 
Boundary reflection is described by a reflection matrix $K_{\pm}(\theta)$ \cite{bib:S88, bib:FLU02} obtained as a diagonal solution of the reflection relation (\ref{refrel_sg}) and (\ref{refrel_susy}). 
Let us note that Hamiltonian and total momentum is obtained from the transfer matrix: 
\begin{equation}
 \cH = \frac{i\gamma}{2\pi a}[\ln T_{\rm R} + \ln T_{\rm L}], 
  \qquad
  \cP = \frac{i\gamma}{2\pi a}[\ln T_{\rm R} - \ln T_{\rm L}]. 
\end{equation}

\ifx10
\end{document}
\fi

\ifx10
\documentclass{article}

%#############
%packages 
%#############
%\usepackage[utf8x]{inputenc}
\usepackage{amsmath, amsthm, amsfonts, amssymb}
\usepackage{bm}
\usepackage{a4wide}
\usepackage{graphicx} % for graphics
\usepackage{color}
\usepackage{euscript}
\usepackage{fancybox}

%#############
%newtheorems
%############
%\renewcommand{\thefootnote}{\fnsymbol{footnote}}
%\theoremstyle{plain}
%\newtheorem{thm}{Theorem}
%\newtheorem{claim}{Claim}
%\newtheorem{propn}{\bfseries Proposition}
%\newtheorem{propnn}{Proposition}
%\renewcommand{\thepropnn}{\arabic{propn}\alph{propnn}}
%\newtheorem{lem}{\bfseries Lemma}
%\newtheorem{cor}{Corollary}
%\newtheorem{conj}{\bfseries Conjecture}
%\newtheorem{defn}{Definition}
%\theoremstyle{remark}
%\newtheorem{rem}{Remark}
%\newtheorem{df}{Definition}
%\newtheorem{th1}[df]{Theorem}
%\newtheorem{lem}[df]{Lemma}
%\newtheorem{conj}[df]{Conjecture}
%\newtheorem{prop}[df]{Proposition}
%\newtheorem{cor}[df]{Corollary}
%\newtheorem{lem2}{Lemma}[section]
%\newtheorem{prop2}{Proposition}[section]
%\newtheorem{cor2}{Corollary}[section]
%\newtheorem{df2}{Definition}[section]

%##############
%newcommands
%#############
\newcommand{\cO}{\mathcal{O}}
\newcommand{\cA}{\mathcal{A}}
\newcommand{\cB}{\mathcal{B}}
\newcommand{\cC}{\mathcal{C}}
\newcommand{\cD}{\mathcal{D}}	
\newcommand{\cF}{\mathcal{F}}
\newcommand{\sP}{\mathsf{P}}
\newcommand{\sB}{\mathsf{B}}
\newcommand{\sY}{\mathsf{Y}}
\newcommand{\sy}{\mathsf{y}}
\newcommand{\sG}{\mathsf{G}}
\newcommand{\sm}{\mathsf{m}}
\newcommand{\sg}{\mathsf{g}}
\newcommand{\cT}{\mathcal{T}}
\newcommand{\cU}{\mathcal{U}}
\newcommand{\cH}{\mathcal{H}}
\newcommand{\cL}{\mathcal{L}}
\newcommand{\bC}{\mathbb{C}}
\newcommand{\txi}{\tilde{\xi}}
\newcommand{\tlambda}{\tilde{\lambda}}
\newcommand{\cI}{\mathcal{I}}
\newcommand{\cR}{\mathcal{R}}
\newcommand{\cS}{\mathcal{S}}
\newcommand{\cN}{\mathcal{N}}
\newcommand{\cM}{\mathcal{M}}
\newcommand{\cG}{\mathcal{G}}
\newcommand{\cK}{\mathcal{K}}
\newcommand{\bZ}{\mathbb{Z}}
\newcommand{\bR}{\mathbb{R}}
\newcommand{\ve}{\varepsilon}
\newcommand{\eH}{\EuScript{H}}
\newcommand{\cP}{\mathcal{P}}
\newcommand{\hT}{\hat{T}}
\newcommand{\htheta}{\hat{\theta}}
\newcommand{\sh}{\,\mathrm{sh}}
\newcommand{\ch}{\,\mathrm{ch}}
\newcommand{\fA}{\mathfrak{A}}
\newcommand{\fa}{\mathfrak{a}}
\newcommand{\fC}{\mathfrak{C}}
\newcommand{\fc}{\mathfrak{c}}
\newcommand{\fD}{\mathfrak{D}}
\newcommand{\fd}{\mathfrak{d}}
\newcommand{\fS}{\mathfrak{S}}
\newcommand{\ba}{\bar{a}}
\newcommand{\bb}{\bar{b}}
\newcommand{\dpsi}{\psi^\dag}
\newcommand{\pa}{a^{(+)}}
\newcommand{\dpa}{a^{(+)\dag}}
\newcommand{\ma}{a^{(-)}}
\newcommand{\dma}{a^{(-)\dag}}
\newcommand{\da}{a^\dag}
\newcommand{\ppsi}{\psi_+}
\newcommand{\dppsi}{\psi_+^\dag}
\newcommand{\mpsi}{\psi_-}
\newcommand{\dmpsi}{\psi_-^\dag}
\newcommand{\vphi}{\varphi}
\newcommand{\pS}{S^+}
\newcommand{\mS}{S^-}
\newcommand{\zS}{S^z}
%%%
%\newcommand{\bm}[1]{\mbox{\boldmath$#1$}}
%\newcommand{\be}{\begin{equation}}
%\newcommand{\ee}{\end{equation}}
%\newcommand{\bea}{\begin{eqnarray}}
%\newcommand{\eea}{\end{eqnarray}}
%\newcommand{\non}{\nonumber}
%\newcommand{\ra}{\rangle}
%\newcommand{\la}{\langle}
%\newcommand{\lam}{\lambda} 
%\newcommand{\Lam}{\Lambda} 
%\newcommand{\tht}{\theta} 
%\newcommand{\al}{\alpha} 
%\newcommand{\bt}{\beta} 
%\newcommand{\gm}{\gamma} 
%\newcommand{\dt}{\delta} 

\begin{document}
\fi

\section{Nonlinear integral equations} \label{sec:nlie}
\subsection{$T$-functions and auxiliary functions}
An eigenvalue of the transfer matrix of the LSSG model with Dirichlet boundary conditions (\ref{T-for-SSG}) was found to be written as a function of Bethe roots \cite{bib:ANS07}: 
\begin{equation} \label{TQ2}
 T_2(\theta) = \lambda_1(\theta) + \lambda_2(\theta) + \lambda_3(\theta), 
\end{equation}
where
\begin{equation}
\begin{split}
 &\lambda_1(\theta) = \sinh\tfrac{\gamma}{\pi}(2\theta - 2i\pi)
 B_-(\theta - \tfrac{i\pi}{2}) B_-(\theta + \tfrac{i\pi}{2})
 \phi(\theta - \tfrac{3i\pi}{2}) \phi(\theta - \tfrac{i\pi}{2}) 
 \frac{Q(\theta + \frac{3i\pi}{2})}{Q(\theta - \frac{i\pi}{2})}, 
\\
 &\lambda_2(\theta) = \sinh\tfrac{\gamma}{\pi}(2\theta)
 B_+(\theta - \tfrac{i\pi}{2}) B_-(\theta + \tfrac{i\pi}{2})
 \phi(\theta - \tfrac{i\pi}{2}) \phi(\theta + \tfrac{i\pi}{2})
 \frac{Q(\theta + \frac{3i\pi}{2}) Q(\theta - \frac{3i\pi}{2})}{Q(\theta - \frac{i\pi}{2})Q(\theta + \frac{i\pi}{2})}, 
\\
 &\lambda_3(\theta) = \sinh\tfrac{\gamma}{\pi}(2\theta + 2i\pi)
 B_+(\theta - \tfrac{i\pi}{2}) B_+(\theta + \tfrac{i\pi}{2})
 \phi(\theta + \tfrac{3i\pi}{2}) \phi(\theta + \tfrac{i\pi}{2})
 \frac{Q(\theta - \frac{3i\pi}{2})}{Q(\theta + \frac{i\pi}{2})}. 
\end{split}
\end{equation}
A function $\phi(\theta)$ gives a phase shift given by 
\begin{equation}
 \phi(\theta) = \sinh^N\tfrac{\gamma}{\pi}(\theta - \Theta) \sinh^N\tfrac{\gamma}{\pi}(\theta + \Theta)
\end{equation}
and functions $B_{\pm}(\theta)$ come from boundary effects which depend on boundary parameters as 
\begin{equation}
 B_{\pm}(\theta)
  =
  \sinh\tfrac{\gamma}{\pi}(\theta \pm \tfrac{i\pi H_+}{2}) \sinh\tfrac{\gamma}{\pi}(\theta \pm \tfrac{i\pi H_-}{2}). 
\end{equation}
Bethe-root dependence shows up through a function $Q(\theta)$: 
\begin{equation}
 Q(\theta) = \prod_{j=1}^M \sinh\tfrac{\gamma}{\pi}(\theta - \theta_j) 
  \sinh\tfrac{\gamma}{\pi}(\theta + \theta_j), 
\end{equation}
where $\theta_j$ is a Bethe root. 

Another transfer matrix is defined for the LSSG model whose eigenvalue is given by 
\begin{equation} \label{TQ1}
 T_1(\theta) = l_1(\theta) + l_2(\theta), 
\end{equation}
where 
\begin{equation}
\begin{split}
 &l_1(\theta) = \sinh\tfrac{\gamma}{\pi}(2\theta + i\pi)
  B_+(\theta) \phi(\theta + i\pi) \frac{Q(\theta - i\pi)}{Q(\theta)}, 
\\
 &l_2(\theta) = \sinh\tfrac{\gamma}{\pi}(2\theta - i\pi)
 B_-(\theta) \phi(\theta - i\pi) \frac{Q(\theta + i\pi)}{Q(\theta)}. 
\end{split}
\end{equation}
Both functions $T_1(\theta)$ and $T_2(\theta)$ are symmetric with respect to a sign of a Bethe root $\theta_j \leftrightarrow -\theta_j$, {\it i.e.} Bethe roots symmetrically locate to the origin of a complex plane. 

Auxiliary functions are defined from $T_2(\theta)$ as 
\begin{equation}
\begin{split}
 &b(\theta) = \frac{\lambda_1(\theta) + \lambda_2(\theta)}{\lambda_3(\theta)}, 
 \qquad 
 \bar{b}(\theta) = \frac{\lambda_3(\theta) + \lambda_2(\theta)}{\lambda_1(\theta)} = b(-\theta), 
\\
 &B(\theta) = 1 + b(\theta),
 \qquad 
 \bar{B}(\theta) = 1 + \bar{b}(\theta). 
\end{split}
\end{equation}
Similarly, we define 
\begin{equation}
\begin{split}
 &a(\theta) = \frac{l_2(\theta)}{l_1(\theta)},
 \qquad
 \bar{a}(\theta) = \frac{l_1(\theta)}{l_2(\theta)} = a(-\theta),
\\
 &A(\theta) = 1 + a(\theta),
 \qquad
 \bar{A}(\theta) = 1 + \bar{a}(\theta). 
\end{split}
\end{equation}
The function $A(\theta)$ has zeros at positions of roots $\theta = \theta_k$, while $B(\theta)$ at positions which become string centers $\theta = \theta_k \pm \frac{i\pi}{2}$ in a large-volume limit \cite{bib:T82}. Therefore, $\ln b(\theta)$ and $\ln a(\theta)$ are interpreted as ``counting functions'' of real roots and two-string roots, respectively. 

Based on algebraic structure of integrable scattering theories, the $T$-system and the $Y$-systems have been developed \cite{bib:Z91, bib:RVT93}. These systems provide a systematic way to connect different types of $T$-functions, {\it e.g.}
\begin{equation} \label{fusion}
\begin{split}
 &T_1(\theta - \tfrac{i\pi}{2}) T_1(\theta + \tfrac{i\pi}{2})
 = f(\theta) + T_0(\theta) T_2(\theta), 
\\
 &T_0(\theta) = \sinh \tfrac{\gamma}{\pi}(2\theta), 
\end{split}
\end{equation}
and $Y$-functions {\it e.g.}
\begin{align} 
%\begin{split}
 &y(\theta) = \frac{T_0(\theta) T_2(\theta)}{f(\theta)}, \label{T-Yrel}
 \\
 &T_1(\theta - \tfrac{i\pi}{2}) T_1(\theta + \tfrac{i\pi}{2}) = f(\theta) Y(\theta), \label{T-Yrel2}
%\end{split}
\end{align}
where $Y(\theta) = 1 + y(\theta)$ and $f(\theta) = l_2(\theta - \tfrac{i\pi}{2}) l_1(\theta + \tfrac{i\pi}{2})$. 

%%%%%%%%%%%%%%%%%%%%%%%%%%%%%%
\subsection{Classification of roots and holes}
%In order to derive NLIEs for logarithms of auxiliary functions, we choose branch cuts as in Figure \ref{}. 
A logarithm of each function in auxiliary functions belongs to different Riemann surface depending on an imaginary value of a root or a hole, and therefore we classify roots in the following way: 
\begin{itemize}
\item Inner roots $c^{\rm IN}_j$ ($j \in \{1,\dots,M_{C^{\rm IN}}\}$) s.t. $|{\rm Im}\,c^{\rm IN}_j| \le \frac{\pi}{2} + \epsilon$ 
\item Close roots $c_j$ ($j \in \{1,\dots,M_C\}$) s.t. $\frac{\pi}{2} + \epsilon <|{\rm Im}\,c_j| < \frac{3\pi}{2}$ 
\item Wide roots $w_j$ ($j \in \{1,\dots,M_W\}$) s.t. $\frac{3\pi}{2} < |{\rm Im}\,w_j| \le \frac{\pi^2}{2\gamma}$ 
%\item Self-conjugate roots: $|{\rm Im}x_j| = \frac{\pi}{2}$ ($j=1,2,\dots,M_{SC}$)	
\end{itemize}
An infinitesimal $\epsilon$ is introduced for two-string roots to be classified into inner roots, {\it i.e.} it is chosen to be greater than root-deviations from two-string roots. Wide roots s.t. $|{\rm Im}\,w_j| = \frac{\pi^2}{2\gamma}$ are called self-conjugate roots, as their complex conjugates are themselves. Note that any complex roots appear in pairs with their complex conjugates except for real and self-conjugate ones. 

Quantization conditions are given by the following relations: 
\begin{align}
 &{\rm Im}\,\ln b(c_j^{{\rm IN}\uparrow} - \tfrac{i\pi}{2}) = 2\pi (I_{c_j^{{\rm IN}\uparrow}} - \tfrac{1}{2}), 
 %\qquad j \in \{1,\dots,M_{C^{\rm IN}}/2\}, 
 \\
 &{\rm Im}\,\ln b(c_j^{\uparrow} - \tfrac{i\pi}{2}) = 2\pi (I_{c_j^{\uparrow}} - \tfrac{1}{2}), 
 %\hspace{13mm} j \in \{1,\dots,M_C/2\}, 
 \\
 &{\rm Im}\,\ln b(w_j^{\uparrow} - \tfrac{i\pi}{2}) = 2\pi (I_{w_j^{\uparrow}} - \tfrac{1}{2}), 
 %\hspace{11mm} j \in \{1,\dots,M_W/2\}, 
\end{align}
where $I_{\theta_j^{\uparrow}}$ ($\theta \in \{c^{\rm IN}, c, w\}$) is a quantum number which takes an integer. Here we introduced a new notation $\theta_j^{\uparrow}$ for a root with a positive imaginary part. For simplicity, we call a shifted root $\tilde{\theta}^{\uparrow}_j = \theta^{\uparrow}_j - \frac{i\pi}{2}$ an effective roots. 
Similarly, quantization conditions for holes and type-$1$ holes are respectively given by 
\begin{align}
 &{\rm Im}\,\ln b(h_j) = 2\pi (I_{h_j} - \tfrac{1}{2}), 
 \hspace{12mm} j \in \{1,\dots,N_H\}, \label{quant_cond_h}
 \\
 &{\rm Im}\,\ln a(h_j) = 2\pi (I_{h_j} - \tfrac{1}{2}), 
 \qquad j \in \{1,\dots,N_1\}. 
\end{align}

In an increasing sequence of quantum numbers with respect to $j$, a root or a hole which makes $I_{\theta_j} < I_{\theta_{j-1}}$ is called a special object. Here we denote special roots $s_j$ and $s_j^R$ whose quantization condition is given by 
\begin{equation}
\begin{split}
 &{\rm Im} \ln b(\tilde{s}_j^{\uparrow}) = 2\pi (I_{s_j^{\uparrow}} - \tfrac{1}{2}), \qquad
 j \in \{1,\dots,N_S\}, 
 \\
 &{\rm Im} \ln a(s_j^{R \uparrow}) = 2\pi (I_{s_j^{R \uparrow}} - \tfrac{1}{2}), \qquad
 j \in \{1,\dots,N^{R}_S\}
\end{split}
\end{equation}
while a special hole and a type-$1$ special hole by $v_j$ and $v_j^{R}$, respectively, whose quantization conditions are given by 
\begin{equation}
\begin{split}
 &{\rm Im} \ln b(v_j) = 2\pi (I_{v_j} - \tfrac{1}{2}), \qquad
 j \in \{1,\dots,N_V\}, 
 \\
 &{\rm Im} \ln a(v^R_j) = 2\pi (I_{v^R_j} - \tfrac{1}{2}), \qquad
 j \in \{1,\dots,N^{R}_V\}
\end{split}
\end{equation}

%%%%%%%%%%%%%%%%%%%%%%%%%%%%%%%%%%%%
\subsection{Cauchy theorem for $T$-functions}
Derivation of NLIEs for the ground state of the LSSG model with Dirichlet boundaries has been closely discussed in \cite{bib:ANS07}. 
Here we derive NLIEs for an arbitrary excited state of the LSSG model. We do not assume string-like distribution of Bethe roots. 

% figure
\begin{figure}
\begin{center}
 \includegraphics[scale=0.75]{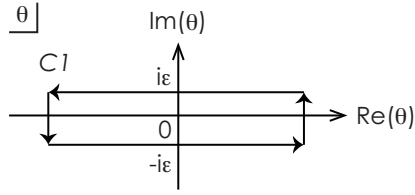}
 \caption{A contour $\mathcal{C}_1$ is taken to surround the real axis.}
 \label{fig:contour1}
\end{center}
\end{figure}
Nontrivial equations can be derived from analyticity structure of the $T$-functions. Since the function $T_2(\theta)$ is analytic and nonzero (ANZ) around the real axis of the complex plane except for the origin and positions of holes, we have the following equation as a result of the Cauchy theorem: 
\begin{equation} \label{CauchyT2}
\oint_{\cC_1} d\theta\; e^{ik\theta} [\ln T_2(\theta)]'' 
 = \frac{2\pi k}{1 - e^{-\pi k}}
 \Big(1 + \sum_{h_j\in \mathbb{R}} e^{ik h_j}\Big), 
\end{equation}
where a contour $\cC_1$ is taken as Figure \ref{fig:contour1}. This is an equation for $B(\theta)$, $\bar{B}(\theta)$, and $y(\theta)$ since $T_2(\theta)$ is expressed by the following two forms besides (\ref{T-Yrel}): 
\begin{align}
 T_2(\theta)
 &= t_+(\theta) \frac{Q(\theta - \frac{3i\pi}{2})}{Q(\theta + \frac{i\pi}{2})} B(\theta) \label{tplus}
 \\
 &= t_-(\theta) \frac{Q(\theta + \frac{3i\pi}{2})}{Q(\theta - \frac{i\pi}{2})} \bar{B}(\theta), \label{tminus}
\end{align}
where 
\begin{equation} \label{tpm}
 t_{\pm}(\theta)=\sinh\tfrac{\gamma}{\pi}(2\theta\pm 2i\pi)
  B_{\pm} \left(\theta - \tfrac{i\pi}{2}\right) 
  B_{\pm} \left(\theta + \tfrac{i\pi}{2}\right)
  \phi\left(\theta \pm \tfrac{3i\pi}{2}\right) 
  \phi\left(\theta \pm \tfrac{i\pi}{2}\right). 
\end{equation}

\begin{figure}
\begin{center}
 \includegraphics[scale=0.75]{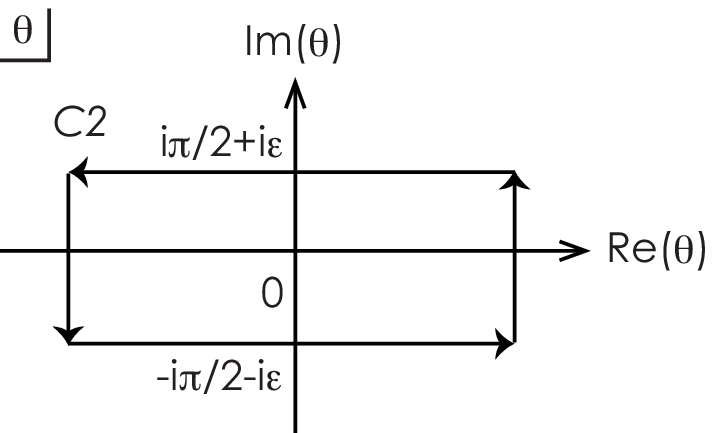}
 \caption{A contour $\mathcal{C}_2$.}
 \label{fig:contour2}
\end{center}
\end{figure}

Another nontrivial equation is derived from ANZ property of $T_1(\theta)$ in ${\rm Im}\theta \in [-\frac{\pi}{2}, \frac{\pi}{2})$ except for the origin and positions of type-$1$ holes. The function $T_1(\theta)$ shows up in the auxiliary function $b(\theta)$ through
\begin{equation}
 b(\theta) 
  = 
  \frac{T_1(\theta - \frac{i\pi}{2})}{\sinh\frac{\gamma}{\pi}(2\theta + 2i\pi)} 
  \frac{\phi(\theta - \frac{i\pi}{2})}{\phi(\theta + \frac{i\pi}{2}) \phi(\theta + \frac{3i\pi}{2})} 
  \frac{B_-(\theta + \frac{i\pi}{2})}{B_+(\theta - \frac{i\pi}{2}) B_+(\theta + \frac{i\pi}{2})} 
  \frac{Q(\theta + \frac{3i\pi}{2})}{Q(\theta - \frac{3i\pi}{2})}. 
\end{equation}
Applying the Cauchy theorem, we have the following equation (Figure \ref{fig:contour2}): 
\begin{equation} \label{cauchy_t1}
 \oint_{\cC_2} d\theta\; e^{ik\theta} [\ln T_1(\theta)]'' 
 = \frac{2\pi k}{1 - e^{-\pi k}}
 \Big(1 + \sum_{{\rm Im}h^{(1)}_j\in [-\frac{\pi}{2}, \frac{\pi}{2})} e^{ik h^{(1)}_j}\Big) 
\end{equation} 
which gives an NLIE for $b(\theta)$. 
% figure

Thus a set of NLIEs is derived for the LSSG model with Dirichlet boundaries as follows: 
\begin{align}
 \ln b(\theta) = 
 &\int_{-\infty}^{\infty} d\theta'\; G(\theta - \theta' - i\epsilon) \ln B(\theta' + i\epsilon)
 - \int_{-\infty}^{\infty} d\theta'\; G(\theta -\theta' + i\epsilon) \ln \bar{B}(\theta' - i\epsilon) \nonumber \\
 &+ \int_{-\infty}^{\infty} d\theta'\; G_K(\theta - \theta' - \textstyle\frac{i\pi}{2} + i\epsilon) \ln Y(\theta' - i\epsilon) 
 + i D_{\rm bulk}(\theta) + i D_{\rm B}(\theta) + i D(\theta) \nonumber \\
 &+ C_b^{(1)} \theta + C_b^{(2)} \label{NLIE_b}\\
 \ln y(\theta) = 
 &\int_{-\infty}^{\infty} d\theta'\; G_K(\theta - \theta' + {\textstyle\frac{i\pi}{2}} - i\epsilon) \ln B(\theta' + i\epsilon) 
 + \int_{-\infty}^{\infty} d\theta'\; G_K(\theta - \theta' -{\textstyle\frac{i\pi}{2}} + i\epsilon) \ln \bar{B}(\theta' - i\epsilon) \nonumber \\
 &+ i D_{\rm SB}(\theta) + i D_K(\theta) + C_y^{(1)} \theta + C_y^{(2)}, \label{NLIE_y}
\end{align}
where $C_b^{(i)}$ and $C_y^{(i)}$ ($i \in \{1,2\}$) are integration constants which are determined by asymptotic analysis of NLIEs (Appendix \ref{sec:int_const}). Functions $G(\theta)$ and $G_K(\theta)$ are given by 
\begin{align}
 G(\theta) = \int_{-\infty}^{\infty} \frac{dk}{2\pi} 
 \frac{e^{-ik\theta} \sinh(\frac{\pi}{\gamma} - 3)\frac{\pi k}{2}}{2 \cosh\frac{\pi k}{2} \sinh(\frac{\pi}{\gamma} - 2)\frac{\pi k}{2}},
 \qquad
 G_K(\theta) = \int_{-\infty}^{\infty} \frac{dk}{2\pi}
 \frac{e^{-ik\theta}}{2 \cosh\frac{\pi k}{2}}, 
\end{align}
which correspond to soliton-soliton scattering factors. Indeed, $G(\theta)$ is nothing but the bulk scattering amplitude of the SG model (\ref{SGamp}). 
A bulk phase shift shows up in $D_{\rm bulk}(\theta)$ as 
\begin{equation}
 D_{\rm bulk}(\theta) = 2N \arctan\frac{\sinh \theta}{\cosh \Theta}. 
\end{equation}
A particle source term $D(\theta)$ is given by 
\begin{equation}
\begin{split}
 &D(\theta) = \sum_j c_j \{g_{(j)}(\theta - \tilde{\theta}_j) + g_{(j)}(\theta + \tilde{\theta}_j)\}, \\
 &g(\theta) = 2\gamma \int_0^{\infty} d\theta'\; G(\theta'), 
 \qquad 
 g_K(\theta) = 2\gamma \int_0^{\infty} d\theta'\; G_K(\theta'), 
\end{split}
\end{equation}
where $\theta_j$ is a Bethe root. A function $g_{(j)}$ is defined for each object differently: 
\begin{equation}
g_{(j)}(\theta) = 
	\begin{cases}
		g_{\rm II}(\theta) = g(\theta) + g(\theta - i\pi\, {\rm sign}({\rm Im}\, \theta)) & \text{for wide roots} \\
		g(\theta + i\epsilon) + g(\theta - i\epsilon) & \text{for specials} \\
		g_K(\theta) & \text{for type-$1$ holes} \\
		g(\theta) & \text{otherwise}, 
	\end{cases}
\end{equation}
together with a choice of $c_j$: 
\begin{equation}
c_{j} = 
	\begin{cases}
		+1 & \text{for holes} \\
		-1 & \text{otherwise}. 
	\end{cases}
\end{equation}
A kink source term $D_K(\theta)$ is given by 
\begin{equation}
\begin{split}
 &D_K(\theta) = \mathop{\lim}_{\epsilon \to +0} \widetilde{D}_K(\theta + \textstyle\frac{i\pi}{2} - i\epsilon) \\
 &\widetilde{D}_K(\theta) = \sum_j c_j \{g^{(1)}_{(j)}(\theta - \tilde{\theta}_j) + g^{(1)}_{(j)}(\theta + \tilde{\theta}_j)\}, 
\end{split}
\end{equation}
where $g^{(1)}_{(j)}(\theta)$ are 
\begin{equation}
g^{(1)}_{(j)}(\theta) = 
	\begin{cases}
		(g_K)_{\rm II}(\theta) = g_K(\theta) + g_K(\theta - i\pi\,{\rm sign}({\rm Im}\, \theta)\,) = 0  & \text{for wide roots} \\
		g_K(\theta + i\epsilon) + g_K(\theta - i\epsilon) & \text{for specials} \\
		g_K(\theta) & \text{otherwise}, 
	\end{cases}
\end{equation}
which means no contribution from wide roots to a kink source term. 
Functions $D_{\rm B}(\theta)$ and $D_{\rm SB}(\theta)$ are boundary terms which we discuss in the next subsection.

For later use, we also derive a NLIE for the auxiliary function $a(\theta)$. Keeping in our mind that real zeros of $T_1(\theta)$ consists of type-$1$ holes which results in (\ref{cauchy_t1}), we obtain 
\begin{equation} \label{NLIE_a}
\begin{split}
 \ln a(\theta) &= 
  \int_{-\infty}^{\infty} d\theta'\;
  G_a(\theta - \theta' + i\epsilon) \ln A(\theta' - i\epsilon)
  - \int_{-\infty}^{\infty} d\theta'\;
  G_a(\theta - \theta' - i\epsilon) \ln \bar{A}(\theta' + i\epsilon)
  \\
 &+ iD^{(a)}_{\rm bulk}(\theta) + iD^{(a)}_{\rm B}(\theta) + iD_{\rm a}(\theta) + C_a, 
\end{split}
\end{equation}
where $C_a$ is an integration constant derived in Appendix \ref{sec:int_const}. A function $G_a(\theta)$ is given by 
\begin{equation}
 G_a(\theta) = \int_{-\infty}^{\infty} \frac{dk}{2\pi}
  \frac{e^{-ik\theta} \sinh(\frac{\pi}{\gamma} - 2)\frac{\pi k}{2}}{2\cosh\frac{\pi k}{2} \sinh(\frac{\pi}{\gamma} - 1)\frac{\pi k}{2}}. 
\end{equation}
A boundary term $iD^{(a)}_{\rm B}(\theta) = F_a(\theta;H_+) + F_a(\theta;H_-) + J_a(\theta)$ depends on boundary parameters: 
\begin{equation}
 F_a(\theta;H) = \int_0^{\infty} d\theta'\; \int_{-\infty}^{\infty} dk\;
  e^{-ik\theta'}
  \frac{({\rm sgn}(H)\,\frac{\pi}{\gamma} - H) \frac{\pi k}{2}}{2\cosh\frac{\pi k}{2} \sinh(\frac{\pi}{\gamma} - 1)\frac{\pi k}{2}}, 
\end{equation}
while a boundary-parameter-independent term $J_a(\theta)$ is given by
\begin{equation}
 J_a(\theta) = \int_0^{\infty} d\theta'\; \int_{-\infty}^{\infty} dk\;
  e^{-ik\theta'}
  \frac{\cosh\frac{\pi k}{4} \sinh(\frac{\pi}{\gamma} - 2)\frac{\pi k}{4}}{\cosh \frac{\pi k}{2} \sinh(\frac{\pi}{\gamma} - 1)\frac{\pi k}{4}}. 
\end{equation}
Particle source terms are written as 
\begin{equation} \label{source-a}
 D_{\rm a}(\theta) = \sum_j c^{(a)}_j 
 \{
 g_{(j)}^{(a)}(\theta - \theta_j) + g_{(j)}^{(a)}(\theta + \theta_j)
 \}, 
\end{equation}
where $c_j^{(a)}$ is defined by 
\begin{equation}
 c_j^{(a)}
 =
 \begin{cases}
  1 & \text{for type-$1$ holes} \\
  -1 & \text{otherwise}. 
 \end{cases} 
\end{equation}
$g_{(j)}^{(a)}(\theta)$ is a function differently defined for each root or hole: 
\begin{equation}
 g_{(j)}^{(a)}(\theta) = 
 \begin{cases}
  (g_a)_{\rm II}(\theta) = g_a(\theta) + g_a(\theta - i\pi {\rm sgn}({\rm Im}\,\theta)) & \text{for roots satisfying $|{\rm Im}\,\theta_j| > \pi$} \\
  g_a(\theta + i\epsilon) + g_A(\theta - i\epsilon) & \text{for specials} \\
  g_a(\theta) & \text{otherwise}, 
  \end{cases}
\end{equation}
where 
\begin{equation}
 g_a(\theta) = 2\gamma \int_0^{\infty} d\theta'\; G_a(\theta'). 
\end{equation}
A summation in (\ref{source-a}) is taken over $j$ such that $\theta_j$ is a type-$1$ hole, a real special object, or a complex root. A bulk term is obtained as 
\begin{equation}
 D^{(a)}_{\rm bulk}(\theta)
  =
  N \{g_a(\theta - i\Theta) + g_a(\theta + i\Theta)\}. 
\end{equation}

%%%%%%%%%%%%%%%%%%%%%%%%%%%%%%%
\subsection{Boundary dependence of NLIEs}
Boundary dependence of NLIEs emerges through branch cuts of logarithms. The following integral often appears in NLIEs: 
\begin{equation}
 \int_{-\infty}^{\infty} d\theta\; e^{ik\theta} [\ln \sinh(\theta- i\alpha)]'' 
  =
  \frac{2\pi k}{1 - e^{-\frac{\pi^2}{\gamma}k}} e^{-(\alpha - \frac{\pi^2}{\gamma}n)k}, 
\end{equation}
in which a branch cut is characterized by an integer $n$ s.t. $0 < {\rm Re}(\alpha - \pi n ) \le \pi$. Boundary parameters $H_{\pm}$ appear through the functions $B_{\pm}(\theta)$ as in forms of $B_{\pm}(\theta \pm \frac{i\pi}{2})$, and then we need to distinguish three regimes for each $H_{\pm}$: (a) $1 < H_{\pm} \leq \frac{2\pi}{\gamma} - 1$; (b) $-1 < H_{\pm} \leq 1$; (c) $-\frac{2\pi}{\gamma} + 1 < H_{\pm} \leq -1$. 

Boundary terms in NLIEs $D_{\rm B}(\theta)$ and $D_{\rm SB}(\theta)$ consist of right-boundary parts and left-boundary parts: 
\begin{align} 
 &D_{\rm B}(\theta) = F(\theta;H_+) + F(\theta;H_-) + J(\theta) \label{boundary_term_sg}
 \\
 &D_{\rm SB}(\theta) = F_y(\theta;H_+) + F_y(\theta;H_-) + J_K(\theta), \label{boundary_term_susy}
\end{align}
where $J(\theta)$ and $J_K(\theta)$ do not depend on boundary parameters: 
\begin{align}
 &J(\theta) = \int_0^{\infty} d\theta' \int_{-\infty}^{\infty} dk\; e^{-ik\theta'}
\frac{\cosh\frac{\pi k}{4} \sinh (\frac{\pi}{\gamma} - 3) \frac{\pi k}{4}}{\cosh\frac{\pi k}{2} \sinh(\frac{\pi}{\gamma} - 2) \frac{\pi k}{2}}, 
 \\
 &J_K(\theta) = 2 \widetilde{g}_K(\theta) = \underset{\epsilon \rightarrow +0}{\lim} 2g_K(\theta + \tfrac{i\pi}{2} - i\epsilon). 
\end{align}
Boundary-parameter dependent parts $F(\theta;H)$ and $F_y(\theta;H)$ have different forms for the three regimes (a), (b), and (c). 
\vspace{2mm}
\par\noindent
\underline{\bf{Regime (a)}}
\begin{align}
 &F(\theta;H) = \int_0^{\infty} d\theta' \int_{-\infty}^{\infty} dk\; e^{-ik\theta'}
 \frac{\sinh (\frac{\pi}{\gamma} - H) \frac{\pi k}{2}}{2\cosh\frac{\pi k}{2} \sinh (\frac{\pi}{\gamma} - 2) \frac{\pi k}{2}}, \label{regimea_b}
 \\
 &F_y(\theta;H) = 0. \label{regimea_y}
\end{align}
\underline{\bf{Regime (b)}}
\begin{align}
 &F(\theta;H) = -\int_0^{\infty} d\theta' \int_{-\infty}^{\infty} dk\; e^{-ik\theta'}
 \frac{\sinh(\frac{\pi}{\gamma} + \pi H - 2) \frac{\pi k}{2}} {2\cosh\frac{\pi k}{2} \sinh (\frac{\pi}{\gamma} - 2) \frac{\pi k}{2}}, \label{regimeb_b}
 \\
 &F_y(\theta;H) = \widetilde{g}_K(\theta - \tfrac{i\pi(1-H)}{2}) + \widetilde{g}_K(\theta + \tfrac{i\pi(1-H)}{2}). \label{regimeb_y}
\end{align}
\underline{\bf{Regime (c)}}
\begin{align}
 &F(\theta;H) = -\int_0^{\infty} d\theta' \int_{-\infty}^{\infty} \frac{dk}{2\pi}\; e^{-ik\theta'}
 \frac{\sinh (\frac{\pi}{\gamma} + H) \frac{\pi k}{2}}{2\cosh\frac{\pi k}{2} \sinh (\frac{\pi}{\gamma} - 2) \frac{\pi k}{2}}, \label{regimec_b}
 \\
 &F_y(\theta;H) = 0. \label{regimec_y}
\end{align}

NLIEs of the BSSG$^+$ model are obtained through lattice regularization of light-cone. The original continuum theory is recovered in the scaling limit \cite{bib:IO95}. Since parameters concerning the scaling limit appear only through a bulk term, NLIEs for the BSSG$^+$ model are obtained just by the following replacement: 
\begin{equation}
 2 N \arctan\frac{\sinh \theta}{\cosh \Theta} 
  \to
  2i m_0 L \sinh\theta. 
\end{equation}
%where $\theta = \frac{\pi x}{\gamma}$ is rapidity of a particle in the BSSG theory. 

%%%%%%%%%%%%%%%%%%%%%%%%%%%%
\subsection{Eigenenergies}
\begin{figure}
\begin{center}
 \includegraphics[scale=0.5]{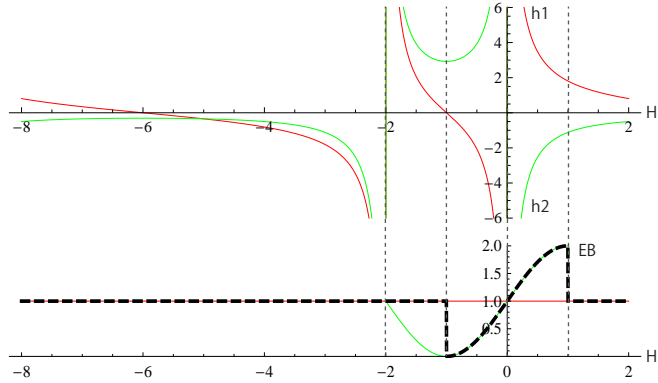}
 \caption{Boundary energy is depicted by a bold dotted line as a function of a boundary parameter $H$ for anisotropy $\gamma = \frac{\pi}{5}$. A mass parameter $m_0$ is taken to be $1$. The upper figure is behavior of boundary magnetic fields.}
 \label{fig:boundary_energy}
\end{center}
\end{figure}

By definition, an eigenenergy of the lattice-regularized BSSG$^+$ model is obtained from the function $T_2(\theta)$ \cite{bib:MNR90}: 
\begin{equation}
 E = 
  \frac{1}{4ia} 
  \left(
   \frac{d}{d\theta} \ln T_2(\theta)\Big|_{\theta = \Theta + \frac{i\pi}{2}} - \frac{d}{d\theta} \ln T_2(\theta)\Big|_{\theta = \Theta - \frac{i\pi}{2}}
  \right). 
\end{equation}
Using a fusion relation (\ref{T-Yrel}) and a NLIE (\ref{NLIE_y}), we obtain an eigenenergy of the BSSG$^+$ model in the scaling limit: 
\begin{equation} \label{energy}
 E = E_{\rm bulk} + E_{\rm B} + E_{\rm ex} + E_{\rm C}, 
\end{equation}
where bulk energy $E_{\rm bulk}$, excitation energy $E_{\rm ex}$, and Casimir energy $E_{\rm C}$ is given by 
\begin{align}
 &E_{\rm bulk} = 0, \\
 &E_{\rm ex} = m_0 \sum_{j=1}^{N_H} \cosh h_{j}
 - m_0 \sum_{j=1}^{M_C} \cosh \tilde{c}_j, \\
 &E_C = \frac{m_0}{2\pi}\, {\rm Im}
 \int_{-\infty - i\epsilon}^{\infty - i\epsilon} d\theta\,
 e^{-\theta} \ln \bar{B}(\theta). 
\end{align}
Boundary energy is given by a function of boundary parameters: 
\begin{align}
 &E_{\rm B} = m_0 + E_{\rm b}(H_+) + E_{\rm b}(H_-), \label{boundary-e} \\
 &E_{\rm b}(H) = 
 \begin{cases}
  0 & |H| > 1, \\
  m_0 \cos\tfrac{\pi(1-H)}{2} & |H| < 1, 
 \end{cases} 
\end{align}
whose behavior is shown in Figure \ref{fig:boundary_energy}. We expect appearance of a boundary bound state at $H = \pm1$ which causes a gap in a boundary energy function. 
% figure
%

%%%%%%%%%%%%%%%%%%%%%%%%%%%%%%%%
\subsection{Restriction on excitations}
Allowed excitations of the BSSG$^+$ model can be discussed by counting equations derived from NLIEs. 
Counting equations relate the numbers of different types of particles, {\it i.e.} the numbers of excitation particles are not arbitrary but restricted by counting equations. 

\subsubsection{Counting equations}
A counting equation for holes is derived by comparing asymptotic behaviors of both sides of the NLIE (\ref{NLIE_b}). As was discussed in Appendix \ref{sec:int_const}, we obtain 
\begin{equation} \label{counting-b}
\begin{split}
 N_H - 2(N_S + N_V) &= 2S^{\rm tot} + M_C + 2M_W - \delta_B \\
 &+ \tfrac{1}{2} ({\rm sgn}(1-H_+) + {\rm sgn}(1-H_-) + {\rm sgn}(1+H_+) + {\rm sgn}(1+H_-)), 
\end{split}
\end{equation}
where $\delta_B$ is defined by (\ref{def_deltaB}). 
There exists another counting equation for type-$1$ holes, which is derived from the NLIE for the auxiliary function $a(\theta)$ (\ref{NLIE_a}): 
\begin{equation} \label{counting-a}
 N_1 - 2(N_S^R + N_V^R) = S^{\rm tot} - M_R + M_{C>\pi} + M_W 
  + \tfrac{1}{2}({\rm sgn}(H_+) + {\rm sgn}(H_-)). 
\end{equation}
\ifx10
\subsubsection{Counting functions}
We first introduce counting functions. Here we use two types of counting functions; Remind that the auxiliary function $B(x)$ has zeros at positions  which becomes real string centers in a large volume limit and holes on positive half of the real axis. Then a function $\lfloor\frac{1}{2\pi} {\rm Im}\, \ln b(x)  + \frac{1}{2}\rfloor$ counts, in a large volume limit, quantum numbers of two-strings and holes whose maximum value is given in $x \to \infty$: 
\begin{equation} \label{basym}
 \left\lfloor
  \tfrac{1}{2\pi} {\rm Im}\, \ln b(\infty) + \tfrac{1}{2}
 \right\rfloor
 = I_{\rm max} 
 = M_2 + N_H - 2(N_S + N_V). 
\end{equation}
Here we denoted the number of two-string centers by $M_2$. 

The auxiliary function $A(x)$ has zeros at positions which becomes real roots in a large volume limit and type-$1$ holes on a positive half of the real axis. Similarly, we introduce a counting function of real roots and type-$1$ holes by $\lfloor \frac{1}{2\pi} {\rm Im}\, \ln a(x) + \frac{1}{2} \rfloor$ whose maximum value is given by 
\begin{equation} \label{aasym}
  \left\lfloor
  \tfrac{1}{2\pi} {\rm Im}\, \ln a(\infty) + \tfrac{1}{2}
 \right\rfloor
 = I_{\rm max}^{(1)} 
 = M_R + N_1 - 2(N_S^R + N_V^R).  
\end{equation}
A parameter $M_R$ denotes the number of real roots, while $N_S^R$ and $N_V^R$ are the numbers of real special objects.

\subsubsection{Counting equations}
Counting equations are derived by evaluating left-hand sides of (\ref{basym}) and (\ref{aasym}). By choosing a branch cut as in Figure \ref{}, the definition of $a(x)$ leads to 
\begin{equation}
 {\rm Im}\, \ln a(\infty) = 2\pi (N -M_I - M_{C<\pi} + \tfrac{1}{2}
  + \tfrac{1}{2}\,{\rm sgn}(H_+) + \tfrac{1}{2}\,{\rm sgn}(H_-))
  - 2\omega, 
\end{equation}
where $M_{C<\pi}$ is the number of close roots whose imaginary parts are less than $\pi$. A parameter $\omega$ is introduced in \ref{}. Thus we obtain a counting equation for type-$1$ holes: 
\begin{equation} \label{sum1}
 N_1 - 2 (N_S^R + N_V^R) 
  = S - M_R + M_{C > \pi} + M_W 
  + \tfrac{1}{2}\{{\rm sgn}(H_+) + {\rm sgn}(H_-)\} 
  - \left\lfloor \tfrac{\omega}{\pi} \right\rfloor, 
\end{equation}
where $M_{C>\pi}$ is the number of close roots whose imaginary parts are greater than $\pi$. 

On the other hand, the definition of $b(x)$ leads to 
\begin{equation}
\begin{split}
 {\rm Im} \ln b(\infty)
 &= 2\pi(
 2N - 2M_I - M_2 - M_C + 1 + \tfrac{\delta_a}{2} - \lfloor \tfrac{\omega}{2\pi} + \tfrac{\delta_a}{2} \rfloor\\
 &+ \tfrac{1}{2}\,{\rm sgn}(H_+-1) + \tfrac{1}{2}\,{\rm sgn}(H_++1) + \tfrac{1}{2}{\rm sgn}(H_--1) + \tfrac{1}{2}\,{\rm sgn}(H_-+1) 
 )
 - 3\omega, 
\end{split}
\end{equation}
where $\delta_a$ is given by 
\begin{equation}
 \delta_a = 
  \begin{cases}
   0 & \cos\omega > 0, \\
   1 & \cos\omega < 0. 
   \end{cases}
\end{equation}
Thus we obtain a counting equation for holes: 
\begin{equation} \label{sum2}
\begin{split}
 N_H - 2(N_S + N_V)
 &= 2S + M_C + 2M_W - \lfloor \tfrac{\omega}{2\pi} + \tfrac{\delta_a}{2} \rfloor - \lfloor \tfrac{3\omega}{2\pi} - \tfrac{\delta_a}{2} - \tfrac{1}{2} \rfloor\\
 &+ \tfrac{1}{2}\{{\rm sgn}(H_+ - 1) + {\rm sgn}(H_+ + 1) + {\rm sgn}(H_- - 1) + {\rm sgn}(H_- + 1)\}. 
\end{split}
\end{equation}
\fi 
In the scaling limit, a special object or a self-conjugate root emerges exactly when $\gamma$-dependent terms raise their values by $1$ \cite{bib:DV97}. Therefore, one can regard $N_H^{\rm eff}$ defined by the following as the number of SSG solitons: 
\begin{equation} \label{counting-b*}
 %N_H^{\rm eff} = N_H - 2(N_S + N_V) - 2M_{SC} + \lfloor \tfrac{\omega}{2\pi} + \tfrac{\delta_a}{2} \rfloor + \lfloor \tfrac{3\omega}{2\pi} - \tfrac{\delta_a}{2} - \tfrac{1}{2} \rfloor. 
 N_H^{\rm eff} = N_H - 2(N_S + N_V) - 2M_{SC} -\delta_B. 
\end{equation}
%Similarly, we introduce the number of effective type-$1$ holes as follows: 
%\begin{equation} \label{sum2*}
% N_1^{\rm eff} = N_1 - 2(N_S^R + N_V^R) - M_{SC} + \lfloor \tfrac{\omega}{\pi} \rfloor. 
%\end{equation}
Then the counting equation for holes is written in terms of $N_H^{\rm eff}$: 
\begin{align}
 N_H^{\rm eff} &= 2S^{\rm tot} + M_C + 2(M_W - M_{SC}) \nonumber \\
 &+ \tfrac{1}{2} \{{\rm sgn}(H_+ - 1) + {\rm sgn}(H_+ + 1) + {\rm sgn}(H_- - 1) + {\rm sgn}(H_- + 1)\}. \label{counting-b*} \\
 %N_1^{\rm eff} &= S - M_R + M_{C>\pi} + (M_W - M_{SC})
 %+ \tfrac{1}{2}\{{\rm sgn}(H_+) + {\rm sgn}(H_-)\}. \label{sum2**}
\end{align}

\subsubsection{Allowed excitations}
\begin{figure}
\begin{center}
\subfigure[Analyticity structure of $T_1(\theta)$ (left) and $T_2(\theta)$ (right) $B(\theta)$ for $H_+=1.5$ and $H_-=2.2$.]
{\includegraphics[scale=0.65]{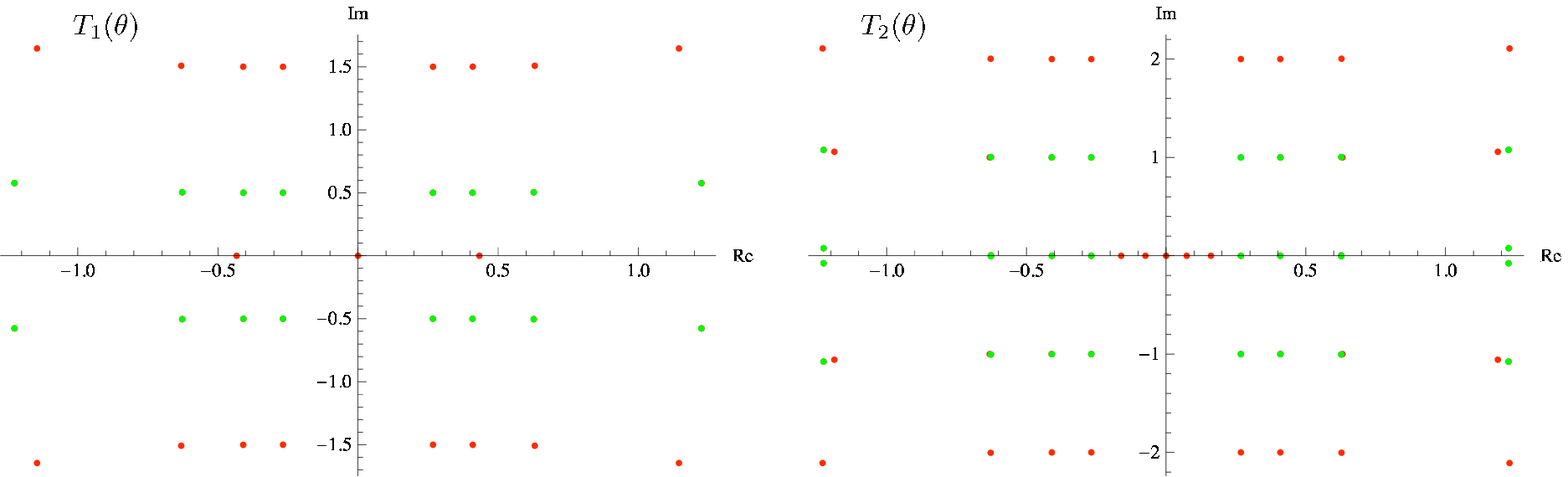}}
\label{fig:zeros-a}
\subfigure[Analyticity structure of $T_1(\theta)$ (left) and $T_2(\theta)$ (right) $B(\theta)$ for $H_+=1.5$ and $H_-=0.3$.]
{\includegraphics[scale=0.65]{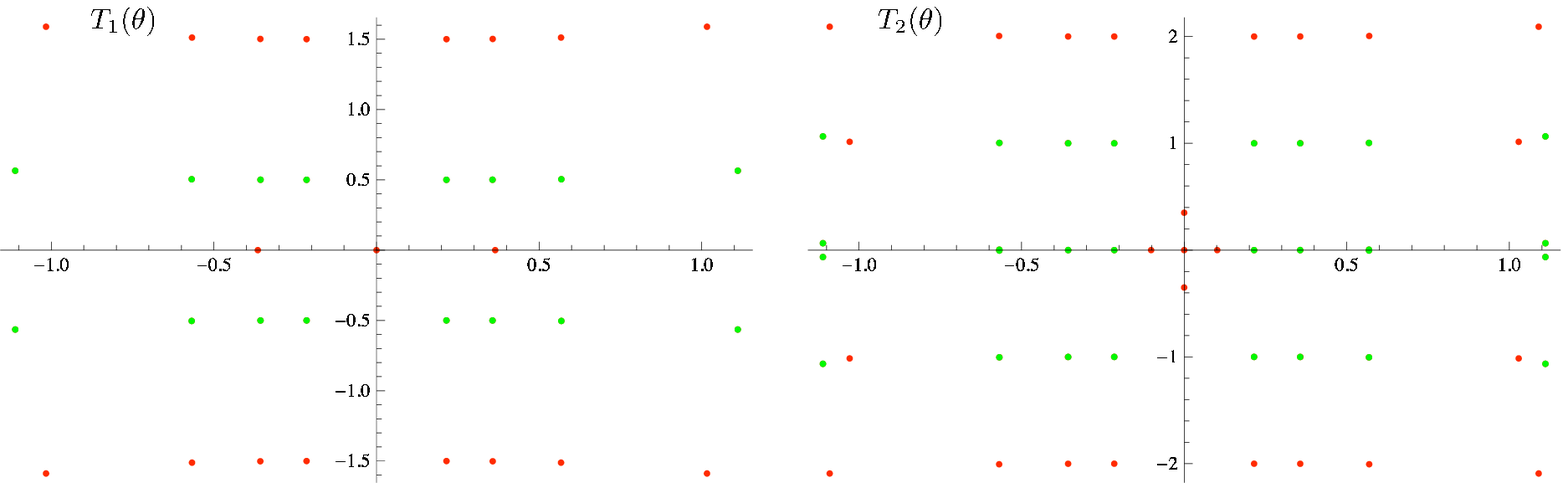}}
\label{fig:zeros-b}
%\subfigure[(left) Zeros of $1+a_1(\theta)$; (right) $B(\theta)$ for $H_+=1.5$ and $H_-=-0.5$.]
%{\includegraphics[scale=0.5]{Hm15_Hp-05}}
%\label{fig:zeros-c}
\subfigure[Analyticity structure of $T_1(\theta)$ (left) and $T_2(\theta)$ (right) $B(\theta)$ for $H_+=1.5$ and $H_-=-1.8$.]
{\includegraphics[scale=0.65]{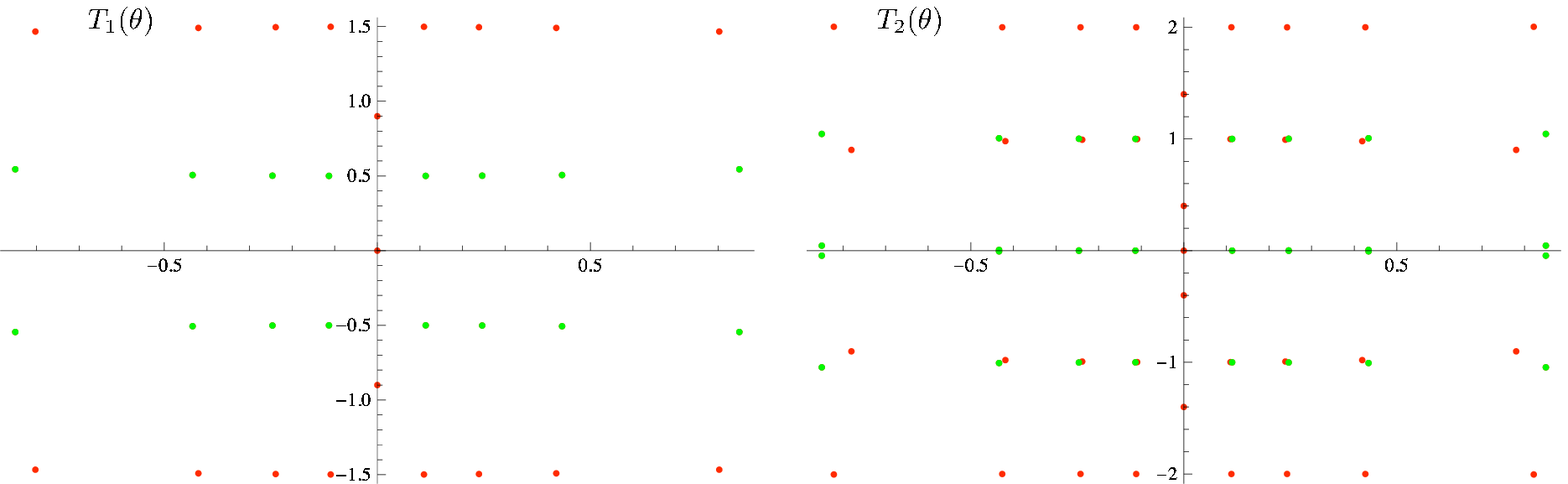}}
\label{fig:zeros-d}
\end{center}
\caption{Analyticity structure of $T_1(\theta)$ and $T_2(\theta)$ is plotted for three regimes of boundary parameters. One of boundary parameters is fixed at $H_+ = 1.5$. Zeros of $T$-functions are depicted by black dots and roots by gray dots. These plots are calculated for a system of length $N=8$ with $4$ pairs of two-string roots in the homogeneous and isotropic limit. } 
 \label{fig:zeros}
\end{figure}

Using (\ref{counting-a}) and (\ref{counting-b*}), we discuss allowed excitations. As we chose the lattice system consisting of an even number of sites, it is obvious that $S^{\rm tot}$ takes only an integer. 
For $H_+>1$ and $H_-<-1$, $N_H^{\rm eff}$ apparently takes an even integer, since close roots always appear in complex conjugate pairs. The counting equations admit a state with no particles and holes under $S^{\rm tot}=0$, that implies the ground state is given by a pure two-string state. 
When $H_+$ crosses $1$ while keeping $H_-<-1$, $N_H^{\rm eff}$ takes an odd integer. Especially a solution $N_H^{\rm eff}=1$ describes a state obtained in Figure \ref{fig:boundary_energy} which corresponds to a one-particle excitation at a boundary. 
%Another solution to (\ref{sum1**}) is given by $M_C=1$, meaning a correct ground state for $0<H_+<1$ includes a boundary bound state. 
Numerical study (Figure \ref{fig:zeros}) shows that rapidity of this boundary bound state seems to be fixed at $\theta = \frac{i\pi(1-H_+)}{2}$ with a small deviation exponentially vanishing as system length increases. 
At $H_+=0$, an energy function is analytically continued in Figure \ref{fig:boundary_energy}. However, for $H_+<0$, the energy function takes negative value implying it gives ground-state energy. 
Together with the fact that $N_1$ takes a different value for positive or negative $H_+$, 
%Since $H_+<0$, a close root renders its imaginary value greater than $\pi$ contributing to the equation (\ref{counting-a}). From this fact, 
we expect that states obtained in this regime belong to a different sector of a superconformal field theory from that for $H_+>0$. 
%Once $H_+$ crosses $-1$, a root at $\widetilde{\theta} = \frac{i\pi(1-H_+)}{2}$ becomes a wide root which does not contribute to energy (\ref{}). 
%
Finally when $H_+$ reaches $-1$, counting equations again admit only an even value for $N_H^{\rm eff}$, meaning that excitations with only an even number of particles are allowed. 
%a solution is obtained as $N_H=N_1=0$, and therefore the ground state is again obtained by a pure two-string state. 
%We obtain that $E_{\rm bulk} + E_{\rm B} + E_{\rm C}$ gives ground state energy for $-1<H_+<0$, while one-particle excitation energy for $-2<H_+<-1$. 
%Finally, when $H_+$ reaches $-2$, a ground state solutions is obtained as $N_H=N_1=0$ but with $S=1$. 

%Due to a symmetry $H_+ \leftrightarrow H_-$, a whole regime of $H_-$ with fixed $H_+$ smaller than $-1$ is discussed in the same way. 
If one starts discussion from $H_{\pm}>1$, the ground state is characterized by two-string roots but with $S^{\rm tot}=-1$. In the realm of a spin chain, it is understood as follows; By polarizing the outermost spins at both ends in the same direction, spins freely interact with their neighbors only on the bulk $N-2$ sites. This gives rise to emergence of a spinon, which results in the ground state with $S^{\rm tot}=-1$. 
Except for $S^{\rm tot}$, discussion for $H_->1$ goes on quite similar to the case of $H_-<-1$ due to a symmetry $H_- \leftrightarrow -H_- -2$. 
%
%Moreover, special care is required for $-2<H_{\pm}<0$, since allowed states strongly depend on 
%
%For $1>H_+>0$, a boundary bound state at $x=\frac{i\pi}{2}(2-H_+)$ is obtained in a ground state. A lowest energy excitation given by $N_H=N_1=1$ occurs at $S=0$. When $H_+$ reaches $0$, an imaginary part of a close root becomes greater than $\pi$. which contributes to the equation (\ref{sum2**}). Then we again obtain the pure two-string ground state for $H_+<-1$. 
For $1>H_{\pm}>0$, the outermost spins are trapped at both boundaries. Each of them can be released separately, resulting in $N_H^{\rm eff}$ with an odd integer. 

Thus, non-holomorphicity of boundary energy shows up due to change of a root configuration of the ground state, which directly affects analyticity structure of $T$-functions and consequently NLIEs. 
%Table \ref{} shows which configurations of roots characterize ground states of the rest regimes. 
We support this statement later in discussion of reflection amplitudes obtained in the IR limit. 
% figure

\ifx10
\end{document}
\fi
\ifx10
\documentclass{article}

%#############
%packages 
%#############
%\usepackage[utf8x]{inputenc}
\usepackage{amsmath, amsthm, amsfonts, amssymb}
\usepackage{bm}
\usepackage{a4wide}
\usepackage{graphicx} % for graphics
\usepackage{color}
\usepackage{euscript}
\usepackage{fancybox}

%#############
%newtheorems
%############
%\renewcommand{\thefootnote}{\fnsymbol{footnote}}
%\theoremstyle{plain}
%\newtheorem{thm}{Theorem}
%\newtheorem{claim}{Claim}
%\newtheorem{propn}{\bfseries Proposition}
%\newtheorem{propnn}{Proposition}
%\renewcommand{\thepropnn}{\arabic{propn}\alph{propnn}}
%\newtheorem{lem}{\bfseries Lemma}
%\newtheorem{cor}{Corollary}
%\newtheorem{conj}{\bfseries Conjecture}
%\newtheorem{defn}{Definition}
%\theoremstyle{remark}
%\newtheorem{rem}{Remark}
%\newtheorem{df}{Definition}
%\newtheorem{th1}[df]{Theorem}
%\newtheorem{lem}[df]{Lemma}
%\newtheorem{conj}[df]{Conjecture}
%\newtheorem{prop}[df]{Proposition}
%\newtheorem{cor}[df]{Corollary}
%\newtheorem{lem2}{Lemma}[section]
%\newtheorem{prop2}{Proposition}[section]
%\newtheorem{cor2}{Corollary}[section]
%\newtheorem{df2}{Definition}[section]

%##############
%newcommands
%#############
\newcommand{\cO}{\mathcal{O}}
\newcommand{\cA}{\mathcal{A}}
\newcommand{\cB}{\mathcal{B}}
\newcommand{\cC}{\mathcal{C}}
\newcommand{\cD}{\mathcal{D}}	
\newcommand{\cF}{\mathcal{F}}
\newcommand{\sP}{\mathsf{P}}
\newcommand{\sB}{\mathsf{B}}
\newcommand{\sY}{\mathsf{Y}}
\newcommand{\sy}{\mathsf{y}}
\newcommand{\sG}{\mathsf{G}}
\newcommand{\sm}{\mathsf{m}}
\newcommand{\sg}{\mathsf{g}}
\newcommand{\cT}{\mathcal{T}}
\newcommand{\cU}{\mathcal{U}}
\newcommand{\cH}{\mathcal{H}}
\newcommand{\cL}{\mathcal{L}}
\newcommand{\bC}{\mathbb{C}}
\newcommand{\txi}{\tilde{\xi}}
\newcommand{\tlambda}{\tilde{\lambda}}
\newcommand{\cI}{\mathcal{I}}
\newcommand{\cR}{\mathcal{R}}
\newcommand{\cS}{\mathcal{S}}
\newcommand{\cN}{\mathcal{N}}
\newcommand{\cM}{\mathcal{M}}
\newcommand{\cG}{\mathcal{G}}
\newcommand{\cK}{\mathcal{K}}
\newcommand{\bZ}{\mathbb{Z}}
\newcommand{\bR}{\mathbb{R}}
\newcommand{\ve}{\varepsilon}
\newcommand{\eH}{\EuScript{H}}
\newcommand{\cP}{\mathcal{P}}
\newcommand{\hT}{\hat{T}}
\newcommand{\htheta}{\hat{\theta}}
\newcommand{\sh}{\,\mathrm{sh}}
\newcommand{\ch}{\,\mathrm{ch}}
\newcommand{\fA}{\mathfrak{A}}
\newcommand{\fa}{\mathfrak{a}}
\newcommand{\fC}{\mathfrak{C}}
\newcommand{\fc}{\mathfrak{c}}
\newcommand{\fD}{\mathfrak{D}}
\newcommand{\fd}{\mathfrak{d}}
\newcommand{\fS}{\mathfrak{S}}
\newcommand{\ba}{\bar{a}}
\newcommand{\bb}{\bar{b}}
\newcommand{\dpsi}{\psi^\dag}
\newcommand{\pa}{a^{(+)}}
\newcommand{\dpa}{a^{(+)\dag}}
\newcommand{\ma}{a^{(-)}}
\newcommand{\dma}{a^{(-)\dag}}
\newcommand{\da}{a^\dag}
\newcommand{\ppsi}{\psi_+}
\newcommand{\dppsi}{\psi_+^\dag}
\newcommand{\mpsi}{\psi_-}
\newcommand{\dmpsi}{\psi_-^\dag}
\newcommand{\vphi}{\varphi}
\newcommand{\pS}{S^+}
\newcommand{\mS}{S^-}
\newcommand{\zS}{S^z}
%%%
%\newcommand{\bm}[1]{\mbox{\boldmath$#1$}}
%\newcommand{\be}{\begin{equation}}
%\newcommand{\ee}{\end{equation}}
%\newcommand{\bea}{\begin{eqnarray}}
%\newcommand{\eea}{\end{eqnarray}}
%\newcommand{\non}{\nonumber}
%\newcommand{\ra}{\rangle}
%\newcommand{\la}{\langle}
%\newcommand{\lam}{\lambda} 
%\newcommand{\Lam}{\Lambda} 
%\newcommand{\tht}{\theta} 
%\newcommand{\al}{\alpha} 
%\newcommand{\bt}{\beta} 
%\newcommand{\gm}{\gamma} 
%\newcommand{\dt}{\delta} 

\begin{document}
\fi

\section{Infrared limit} \label{sec:IRlim}
Scattering and reflection amplitudes can be read off from the IR limit of NLIEs given by $m_0 L \rightarrow \infty$. In this limit, rapidities of bound states are given by positions of poles in a $S$-matrix, while those of boundary bound states locate at positions of poles in a reflection matrix. At the same time, the ground state configuration of Bethe roots forms pure two-strings in an absence of Dirichlet boundaries. 

However, less is known about a ground state and excitation states under Dirichlet boundaries besides emergence of boundary bound states \cite{bib:BPT02}. Therefore, we discuss how presence of boundary bound states affects ground state configurations of Bethe roots from the viewpoint of NLIEs. 

In the IR limit, a simplification occurs in NLIEs, since the terms involving $\ln B(\theta)$ becomes negligibly small \cite{bib:HRS07}. However, the third term in (\ref{NLIE_b}) remains finite and we obtain a set of NLIEs as follows: 
\begin{align} 
 \ln b(\theta) = 
 &\int_{-\infty}^{\infty} d\theta'\; G_K(\theta - \theta' - \textstyle\frac{i\pi}{2} + i\epsilon) \ln Y(\theta' - i\epsilon) + 2im_0L \sinh \theta \nonumber \\
 &+ iD_{\rm B}(\theta) + iD(\theta) + i \pi C^{(2)}_b, \label{NLIE_IR1} \\
 \ln y(\theta) = 
 &iD_{\rm SB}(\theta) + iD_K(\theta) + i\pi C^{(2)}_y. \label{NLIE_IR2}
\end{align}

From the quantization condition for holes (\ref{quant_cond_h}), the following equation holds for any rapidity of holes $h_j$: 
\begin{equation} \label{quant_nlie}
 b(h_j) = -1. 
\end{equation}
On the other hand, a quantization condition in a realm of a quantum field theory has been obtained from a boundary condition for phase shifts \cite{bib:K79, bib:AD84}: 
\begin{equation} \label{quant_qft}
 e^{2im_0L \sinh \theta_j} R(\theta_j;\xi_+) \cdot
  \prod_{l=1 \atop l \neq j}^n S(\theta_j - \theta_l) S(\theta_j + \theta_l) \cdot
  R(\theta_j;\xi_-)
  = 1, 
\end{equation}
where $\theta_j$ is rapidity of a SSG soliton. 
Comparing (\ref{quant_nlie}) with (\ref{quant_qft}), we obtain the following relation between scattering amplitudes and NLIEs: 
\begin{equation}
\begin{split}
 &\int_{-\infty}^{\infty} d\theta' G_K(h_j - \theta' - \tfrac{i\pi}{2} + i\epsilon) \ln Y(\theta' - i\epsilon)
 + iD_{\rm B}(h_j) + iD(h_j) + i\pi (C_b^{(2)} + 1) \\
 & = \ln R(h_j;\xi_+) + \sum_{l=1 \atop l\neq j}^n \left(\ln S(h_j - h_l) + \ln S(h_j + h_l)\right) + \ln R(h_j;\xi_-), 
\end{split}
\end{equation}
which allows us to obtain scattering and reflection amplitudes in terms of lattice parameters. 

Counting equations (\ref{counting-a}) and (\ref{counting-b*}) admit $N_H^{\rm eff} = N_1 = 0$ for $H_{\pm} > 1$, which means no particle is obtained in a ground state. In this regime, both boundary terms belong to the regime (a) of NLIEs. From now on, we focus on the SG part of a reflection amplitude, as the RSOS part does not concern with boundary bound states. Reflection amplitudes for the regime (a) have been derived in \cite{bib:ANS07}: 
\begin{align} 
%\begin{split}
 &\ln R_1^+(\theta)
 = iF^{(a)}(\theta;H), \label{sg-ref1}
 %= \int_{-\infty}^{\infty} dk
 %\frac{e^{-ik\theta} \sinh(\frac{\pi}{\gamma} - H)\frac{\pi k}{2}}{2 \cosh\frac{\pi k}{2} \sinh(\frac{\pi}{\gamma} - 2)\frac{\pi k}{2}}, \label{sg-ref1}
 %\right], 
 %\\
 %&\frac{1}{i} \frac{d}{d\theta} \ln P_0(\theta)
 %= \int_{-\infty}^{\infty} dk\; 
 %\frac{e^{-2ik\theta}}{\cosh^2\frac{\pi k}{2} \cosh^2\pi k}, \label{susy-ref1} 
%\end{split}
 \\
 &\ln R_2(\theta) = iJ(\theta), \label{susy-ref1}
\end{align}
which result in relations obtained in Table \ref{tab:para-rel}. We denote boundary dependent terms for each regime by $F^{(x)}$ and $F_y^{(x)}$ ($x\in\{a,b,c\}$). A factor $2^{-\theta/2\pi}$ obtained in (\ref{ref_susy_int}) does not appear, since it is removed by a similarity transformation. 

When $H_+$ reaches $1$ while keeping $H_-$ greater than $1$, the counting equations are solved as $M_C=1$, showing a no-pairing close root emerges. A boundary term of NLIEs is given by the regime (b) (\ref{regimeb_b}) and (\ref{regimeb_y}) for $H_+$, although the terms of $H_-$ are still given by the regime (a) (\ref{regimea_b}) and (\ref{regimea_y}). 
This change occurs due to emergence of a boundary bound state; Boundary-dependent parts for the regime (b) are expressed through those for the regime (a): 
\begin{equation} \label{Fb}
\begin{split}
 iF^{(b)}(\theta;H) &= \ln R_1^+(\theta) 
 + ig(\theta - \tfrac{i\pi}{2}(1-H)) + ig(\theta + \tfrac{i\pi}{2}(1-H), 
 \\
 iF_y^{(b)}(\theta;H) &= iF_y^{(a)}(\theta;H) + i\widetilde{g}_K(\theta - \tfrac{i\pi}{2}(1-H)) + i\widetilde{g}_K(\theta + \tfrac{i\pi}{2}(1-H)). 
\end{split}
\end{equation}
This state is interpreted as a pure two-string state with an imaginary hole at $\theta = \frac{i\pi}{2}(1-H)$, which is a pole of a reflection amplitude (\ref{sg-ref1}). Indeed, the boundary bootstrap principle leads to the relations (\ref{Fb}) \cite{bib:ZZ79}. Boundary energy in Figure \ref{fig:boundary_energy} also supports this interpretation, by showing an energy gap $E_{\rm B}^{(b)\to(a)} = m_0 \cos\frac{\pi}{2}(1-H)$ at $H=1$. 

For $H_+<0$, we still obtain a solution $M_C=1$ from counting equations, whose imaginary part is, however, greater than $\pi$. Such a root contributes to the counting equation for type-$1$ holes (\ref{counting-a}). Ground-state reflection amplitudes in this regime are obtained as  
\begin{align}
 &\ln R_1^+(\theta)
 = 
 %\int_{-\infty}^{\infty} dk\;
 %\left[
 %\frac{e^{-ik\theta} \sinh(-\frac{\pi}{\gamma} - H + 2)\frac{\pi k}{2}}{2 \cosh\frac{\pi k}{2} \sinh(\frac{\pi}{\gamma} - 2)\frac{\pi k}{2}}, 
 iF^{(b)}(\theta;H) - ig_K(\theta - \tfrac{i\pi H}{2}) - ig_K(\theta + \tfrac{i\pi H}{2}), \label{sg-ref2}
 %\right], 
 %\\
 %&\frac{1}{i} \frac{d}{d\theta} \ln P_0(\theta)
 %= \int_{-\infty}^{\infty} dk\; 
 %\frac{e^{-2ik\theta}}{\cosh^2\frac{\pi k}{2} \cosh^2\pi k}, \label{susy-ref2} 
\end{align}
which require different parameter relations as shown in Table \ref{tab:para-rel}. 
Subsequently, boundary energy becomes negative (Figure \ref{fig:boundary_energy}), giving ground state energy. Thus, a ground state for this regime includes a close root at $\tilde{\theta} = \frac{i\pi}{2}(1+H)$. 
%\begin{equation} \label{Fb*}
%\begin{split}
% F^{(b)}(\theta;H) &= R_1(\theta) 
% - g(\theta - \tfrac{i\pi}{2}(1+H)) - g(\theta + \tfrac{i\pi}{2}(1+H)) 
% + g_K(\theta - \tfrac{i\pi H}{2}) + g_K(\theta + \tfrac{i\pi H}{2}), 
% \\
% F_y^{(b)}(\theta;H) &= F_y^{(a)}(\theta;H) - \widetilde{g}_K(\theta + \tfrac{i\pi}{2}(1+H)) - \widetilde{g}_K(\theta - \tfrac{i\pi}{2}(1+H)), 
%\end{split}
%\end{equation}
Besides this close root, the state includes a type-$1$ hole, implying existence of a non-paring root which affects on RSOS indices. Thus, the state for $H_+<0$ has different RSOS indices from the ground state obtained for $H_+>0$, and we expect a soliton state constructed through a light-cone lattice in this regime belongs to a different sector of a superconformal field theory from that for $H_+>0$.

When $H_+$ reaches $-1$, boundary terms for $H_+$ belong to the regime (c), which are expressed by adding two holes and one type-$1$ hole to the two-string state: 
\begin{equation} \label{Fc}
\begin{split}
 iF^{(c)}(\theta;H) &= iF^{(a)}(\theta;H) + ig(\theta - \tfrac{i\pi}{2}(1-H)) + ig(\theta + \tfrac{i\pi}{2}(1-H)) \\
 &+ ig(\theta + \tfrac{i\pi}{2}(1+H)) + ig(\theta - \tfrac{i\pi}{2}(1+H))
 + ig_K(\theta - \tfrac{i\pi H}{2}) + ig_K(\theta + \tfrac{i\pi H}{2}), 
 \\
 iF_y^{(c)}(\theta;H) &= iF_y^{(a)}(\theta;H). 
\end{split}
\end{equation}
Since boundary terms (\ref{Fb}) describe a ground state for $H_+<0$, we write boundary dependent terms as 
\begin{equation} \label{Fc*}
\begin{split}
 iF^{(c)}(\theta;H) &= iF^{(b)}(\theta;H) + ig(\theta - \tfrac{i\pi}{2}(1+H)) + ig(\theta + \tfrac{i\pi}{2}(1+H)), 
 %&+ g_K(\theta - \tfrac{i\pi H}{2}) + g_K(\theta + \tfrac{i\pi H}{2}), 
 \\
 iF_y^{(c)}(\theta;H) &= iF_y^{(b)}(\theta;H) + i\widetilde{g}_K(\theta + \tfrac{i\pi}{2}(1+H)) + i\widetilde{g}_K(\theta - \tfrac{i\pi}{2}(1+H)). 
\end{split}
\end{equation}
Thus, boundary dependent terms for the regime (c) describe a one-particle excitation state from the ground state including a close root. A reflection amplitude (\ref{sg-ref2}) indeed has a pole at $\theta = -\frac{i\pi}{2}(1+H)$. Figure \ref{fig:boundary_energy} also supports this interpretation which shows an energy gap $E_{\rm B}^{(c) \to (b)} = m_0 \cos\frac{\pi}{2}(1+H)$ at $H=-1$. 

Finally, $H_+$ reaches $-2$ and again we obtain the ground state with no particles nor holes. A reflection amplitude on this ground state is then obtained as 
\begin{align}
 &\ln R_1^+(\theta)
 = 
 %\left[
 iF^{(c)}(\theta;H), \label{sg-ref3}
 %\right], 
 %\\
 %&\frac{1}{i} \frac{d}{d\theta} \ln P_0(\theta)
 %= \int_{-\infty}^{\infty} dk\; 
 %\frac{e^{-2ik\theta}}{\cosh^2\frac{\pi k}{2} \cosh^2\pi k}, \label{susy-ref3} 
\end{align}
by resetting parameter relations as in Table \ref{tab:para-rel}. 
Besides solutions $N_H^{\rm eff} = N_1 = 0$, counting equations admit a solution $M_W = 1$ together with $S = -1$. Taking into account that a wide root gets into a second determination, one can also write (\ref{Fc}) as 
\begin{equation}
\begin{split}
 iF^{(c)}(\theta;H) &= \ln R_1^+(\theta) + ig_K(\theta - \tfrac{i\pi H}{2}) + ig_K(\theta + \tfrac{i\pi H}{2}) \nonumber \\
 &- ig_{\rm II}(\theta - \tfrac{i\pi}{2}(1+H)) - ig_{\rm II}(\theta - \tfrac{i\pi}{2}(1+H)), 
 \\
 iF_y^{(c)}(\theta;H) &= iF_y^{(a)}(\theta;H), 
\end{split}
\end{equation}
by regarding contribution of two holes in (\ref{Fc}) as that of a wide root. 
%Since a wide root has no contribution to energy, this state degenerates with the ground state, which occurs due to a uniaxial symmetry $h_1(H) \leftrightarrow -h_1(H)$, corresponding to spin reverse $S^z \leftrightarrow -S^z$. 
It is natural to consider a state in this regime is obtained just by a soliton-antisoliton translation from the regime $H_+>1$, and therefore, belongs to the same sector of a superconformal field theory. 
%table
\begin{table} \caption{Relations between a lattice boundary parameter and a QFT boundary parameter. } \label{tab:para-rel}
 \begin{center}
 \begin{tabular}{ccc}
  \hline\hline 
  $H>0$ & $0>H>-2$ & $-2>H$ \\
  \hline \\[-2mm]
  $H = -\frac{2\xi}{\pi\lambda} + \frac{1}{\lambda} + 1$
  &
      $H = -\frac{2\xi}{\pi\lambda} + \frac{1}{\lambda} - 1$
      &
	  $H = -\frac{2\xi}{\pi\lambda} - \frac{1}{\lambda} - 3$ \\[2mm] 
  \hline\hline
 \end{tabular}
 \end{center}
\end{table} 

Now let us discuss how symmetries obtained in a discretized system survive after taking the scaling limit. By replacing $H$ of a reflection amplitude on a ground state for $|H+1|>1$ by $-H-\frac{2\pi}{\gamma}-2$, one obtains a antisoliton reflection amplitude (\ref{SG-antiref}): 
\begin{equation}
 iF^{(a)}(\theta;-H-\tfrac{2\pi}{\gamma}-2) = iF^{(c)}(\theta;-H-\tfrac{2\pi}{\gamma}-2) = \ln R_1^-(\theta). 
\end{equation}
The same symmetry is also obtained for $|H+1|<1$ by substituting $H$ by $-H-2$: 
\begin{equation}
\begin{split}
 &iF^{(b)}(\theta;-H-2) - ig_K(\theta - \tfrac{i\pi}{2}(-H-2)) - ig_K(\theta + \tfrac{i\pi}{2}(-H-2)) \\
 &=iF^{(c)}(\theta;-H-2) - ig_K(\theta - \tfrac{i\pi}{2}(-H-2)) - ig_K(\theta + \tfrac{i\pi}{2}(-H-2)) \\
 &= \ln R_1^-(\theta). 
\end{split}
\end{equation}
This implies that a $S^z \leftrightarrow -S^z$-symmetry, {\it i.e.} a soliton-antisoliton symmetry survives even after continualization.

\ifx10
\end{document}
\fi
\ifx10
\documentclass{article}

%#############
%packages 
%#############
%\usepackage[utf8x]{inputenc}
\usepackage{amsmath, amsthm, amsfonts, amssymb}
\usepackage{bm}
\usepackage{a4wide}
\usepackage{graphicx} % for graphics
\usepackage{color}
\usepackage{euscript}
\usepackage{fancybox}

%#############
%newtheorems
%############
%\renewcommand{\thefootnote}{\fnsymbol{footnote}}
%\theoremstyle{plain}
%\newtheorem{thm}{Theorem}
%\newtheorem{claim}{Claim}
%\newtheorem{propn}{\bfseries Proposition}
%\newtheorem{propnn}{Proposition}
%\renewcommand{\thepropnn}{\arabic{propn}\alph{propnn}}
%\newtheorem{lem}{\bfseries Lemma}
%\newtheorem{cor}{Corollary}
%\newtheorem{conj}{\bfseries Conjecture}
%\newtheorem{defn}{Definition}
%\theoremstyle{remark}
%\newtheorem{rem}{Remark}
%\newtheorem{df}{Definition}
%\newtheorem{th1}[df]{Theorem}
%\newtheorem{lem}[df]{Lemma}
%\newtheorem{conj}[df]{Conjecture}
%\newtheorem{prop}[df]{Proposition}
%\newtheorem{cor}[df]{Corollary}
%\newtheorem{lem2}{Lemma}[section]
%\newtheorem{prop2}{Proposition}[section]
%\newtheorem{cor2}{Corollary}[section]
%\newtheorem{df2}{Definition}[section]

%##############
%newcommands
%#############
\newcommand{\cO}{\mathcal{O}}
\newcommand{\cA}{\mathcal{A}}
\newcommand{\cB}{\mathcal{B}}
\newcommand{\cC}{\mathcal{C}}
\newcommand{\cD}{\mathcal{D}}	
\newcommand{\cF}{\mathcal{F}}
\newcommand{\sP}{\mathsf{P}}
\newcommand{\sB}{\mathsf{B}}
\newcommand{\sY}{\mathsf{Y}}
\newcommand{\sy}{\mathsf{y}}
\newcommand{\sG}{\mathsf{G}}
\newcommand{\sm}{\mathsf{m}}
\newcommand{\sg}{\mathsf{g}}
\newcommand{\cT}{\mathcal{T}}
\newcommand{\cU}{\mathcal{U}}
\newcommand{\cH}{\mathcal{H}}
\newcommand{\cL}{\mathcal{L}}
\newcommand{\bC}{\mathbb{C}}
\newcommand{\txi}{\tilde{\xi}}
\newcommand{\tlambda}{\tilde{\lambda}}
\newcommand{\cI}{\mathcal{I}}
\newcommand{\cR}{\mathcal{R}}
\newcommand{\cS}{\mathcal{S}}
\newcommand{\cN}{\mathcal{N}}
\newcommand{\cM}{\mathcal{M}}
\newcommand{\cG}{\mathcal{G}}
\newcommand{\cK}{\mathcal{K}}
\newcommand{\bZ}{\mathbb{Z}}
\newcommand{\bR}{\mathbb{R}}
\newcommand{\ve}{\varepsilon}
\newcommand{\eH}{\EuScript{H}}
\newcommand{\cP}{\mathcal{P}}
\newcommand{\hT}{\hat{T}}
\newcommand{\htheta}{\hat{\theta}}
\newcommand{\sh}{\,\mathrm{sh}}
\newcommand{\ch}{\,\mathrm{ch}}
\newcommand{\fA}{\mathfrak{A}}
\newcommand{\fa}{\mathfrak{a}}
\newcommand{\fC}{\mathfrak{C}}
\newcommand{\fc}{\mathfrak{c}}
\newcommand{\fD}{\mathfrak{D}}
\newcommand{\fd}{\mathfrak{d}}
\newcommand{\fS}{\mathfrak{S}}
\newcommand{\ba}{\bar{a}}
\newcommand{\bb}{\bar{b}}
\newcommand{\dpsi}{\psi^\dag}
\newcommand{\pa}{a^{(+)}}
\newcommand{\dpa}{a^{(+)\dag}}
\newcommand{\ma}{a^{(-)}}
\newcommand{\dma}{a^{(-)\dag}}
\newcommand{\da}{a^\dag}
\newcommand{\ppsi}{\psi_+}
\newcommand{\dppsi}{\psi_+^\dag}
\newcommand{\mpsi}{\psi_-}
\newcommand{\dmpsi}{\psi_-^\dag}
\newcommand{\vphi}{\varphi}
\newcommand{\pS}{S^+}
\newcommand{\mS}{S^-}
\newcommand{\zS}{S^z}
%%%
%\newcommand{\bm}[1]{\mbox{\boldmath$#1$}}
%\newcommand{\be}{\begin{equation}}
%\newcommand{\ee}{\end{equation}}
%\newcommand{\bea}{\begin{eqnarray}}
%\newcommand{\eea}{\end{eqnarray}}
%\newcommand{\non}{\nonumber}
%\newcommand{\ra}{\rangle}
%\newcommand{\la}{\langle}
%\newcommand{\lam}{\lambda} 
%\newcommand{\Lam}{\Lambda} 
%\newcommand{\tht}{\theta} 
%\newcommand{\al}{\alpha} 
%\newcommand{\bt}{\beta} 
%\newcommand{\gm}{\gamma} 
%\newcommand{\dt}{\delta} 

\begin{document}
\fi

\section{Ultraviolet limit} \label{sec:UVlim}
As discussed in Section \ref{sec:ssg}, the SSG model is known to be a perturbed theory of an $\mathcal{N}=1$ superconformal field theory. Conformal invariance is obtained in the UV limit realized by $m_0L \to 0$. From the original SSG model, one obtains a complete space of states of an $\mathcal{N}=1$ superconformal field thoery, while it has been known the R sector cannot be realized through a lattice regularization under the periodic boundary condition. 
In this section, we discuss the UV limit of the SSG model using NLIEs, which are derived via a lattice regularization. Under Dirichlet boundary conditions, NLIEs show dependence on boundary parameters, resulting in different forms. Consequently, we obtained a first evidence that subsectors not being obtained under the periodic boundary condition are realized through lattice regularization. In order to support this statement, we calculate conformal dimensions of eigenstates in each regime of boundary parameters and then show both NS and R sectors can be obtained. 

\subsection{UV behavior of Bethe roots and scaling functions}
In the UV limit, there exist Bethe roots which tend to infinity. Such roots behave as $\theta \sim \hat{\theta} - \ln m_0L$, asymptotizing to infinity as $m_0 L \to 0$ \cite{bib:DV92}. Thus, a scaling function defined by $f^+(\hat{\theta}) = f(\hat{\theta} - \ln m_0L)$ shows a step-function like behavior at $\hat{\theta} \sim \ln m_0L$ \cite{bib:DV92, bib:KBP91, bib:Z90}. Using this function, one can rewrite NLIEs as follows: 
\begin{align}
 \ln b^+(\hat{\theta}) 
 &= \int_{-\infty}^{\infty} d\hat{\theta}'\; G(\hat{\theta} - \hat{\theta}' - i\epsilon) \ln B^+(\hat{\theta}' + i\epsilon)
 - \int_{-\infty}^{\infty} d\hat{\theta}'\; G(\hat{\theta} - \hat{\theta}' + i\epsilon) \ln \bar{B}^+(\hat{\theta}' - i\epsilon) 
 \nonumber \\
 &+ \int_{-\infty}^{\infty} d\hat{\theta}'\; G_K(\hat{\theta} - \hat{\theta}' - \tfrac{i\pi}{2} + i\epsilon) \ln Y^+(\hat{\theta}' - i\epsilon) 
 + ie^{\hat{\theta}} 
 + i\sum_{j} c_j g_{(j)}(\hat{\theta} - \hat{\theta}_j) 
 + i\pi \hat{C}_b, \label{NLIEb-UV} 
 \\
 \ln y^+(\hat{\theta}) 
 &= \int_{-\infty}^{\infty} d\hat{\theta}'\; G_K(\hat{\theta} - \hat{\theta}' + \tfrac{i\pi}{2} - i\epsilon) \ln B^+(\hat{\theta}' + i\epsilon) \nonumber \\
 &+ \int_{-\infty}^{\infty} d\hat{\theta}'\; G_K(\hat{\theta} - \hat{\theta}' - \frac{i\pi}{2} + i\epsilon) \ln\bar{B}^+(\hat{\theta}' - i\epsilon) 
 + i\sum_{j } c_{j} g^{(1)}_{(j)}(\hat{\theta} - \hat{\theta}_j )
 + i\pi \hat{C}_y. \label{NLIEy-UV}
\end{align}
Constants $\hat{C}_b$ and $\hat{C}_y$, which are not necessarily integers, include integration constants and asymptotic values obtained in Appendix \ref{sec:int_const}: 
\begin{align}
 i\pi \hat{C}_b &= i\pi \widetilde{C}_b^{(2)} + iF(\infty; H_+) + iF(\infty; H_-) + iJ(\infty) \nonumber \\
 &+ ig(\infty) (N_H^+ - 2(N_S^+ + N_V^+) - M_C^+ - 2M_W^+) 
 + 2ig(\infty) (N_H^0 - 2(N_S^0 + N_V^0) - M_C^0 - 2M_W^0) \nonumber \\
 &+ ig_K(\infty) N_1^+ + 2ig_K(\infty) N_1^0, 
 \\
 i\pi \hat{C}_y &= i\pi \widetilde{C}_y^{(2)} + iF_y(\infty; H_+) + iF_y(\infty; H_-) + 2ig_K(\infty) \nonumber \\
 &+ ig_K(\infty) (N_H^+ - 2 (N_S^+ + N_V^+) - M_C^+) + 2ig_K(\infty) (N_H^0 - 2 (N_S^0 + N_V^0) - M_C^0), 
\end{align}
where $A^+$ ($A \in \{N_H, N_1, N_S, N_V, M_C, M_W\}$) is a number of roots/holes which tend to infinity, while we denote those which remain finite in the UV limit by $A^0$. 

As energy in the context of conformal field theories is written as a function of system length $L$, we write eigenenergy obtained in the UV limit of (\ref{energy}) as a function of system length: 
\begin{equation} \label{cft-energy}
\begin{split}
%E(0) = E^+(0) + E^-(0) = 2 E^+(0) \\
 &E_{\rm CFT}(L) = E(L) - (E_{\rm bulk} + E_{\rm B}) = E_{\rm ex}(L) + E_C(L), \\
 &E_{\rm ex}(L) = \frac{1}{2L} \sum_{j = 1}^{N_H^+} e^{\hat{h}_j} - \frac{1}{2L} \sum_{j = 1}^{M_C^+} e^{\hat{c}_{j}}, \\
 &E_C(L) = \frac{1}{2\pi L} {\rm Im} \int_{-\infty}^{\infty} d\hat{\theta}\;  e^{\htheta} \ln\bar{B}^+(\hat{\theta}). 
\end{split}
\end{equation}
Although it is cumbersome to calculate these quantities directly, a trick used in \cite{bib:S99} allows us to write eigenenergy in a form which does not depend on Bethe roots (Appendix \ref{sec:uv_der}): 
\begin{equation} \label{UVenergy}
\begin{split}
 E_{\rm CFT}(L) &= \frac{1}{4\pi L} 
 \{ L_+(b^+(\infty)) - L_+(b^+(-\infty)) + L_+(\bar{b}^+(\infty)) - L_+(\bar{b}^+(-\infty)) \\
 &\hspace{13mm}+ L_+(y^+(\infty)) - L_+(y^+(-\infty)) \} \\
 &+ \frac{i}{8\pi L} 
 \Big[
 \{ e^{\htheta} + \sum_j c_j g_{(j)}(\htheta - \htheta_j) + \pi \hat{C}_b \} 
 (\ln B^+(\hat{\theta}) - \ln\bar{B}^+(\hat{\theta}))
 \Big]_{-\infty}^{\infty} \\
 &+ \frac{i}{8\pi L}
 \Big[
 \{ \sum_j c_j g^{(1)}_{(j)}(\htheta - \htheta_j) + \pi \hat{C}_y \}
 \ln Y^+(\hat{\theta})
 \Big]_{-\infty}^{\infty} \\
 &+ \frac{\pi}{L} (I_{N_H^+} - 2(I_{N_S^+} + I_{N_V^+}) - I_{M_C^+} - I_{M_W^+} + I_{N_1^+}) \\
 &- \frac{\pi}{2L} \{ \hat{C}_b (N_H^+ - 2(N_S^+ + N_V^+) - M_C^+ - M_W^+) + \hat{C}_y N_1^+ \}, 
\end{split}
\end{equation}
where $L_+(x)$ is a dilogarithm function defined by 
\begin{equation} \label{dilog}
 L_+(x) = \frac{1}{2} \int_0^x dy\; 
  \left( 
   \frac{\ln (1 + y)}{y} - \frac{\ln y}{1 + y} 
   \right). 
\end{equation}
Asymptotic values in (\ref{UVenergy}) are directly calculated from NLIEs (\ref{NLIEb-UV}) and (\ref{NLIEy-UV}): 
\begin{equation} \label{asym-uv}
\begin{split}
 &b^+(\infty) = 0,
 \hspace{27mm}
 y^+(\infty) = (-1)^{\frac{1}{2}({\rm sgn}(1-H_+) + {\rm sgn}(1-H_-) + {\rm sgn}(1+H_+) + {\rm sgn}(1+H_-))_{{\rm mod}\,2}}, 
 \\
 &b^+(-\infty) = 2e^{3i\rho_+} \cos\rho_+, 
 \qquad
 y_+(-\infty) = \frac{\sin3\rho_+}{\sin\rho_+}, 
\end{split}
\end{equation}
where 
\begin{equation}
\begin{split}
 \rho_+ &= \pi\{
 N_H^0 - 2(N_S^0 + N_V^0) - M_C^0 - 2M_W^0 + N_1^0 + 1 \\
 &\hspace{8mm}
 + \tfrac{1}{3}(n_b(H_+) - n_y(H_+) + n_b(H_-) - n_y(H_-)) + \widetilde{C}_b^{(2)}
 \} 
 \\
 &- \gamma\{
 3(N_H^0 - 2(N_S^0 + N_V^0) - M_C^0 - 2M_W^0) + 2N_1^0 + 3 + H - n_y(H_+) - n_y(H_-) + 2\widetilde{C}_b^{(2)}
 \}. 
\end{split}
\end{equation}
Besides, we obtain the following condition: 
\begin{equation}
 N_H^0 - 2(N_S^0 + N_V^0) - M_C^0 + 1 + \widetilde{C}_y^{(2)} + n_y(H_+) + n_y(H_-) = 0. 
\end{equation}
Asymptotic values of $\bar{b}(\hat{\theta})$ are obtained by taking complex conjugates of $b(\hat{\theta})$. 

\begin{figure}
\begin{center}
 \includegraphics[scale=0.4]{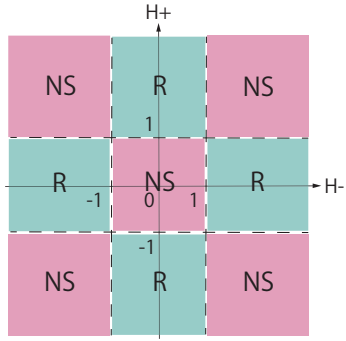}
 \caption{A phase separation between the NS sector and the R sector is obtained for boundary parameters in the UV limit.}
 \label{fig:UV_phase_transition}
\end{center}
\end{figure}

Using (\ref{asym-uv}) and properties of dilogarithm functions (Appendix \ref{sec:dilog}), we finally obtain 
\begin{equation} \label{UVenergy*}
\begin{split}
 E_{\rm CFT}(L) &= \frac{\pi}{2L} \Bigg(\frac{1}{\sqrt{\pi}} (\Phi_+ - \Phi_-) + 
 \sqrt{\frac{\pi - 2\gamma}{\pi}} \Big(\widetilde{C}_b^{(2)} + N_1^0 + 3S^0 + 1 - \frac{1}{2}(n_y(H_+) - n_y(H_-)) \Big)
 \\
 &- \sqrt{\frac{\pi}{\pi - 2\gamma}} \Big(S - \frac{1}{4}({\rm sign}(1-H_+) + {\rm sign}(1-H_-) + {\rm sign}(1+H_+) + {\rm sign}(1+H_-)) \Big)
 \Bigg)^2 
 \\
 &+ \frac{\pi}{16 L} \left(\frac{1}{2}({\rm sgn}(1-H_+) + {\rm sgn}(1-H_-) + {\rm sgn}(1+H_+) + {\rm sgn}(1+H_-))\right)_{{\rm mod}\, 2}
 - \frac{\pi}{16 L} 
 \\
 &+ \frac{\pi}{L} (I_{N_H^+} - 2(I_{N_S^+} + I_{N_V^+}) - I_{M_C^+} - I_{M_W^+})
 - \frac{\pi}{2L} \left((3S^+ + 2N_1^+) S^+ + (S^+ + M_W^+) N_1^+ \right). 
 %-\frac{\pi}{2L} (S^+ + M_W^+) N_1^+. 
\end{split}
\end{equation}
where 
\begin{align}
 &\Phi_{\pm} = \mp \frac{\gamma (H_{\pm} + 1)}{2 \sqrt{\pi - 2\gamma}}, \\
 &2S^\alpha = N_H^\alpha - 2(N_S^\alpha + N_V^\alpha) - M_C^\alpha - 2M_W^\alpha, 
 \qquad 
 \alpha = \{0,+\}, \\
 &S = S^+ + S^0. 
\end{align}
By comparing (\ref{UVenergy*}) with energy obtained in the context of a conformal field theory (\ref{energy_BFB}) and (\ref{energy_BFF}), one obtains a central charge and a conformal dimension as 
\begin{align}
 &c = \frac{3}{2}, \\
 &\Delta = \frac{1}{2}\left(
 \frac{\Phi_+ - \Phi_-}{\sqrt{\pi}} + mR + \frac{n}{R}
 \right)^2 \\
 &\hspace{7mm}+ \frac{1}{16} \left(\frac{1}{2}({\rm sgn}(1-H_+) + {\rm sgn}(1-H_-) + {\rm sgn}(1+H_+) + {\rm sgn}(1+H_-))\right)_{{\rm mod}\, 2}, 
\end{align}
where 
\begin{align}
 &m = -S + \tfrac{1}{4}({\rm sign}(1-H_+) + {\rm sign}(1-H_-) + {\rm sign}(1+H_+) + {\rm sign}(1+H_-)), \label{conf-dim-bm}
 \\
 &n = \widetilde{C}_b^{(2)} + N_1^0 + 3S^0 + 1 - \tfrac{1}{2}(n_y(H_+) + n_y(H_-)), \label{conf-dim-bn}
\end{align}
under a choice of compactification radius $R = \sqrt{\frac{\pi}{\pi - 2\gamma}}$. 
Thus, the theory belongs to the NS sector when $(\frac{1}{2}({\rm sgn}(1-H_+) + {\rm sgn}(1-H_-) + {\rm sgn}(1+H_+) + {\rm sgn}(1+H_-)))_{{\rm mod}\, 2} = 0$ giving $\Delta_{\rm F} = 0$, while it belongs to the R sector when $(\frac{1}{2}({\rm sgn}(1-H_+) + {\rm sgn}(1-H_-) + {\rm sgn}(1+H_+) + {\rm sgn}(1+H_-)))_{{\rm mod}\, 2} = 1$ giving $\Delta_{\rm F} = \frac{1}{16}$. This sector separation with respect to boundary parameters is shown in Figure \ref{fig:UV_phase_transition}. 
A boson part of conformal dimensions is labeled by two indices $(m,n)$. A momentum part (\ref{conf-dim-bn}) must vanish due to Dirichlet boundary conditions. On the other hand, $m$ given by (\ref{conf-dim-bm}) takes either an integer or a half-integer depending on boundary parameters, as $S$ takes only an integer. 
% figure

\ifx10
\end{document}
\fi
\ifx10
\documentclass{article}

%#############
%packages 
%#############
%\usepackage[utf8x]{inputenc}
\usepackage{amsmath, amsthm, amsfonts, amssymb}
\usepackage{bm}
\usepackage{a4wide}
\usepackage{graphicx} % for graphics
\usepackage{color}
\usepackage{euscript}
\usepackage{fancybox}

%#############
%newtheorems
%############
%\renewcommand{\thefootnote}{\fnsymbol{footnote}}
%\theoremstyle{plain}
%\newtheorem{thm}{Theorem}
%\newtheorem{claim}{Claim}
%\newtheorem{propn}{\bfseries Proposition}
%\newtheorem{propnn}{Proposition}
%\renewcommand{\thepropnn}{\arabic{propn}\alph{propnn}}
%\newtheorem{lem}{\bfseries Lemma}
%\newtheorem{cor}{Corollary}
%\newtheorem{conj}{\bfseries Conjecture}
%\newtheorem{defn}{Definition}
%\theoremstyle{remark}
%\newtheorem{rem}{Remark}
%\newtheorem{df}{Definition}
%\newtheorem{th1}[df]{Theorem}
%\newtheorem{lem}[df]{Lemma}
%\newtheorem{conj}[df]{Conjecture}
%\newtheorem{prop}[df]{Proposition}
%\newtheorem{cor}[df]{Corollary}
%\newtheorem{lem2}{Lemma}[section]
%\newtheorem{prop2}{Proposition}[section]
%\newtheorem{cor2}{Corollary}[section]
%\newtheorem{df2}{Definition}[section]

%##############
%newcommands
%#############
\newcommand{\cO}{\mathcal{O}}
\newcommand{\cA}{\mathcal{A}}
\newcommand{\cB}{\mathcal{B}}
\newcommand{\cC}{\mathcal{C}}
\newcommand{\cD}{\mathcal{D}}	
\newcommand{\cF}{\mathcal{F}}
\newcommand{\sP}{\mathsf{P}}
\newcommand{\sB}{\mathsf{B}}
\newcommand{\sY}{\mathsf{Y}}
\newcommand{\sy}{\mathsf{y}}
\newcommand{\sG}{\mathsf{G}}
\newcommand{\sm}{\mathsf{m}}
\newcommand{\sg}{\mathsf{g}}
\newcommand{\cT}{\mathcal{T}}
\newcommand{\cU}{\mathcal{U}}
\newcommand{\cH}{\mathcal{H}}
\newcommand{\cL}{\mathcal{L}}
\newcommand{\bC}{\mathbb{C}}
\newcommand{\txi}{\tilde{\xi}}
\newcommand{\tlambda}{\tilde{\lambda}}
\newcommand{\cI}{\mathcal{I}}
\newcommand{\cR}{\mathcal{R}}
\newcommand{\cS}{\mathcal{S}}
\newcommand{\cN}{\mathcal{N}}
\newcommand{\sM}{\mathsf{M}}
\newcommand{\cG}{\mathcal{G}}
\newcommand{\cK}{\mathcal{K}}
\newcommand{\bZ}{\mathbb{Z}}
\newcommand{\bR}{\mathbb{R}}
\newcommand{\ve}{\varepsilon}
\newcommand{\eH}{\EuScript{H}}
\newcommand{\cP}{\mathcal{P}}
\newcommand{\hT}{\hat{T}}
\newcommand{\htheta}{\hat{\theta}}
\newcommand{\sh}{\,\mathrm{sh}}
\newcommand{\ch}{\,\mathrm{ch}}
\newcommand{\fA}{\mathfrak{A}}
\newcommand{\fa}{\mathfrak{a}}
\newcommand{\fC}{\mathfrak{C}}
\newcommand{\fc}{\mathfrak{c}}
\newcommand{\fD}{\mathfrak{D}}
\newcommand{\fd}{\mathfrak{d}}
\newcommand{\fS}{\mathfrak{S}}
\newcommand{\ba}{\bar{a}}
\newcommand{\bb}{\bar{b}}
\newcommand{\dpsi}{\psi^\dag}
\newcommand{\pa}{a^{(+)}}
\newcommand{\dpa}{a^{(+)\dag}}
\newcommand{\ma}{a^{(-)}}
\newcommand{\dma}{a^{(-)\dag}}
\newcommand{\da}{a^\dag}
\newcommand{\ppsi}{\psi_+}
\newcommand{\dppsi}{\psi_+^\dag}
\newcommand{\mpsi}{\psi_-}
\newcommand{\dmpsi}{\psi_-^\dag}
\newcommand{\vphi}{\varphi}
\newcommand{\bpsi}{\bar{\psi}}
\newcommand{\pS}{S^+}
\newcommand{\mS}{S^-}
\newcommand{\zS}{S^z}
%%%
%\newcommand{\bm}[1]{\mbox{\boldmath$#1$}}
%\newcommand{\be}{\begin{equation}}
%\newcommand{\ee}{\end{equation}}
%\newcommand{\bea}{\begin{eqnarray}}
%\newcommand{\eea}{\end{eqnarray}}
%\newcommand{\non}{\nonumber}
%\newcommand{\ra}{\rangle}
%\newcommand{\la}{\langle}
%\newcommand{\lam}{\lambda} 
%\newcommand{\Lam}{\Lambda} 
%\newcommand{\tht}{\theta} 
%\newcommand{\al}{\alpha} 
%\newcommand{\bt}{\beta} 
%\newcommand{\gm}{\gamma} 
%\newcommand{\dt}{\delta} 

%#############
%document
%#############
\begin{document}
\fi
\section{Concluding remarks}
We have been discussed subspaces of the SSG model with Dirichelt boundaries obtained through a light-cone lattice, on which we derived NLIEs from the corresponding  Zamolodchikov-Fateev spin-$1$ XXZ chain with boundary magnetic field. The most important result in this paper is the fact that Dirichlet boundaries allows us to obtain the R sector, which cannot be obtained from the periodic system. 
%Sector separation of a superconformal field theory coincides with domain separation occurring in boundary magnetic field of the corresponding spin chain. This implies symmetries that a boundary magnetic field has still survive even after the scaling limit. 
According to UV analysis, it is the NLIE for $y(\theta)$ that determines which sector is realized from the LSSG model. On the other hand, a winding number $m$ is determined by $b(\theta)$ in such a way that takes an integer for the NS sector and a half integer for the R sector. At a separation point of these two sectors, an energy gap has been obtained (Figure \ref{fig:boundary_energy}). 

Counting equations (\ref{counting-a}) and (\ref{counting-b*}) also show existence of sector separations with respect to boundary parameters; Either an even number or an odd number of particles are allowed to exist depending on boundary parameters. In principle, a light-cone regularized quantum field theory consists of an even number of sites in order to generate a pair of a right-mover and a left-mover, on which only an even number of excitation particles are allowed to exist. However, in connection with allowed excitations in the boundary SG model \cite{bib:ABPR08}, strong enough boundary field arrests a particle, making a system effectively consisting of an odd number of sites, and then an excitation state with an odd number of particles is also obtained. 

In the IR limit, we have analyzed difference of boundary terms in NLIEs in a realm of boundary bootstrap principle. Mathematically, difference in boundary terms originates in change of analyticity structure of $T$-functions. Boundary bootstrap approach tells that this change occurs due to emergence of a boundary bound state. According to discussion in Section \ref{sec:IRlim}, the symmetries obtained in a corresponding spin chain are preserved in this limit. However, the given interpretation is incomplete, since it is the SUSY part which brings the phase separation and we did not discuss it yet. 

If both statements for the UV limit and the IR limit are correct, there seems to be hidden symmetry between the ground state and the boundary excitation state. In order to make it clear how phase separations vary from the UV limit to the IR limit, analysis on intermediate volume would be important. 
Another interesting future problem is to construct full regime of the SSG model from the spin chain. Recently, a supercharge defined on a spin chain is studied in connection with integrability of a system \cite{bib:HF12, bib:H13}. This supercharge adds one site to a system, {\it i.e.} it makes a system consisting of an even number of sites to that of an odd number of sites. If a supercharge defined on a lattice is correctly identified with that originally defined on a continuum theory, we may obtain subspaces of quantum field theories which cannot be obtained by a conventional method.

\ifx10
\end{document}
\fi
\section*{Acknowledgements}

The author is grateful to C. Ahn, P. Dorey, F. G\"{o}hmann, A. Kl\"{u}mper, and J. Suzuki for helpful discussions and comments. We acknowledge JSPS Research Fellowship for Young Scientists for supporting the beginning of this work. This research is also partially supported by the Aihara Innovative Mathematical Modelling Project, the Japan Society for the Promotion of Science (JSPS) through the ``Funding Program for World-Leading Innovative R\&D on Science and Technology (FIRST Program)'', initiated by the Council for Science and Technology Policy (CSTP).

%Appendix
\appendix
\ifx10
\documentclass{article}

%#############
%packages 
%#############
%\usepackage[utf8x]{inputenc}
\usepackage{amsmath, amsthm, amsfonts, amssymb}
\usepackage{bm}
\usepackage{a4wide}
\usepackage{graphicx} % for graphics
\usepackage{color}
\usepackage{euscript}
\usepackage{fancybox}

%#############
%newtheorems
%############
%\renewcommand{\thefootnote}{\fnsymbol{footnote}}
%\theoremstyle{plain}
%\newtheorem{thm}{Theorem}
%\newtheorem{claim}{Claim}
%\newtheorem{propn}{\bfseries Proposition}
%\newtheorem{propnn}{Proposition}
%\renewcommand{\thepropnn}{\arabic{propn}\alph{propnn}}
%\newtheorem{lem}{\bfseries Lemma}
%\newtheorem{cor}{Corollary}
%\newtheorem{conj}{\bfseries Conjecture}
%\newtheorem{defn}{Definition}
%\theoremstyle{remark}
%\newtheorem{rem}{Remark}
%\newtheorem{df}{Definition}
%\newtheorem{th1}[df]{Theorem}
%\newtheorem{lem}[df]{Lemma}
%\newtheorem{conj}[df]{Conjecture}
%\newtheorem{prop}[df]{Proposition}
%\newtheorem{cor}[df]{Corollary}
%\newtheorem{lem2}{Lemma}[section]
%\newtheorem{prop2}{Proposition}[section]
%\newtheorem{cor2}{Corollary}[section]
%\newtheorem{df2}{Definition}[section]

%##############
%newcommands
%#############
\newcommand{\cO}{\mathcal{O}}
\newcommand{\cA}{\mathcal{A}}
\newcommand{\cB}{\mathcal{B}}
\newcommand{\cC}{\mathcal{C}}
\newcommand{\cD}{\mathcal{D}}	
\newcommand{\cF}{\mathcal{F}}
\newcommand{\sP}{\mathsf{P}}
\newcommand{\sB}{\mathsf{B}}
\newcommand{\sY}{\mathsf{Y}}
\newcommand{\sy}{\mathsf{y}}
\newcommand{\sG}{\mathsf{G}}
\newcommand{\sm}{\mathsf{m}}
\newcommand{\sg}{\mathsf{g}}
\newcommand{\cT}{\mathcal{T}}
\newcommand{\cU}{\mathcal{U}}
\newcommand{\cH}{\mathcal{H}}
\newcommand{\cL}{\mathcal{L}}
\newcommand{\bC}{\mathbb{C}}
\newcommand{\txi}{\tilde{\xi}}
\newcommand{\tlambda}{\tilde{\lambda}}
\newcommand{\cI}{\mathcal{I}}
\newcommand{\cR}{\mathcal{R}}
\newcommand{\cS}{\mathcal{S}}
\newcommand{\cN}{\mathcal{N}}
\newcommand{\sM}{\mathsf{M}}
\newcommand{\cG}{\mathcal{G}}
\newcommand{\cK}{\mathcal{K}}
\newcommand{\bZ}{\mathbb{Z}}
\newcommand{\bR}{\mathbb{R}}
\newcommand{\ve}{\varepsilon}
\newcommand{\eH}{\EuScript{H}}
\newcommand{\cP}{\mathcal{P}}
\newcommand{\hT}{\hat{T}}
\newcommand{\htheta}{\hat{\theta}}
\newcommand{\sh}{\,\mathrm{sh}}
\newcommand{\ch}{\,\mathrm{ch}}
\newcommand{\fA}{\mathfrak{A}}
\newcommand{\fa}{\mathfrak{a}}
\newcommand{\fC}{\mathfrak{C}}
\newcommand{\fc}{\mathfrak{c}}
\newcommand{\fD}{\mathfrak{D}}
\newcommand{\fd}{\mathfrak{d}}
\newcommand{\fS}{\mathfrak{S}}
\newcommand{\ba}{\bar{a}}
\newcommand{\bb}{\bar{b}}
\newcommand{\dpsi}{\psi^\dag}
\newcommand{\pa}{a^{(+)}}
\newcommand{\dpa}{a^{(+)\dag}}
\newcommand{\ma}{a^{(-)}}
\newcommand{\dma}{a^{(-)\dag}}
\newcommand{\da}{a^\dag}
\newcommand{\ppsi}{\psi_+}
\newcommand{\dppsi}{\psi_+^\dag}
\newcommand{\mpsi}{\psi_-}
\newcommand{\dmpsi}{\psi_-^\dag}
\newcommand{\vphi}{\varphi}
\newcommand{\bpsi}{\bar{\psi}}
\newcommand{\pS}{S^+}
\newcommand{\mS}{S^-}
\newcommand{\zS}{S^z}
%%%
%\newcommand{\bm}[1]{\mbox{\boldmath$#1$}}
%\newcommand{\be}{\begin{equation}}
%\newcommand{\ee}{\end{equation}}
%\newcommand{\bea}{\begin{eqnarray}}
%\newcommand{\eea}{\end{eqnarray}}
%\newcommand{\non}{\nonumber}
%\newcommand{\ra}{\rangle}
%\newcommand{\la}{\langle}
%\newcommand{\lam}{\lambda} 
%\newcommand{\Lam}{\Lambda} 
%\newcommand{\tht}{\theta} 
%\newcommand{\al}{\alpha} 
%\newcommand{\bt}{\beta} 
%\newcommand{\gm}{\gamma} 
%\newcommand{\dt}{\delta} 

%#############
%document
%#############
\begin{document}
\fi

\section{Asymptotic behaviors of NLIEs} \label{sec:int_const}
Integration constants of (\ref{NLIE_b}) and (\ref{NLIE_y}) are determined from asymptotic behaviors of NLIEs. From the definitions, auxiliary functions $b(\theta)$ and $y(\theta)$ behave as 
\begin{align}
 &b(\infty) = e^{-2i\omega} + e^{-4i\omega}, \label{asym_b} \\ 
 &B(\infty) = 1 + e^{-2i\omega} + e^{-4i\omega}, \label{asym_B} \\
 &y(\infty) = e^{2i\omega} + 1 + e^{-2i\omega}, \label{asym_y} \\
 &Y(\infty) = e^{2i\omega} + 2 + e^{-2i\omega}. \label{asym_Y}
\end{align}
Here we set $\omega = \gamma(2S^{\rm tot} + H + 1)$ by defining total spin $S^{\rm tot} = N - M$ and an averaged boundary parameter $H = \frac{H_+ + H_-}{2}$. Apparently, left-hand sides of NLIEs remain finite, while linear terms $C_b^{(1)} \theta$ and $C_y^{(1)} \theta$ go to infinity as $\theta \to \infty$, which results in $C_b^{(1)} = C_y^{(1)} = 0$. 

Right-hand sides of NLIEs at $x \to \infty$ are evaluated from the following asymptotic behaviors: 
\begin{equation} \label{asym}
\begin{split}
 &g(\infty) = \pi G(\infty) 
 = \frac{\pi}{2} \frac{\pi - 3\gamma}{\pi - 2\gamma}, \qquad
 g_K(\infty) = \pi G_K(\infty)
 = \frac{\pi}{2}, \\
 &J(\infty) = \pi \frac{\pi - 3\gamma}{\pi - 2\gamma}, \hspace{23mm}
 J_K(\infty) = \pi. 
\end{split}
\end{equation}
Boundary terms $F(\theta;H)$ and $F_y(\theta;H)$ in the regime (a) behave as 
\begin{equation} \label{basym-a}
 F(\infty;H) = \frac{\pi}{2} \frac{\pi - \gamma H}{\pi - 2\gamma}, \qquad
  F_y(\infty;H) = 0, 
\end{equation}
while for the regime (b): 
\begin{equation} \label{basym-b}
 F(\infty;H) = \frac{\pi}{2} \frac{-\pi - \gamma H + 2 \gamma}{\pi - 2\gamma}, \qquad
  F_y(\infty;H) = \pi,
\end{equation}
and then for the regime (c) we have 
\begin{equation} \label{basym-c}
 F(\infty;H) = \frac{\pi}{2} \frac{-\pi - \gamma H}{\pi - 2\gamma}, \qquad
  F_y(\infty;H) = 0. 
\end{equation}

Substituting (\ref{asym_B}), (\ref{asym_y}), (\ref{asym}), and (\ref{basym-a})-(\ref{basym-c}) into the NLIE for $\ln y(\theta)$, the integration constant $C_y^{(2)}$ is determined as 
\begin{equation}
 C_y^{(2)} = i\pi \widetilde{C}_y^{(2)}
  = -i\pi [N_H - 2 (N_S + N_V) - M_C + 1 + n_y(H_+) + n_y(H_-)]_{{\rm mod}\,2}, 
\end{equation}
where 
\begin{equation}
 n_y(H) = 
  \begin{cases}
   0 & |H| > 1, \\
   1 & |H| < 1. 
   \end{cases}
\end{equation}
The integration constant $C_b^{(2)}$ is obtained from (\ref{asym_b}), (\ref{asym_B}), (\ref{asym_Y}), (\ref{asym}), and (\ref{basym-a})-(\ref{basym-c}): 
\begin{equation} \label{Cb_const}
 C_b^{(2)} = i\pi \widetilde{C}_b^{(2)}
  = -i\pi [2S^{\rm tot} + N + N_1 + 1 - \delta_b + n_b(H_+) + n_b(H_-)]_{{\rm mod}\,2}, 
\end{equation}
where 
\begin{equation}
 n_b(H) = 
  \begin{cases}
   \frac{3}{2} & H > 1, \\
   -\frac{1}{2} & |H| < 1, \\
   -\frac{3}{2} & -1 > H 
  \end{cases}
\end{equation}
and 
\begin{equation}
 \delta_b = 
  \begin{cases}
   0 & \cos\omega > 0, \\
   1 & \cos\omega < 0. 
  \end{cases}
\end{equation}
Here we used a notation $\lfloor*\rfloor$ which means the largest integer part of $*$. 

Besides (\ref{Cb_const}), we obtain a counting equation for holes: 
\begin{equation}
\begin{split}
 N_H - 2(N_S + N_V) &= 2S^{\rm tot} + M_C + 2M_W -\delta_B \\
 &+ \tfrac{1}{2}({\rm sgn}(1-H_+) + {\rm sgn}(1 + H_+) + {\rm sgn}(1-H_-) + {\rm sgn}(1+H_-)), 
\end{split}
\end{equation}
where 
\begin{equation} \label{def_deltaB}
 \delta_B = 
  \begin{cases}
   0 & 1 + 2\cos2\omega > 0 \\
   1 & 1 + 2\cos2\omega < 0. 
   \end{cases}
\end{equation}

In order to derive a counting equation for type-$1$ holes, we need to derive a NLIE for the auxiliary function $a(\theta)$. 
\ifx10
Keeping in our mind that real zeros of $T_1(\theta)$ consists of type-$1$ holes, we obtain 
\begin{equation} \label{NLIEa}
\begin{split}
 \ln a(\theta) &= 
  \int_{-\infty}^{\infty} d\theta'\;
  G_a(\theta - \theta' + i\epsilon) \ln A(\theta' - i\epsilon)
  - \int_{-\infty}^{\infty} d\theta'\;
  G_a(\theta - \theta' - i\epsilon) \ln \bar{A}(\theta' + i\epsilon)
  \\
 &+ iD^{(a)}_{\rm bulk}(\theta) + iD^{(a)}_{\rm B}(\theta) + iD_{\rm a}(\theta) + C_a, 
\end{split}
\end{equation}
where 
\begin{equation}
 G_a(\theta) = \int_{-\infty}^{\infty} \frac{dk}{2\pi}
  \frac{e^{-ik\theta} \sinh(\frac{\pi}{\gamma} - 2)\frac{\pi k}{2}}{2\cosh\frac{\pi k}{2} \sinh(\frac{\pi}{\gamma} - 1)\frac{\pi k}{2}}. 
\end{equation}
Boundary terms $iD^{(a)}_{\rm B}(\theta) = F_a(\theta;H_+) + F_a(\theta;H_-) + J_a(\theta)$ depend on boundary parameters and have different forms for positive $H$: 
\begin{equation}
 F_a(\theta;H) = \int d\theta\; \int_{-\infty}^{\infty} dk\;
  e^{-ik\theta}
  \frac{(\frac{\pi}{\gamma} - H) \frac{\pi k}{2}}{2\cosh\frac{\pi k}{2} \sinh(\frac{\pi}{\gamma} - 1)\frac{\pi k}{2}} 
\end{equation}
and negative $H$: 
\begin{equation}
 F_a(\theta;H) = -\int d\theta\; \int_{-\infty}^{\infty} dk\;
  e^{-ik\theta}
  \frac{(\frac{\pi}{\gamma} + H) \frac{\pi k}{2}}{2\cosh\frac{\pi k}{2} \sinh(\frac{\pi}{\gamma} - 1)\frac{\pi k}{2}} 
\end{equation}
A boundary independent term $J_a(\theta)$ is given by
\begin{equation}
 J_a(\theta) = \int d\theta\; \int_{-\infty}^{\infty} dk\;
  e^{-ik\theta}
  \frac{\cosh\frac{\pi k}{4} \sinh(\frac{\pi}{\gamma} - 2)\frac{\pi k}{4}}{\cosh \frac{\pi k}{2} \sinh(\frac{\pi}{\gamma} - 1)\frac{\pi k}{4}}. 
\end{equation}
Particle source terms are given by 
\begin{equation} \label{source-a}
\begin{split}
 &D_{\rm a}(\theta) = \sum_j c^{(a)}_j 
 \{
 g_{(j)}^{(a)}(\theta - \theta_j) + g_{(j)}^{(a)}(\theta + \theta_j)
 \}, \\
 &c_j^{(a)}
 =
 \begin{cases}
  1 & \text{for type-$1$ holes} \\
  -1 & \text{otherwise}, 
 \end{cases} \\
 &g_{(j)}^{(a)}(\theta) = 
 \begin{cases}
  (g_a)_{\rm II}(\theta) = g_a(\theta) + g_a(\theta - i\pi {\rm sgn}({\rm Im}\,\theta)) & \text{for roots satisfying $|{\rm Im}\,\theta_j| > \pi$} \\
  g_a(\theta + i\epsilon) + g_A(\theta - i\epsilon) & \text{for specials} \\
  g_a(\theta) & \text{otherwise}, 
  \end{cases}
\end{split}
\end{equation}
where 
\begin{equation}
 g_a(\theta) = 2\gamma \int d\theta\; G_a(\theta). 
\end{equation}
A bulk term is very similar to source terms, obtained as 
\begin{equation}
 D^{(a)}_{\rm bulk}(\theta)
  =
  N \{g_a(\theta - i\Theta) + g_a(\theta + i\Theta)\}. 
\end{equation}
A summation in (\ref{source-a}) is taken over $j$ such that $\theta_j$ is a type-$1$ hole, a real special object, or a complex root. 
\fi
Since the auxiliary function $a(\theta)$ asymptotically behaves as $a(\infty) = e^{-2i\omega}$, we obtain the following counting equation for type-$1$ holes by comparing both sides of a NLIE (\ref{NLIE_a}): 
\begin{equation} \label{counting_n1}
 N_1 - 2(N_S^R + N_V^R) = S^{\rm tot} - M_R + M_{C>\pi} + M_W + \tfrac{1}{2}({\rm sgn}(H_+) + {\rm sgn}(H_-)), 
\end{equation}
where $M_{C>\pi}$ represents the number of roots which satisfy $|{\rm Im}\,\theta_j| > \pi$. 
An integration constant $C_a$ is also determined as follows: 
and an integration constant: 
\begin{equation} \label{int_const_a}
 C_a = i\pi\widetilde{C}_a
  = -i\pi[
  2S^{\rm tot} + 1 + {\rm sgn}(H_+) + {\rm sgn}(H_-)
  ]_{{\rm mod}\, 2}. 
\end{equation}
In derivation of (\ref{counting_n1}) and (\ref{int_const_a}), we used the following asymptotic behaviors: 
\begin{align}
 &g_a(\infty) = \pi G_a(\infty) = \frac{\pi}{2} \frac{\pi - 2\gamma}{\pi - \gamma}, 
 \\
 &J_a(\infty) = \pi \frac{\pi - 2\gamma}{\pi - \gamma}, 
 \qquad
 F_a(\infty;H) = \frac{\pi}{2} \frac{{\rm sgn}(H)\, \pi - \gamma H}{\pi - \gamma} 
\end{align}
and a relation $M = M_R + M_C + M_W$.

\ifx10
\end{document}
\fi
\ifx10
\documentclass{article}

%#############
%packages 
%#############
%\usepackage[utf8x]{inputenc}
\usepackage{amsmath, amsthm, amsfonts, amssymb}
\usepackage{bm}
\usepackage{a4wide}
\usepackage{graphicx} % for graphics
\usepackage{color}
\usepackage{euscript}
\usepackage{fancybox}

%#############
%newtheorems
%############
%\renewcommand{\thefootnote}{\fnsymbol{footnote}}
%\theoremstyle{plain}
%\newtheorem{thm}{Theorem}
%\newtheorem{claim}{Claim}
%\newtheorem{propn}{\bfseries Proposition}
%\newtheorem{propnn}{Proposition}
%\renewcommand{\thepropnn}{\arabic{propn}\alph{propnn}}
%\newtheorem{lem}{\bfseries Lemma}
%\newtheorem{cor}{Corollary}
%\newtheorem{conj}{\bfseries Conjecture}
%\newtheorem{defn}{Definition}
%\theoremstyle{remark}
%\newtheorem{rem}{Remark}
%\newtheorem{df}{Definition}
%\newtheorem{th1}[df]{Theorem}
%\newtheorem{lem}[df]{Lemma}
%\newtheorem{conj}[df]{Conjecture}
%\newtheorem{prop}[df]{Proposition}
%\newtheorem{cor}[df]{Corollary}
%\newtheorem{lem2}{Lemma}[section]
%\newtheorem{prop2}{Proposition}[section]
%\newtheorem{cor2}{Corollary}[section]
%\newtheorem{df2}{Definition}[section]

%##############
%newcommands
%#############
\newcommand{\cO}{\mathcal{O}}
\newcommand{\cA}{\mathcal{A}}
\newcommand{\cB}{\mathcal{B}}
\newcommand{\cC}{\mathcal{C}}
\newcommand{\cD}{\mathcal{D}}	
\newcommand{\cF}{\mathcal{F}}
\newcommand{\sP}{\mathsf{P}}
\newcommand{\sB}{\mathsf{B}}
\newcommand{\sY}{\mathsf{Y}}
\newcommand{\sy}{\mathsf{y}}
\newcommand{\sG}{\mathsf{G}}
\newcommand{\sm}{\mathsf{m}}
\newcommand{\sg}{\mathsf{g}}
\newcommand{\cT}{\mathcal{T}}
\newcommand{\cU}{\mathcal{U}}
\newcommand{\cH}{\mathcal{H}}
\newcommand{\cL}{\mathcal{L}}
\newcommand{\bC}{\mathbb{C}}
\newcommand{\txi}{\tilde{\xi}}
\newcommand{\tlambda}{\tilde{\lambda}}
\newcommand{\cI}{\mathcal{I}}
\newcommand{\cR}{\mathcal{R}}
\newcommand{\cS}{\mathcal{S}}
\newcommand{\cN}{\mathcal{N}}
\newcommand{\cM}{\mathcal{M}}
\newcommand{\cG}{\mathcal{G}}
\newcommand{\cK}{\mathcal{K}}
\newcommand{\bZ}{\mathbb{Z}}
\newcommand{\bR}{\mathbb{R}}
\newcommand{\ve}{\varepsilon}
\newcommand{\eH}{\EuScript{H}}
\newcommand{\cP}{\mathcal{P}}
\newcommand{\hT}{\hat{T}}
\newcommand{\htheta}{\hat{\theta}}
\newcommand{\sh}{\,\mathrm{sh}}
\newcommand{\ch}{\,\mathrm{ch}}
\newcommand{\fA}{\mathfrak{A}}
\newcommand{\fa}{\mathfrak{a}}
\newcommand{\fC}{\mathfrak{C}}
\newcommand{\fc}{\mathfrak{c}}
\newcommand{\fD}{\mathfrak{D}}
\newcommand{\fd}{\mathfrak{d}}
\newcommand{\fS}{\mathfrak{S}}
\newcommand{\ba}{\bar{a}}
\newcommand{\bb}{\bar{b}}
\newcommand{\dpsi}{\psi^\dag}
\newcommand{\pa}{a^{(+)}}
\newcommand{\dpa}{a^{(+)\dag}}
\newcommand{\ma}{a^{(-)}}
\newcommand{\dma}{a^{(-)\dag}}
\newcommand{\da}{a^\dag}
\newcommand{\ppsi}{\psi_+}
\newcommand{\dppsi}{\psi_+^\dag}
\newcommand{\mpsi}{\psi_-}
\newcommand{\dmpsi}{\psi_-^\dag}
\newcommand{\vphi}{\varphi}
\newcommand{\pS}{S^+}
\newcommand{\mS}{S^-}
\newcommand{\zS}{S^z}
%%%
%\newcommand{\bm}[1]{\mbox{\boldmath$#1$}}
%\newcommand{\be}{\begin{equation}}
%\newcommand{\ee}{\end{equation}}
%\newcommand{\bea}{\begin{eqnarray}}
%\newcommand{\eea}{\end{eqnarray}}
%\newcommand{\non}{\nonumber}
%\newcommand{\ra}{\rangle}
%\newcommand{\la}{\langle}
%\newcommand{\lam}{\lambda} 
%\newcommand{\Lam}{\Lambda} 
%\newcommand{\tht}{\theta} 
%\newcommand{\al}{\alpha} 
%\newcommand{\bt}{\beta} 
%\newcommand{\gm}{\gamma} 
%\newcommand{\dt}{\delta} 

\begin{document}
\fi

\section{Evaluation of eigenenergy from NLIEs} \label{sec:uv_der}
Eigenenergy is evaluated from NLIEs without solving them. This technique was first introduced in \cite{bib:KBP91} and widely used for analytical calculation of $\mathcal{O}(N^{-1})$-corrections. Let us rewrite NLIEs (\ref{NLIEb-UV}) and (\ref{NLIEy-UV}) in a vector form: 
\begin{equation} \label{nlie-int}
 \bm{lb}^+(\hat{\theta}) = \mathcal{G}*\bm{lB}^+(\hat{\theta}) + i\bm{g}(\hat{\theta}), 
\end{equation}
where 
\begin{align}
 &\bm{lb}^+(\hat{\theta}) = 
 \begin{pmatrix}
  \ln b^+(\hat{\theta}) \\ \ln \bar{b}^+(\hat{\theta}) \\ \ln y^+(\hat{\theta})
 \end{pmatrix}, 
 \qquad
 \bm{lB}^+(\hat{\theta}) = 
 \begin{pmatrix}
  \ln B^+(\hat{\theta}) \\ \ln \bar{B}^+(\hat{\theta}) \\ \ln Y^+(\hat{\theta})
 \end{pmatrix}, 
 \\
 &\bm{g}(\hat{\theta}) = 
 \begin{pmatrix}
  e^{\hat{\theta}} + \sum_j c_j g_{(j)} (\hat{\theta} - \hat{\theta}_j) + \pi \hat{C}_b \\
  -e^{\hat{\theta}} - \sum_j c_j g_{(j)} (\hat{\theta} - \hat{\theta}_j) - \pi \hat{C}_b \\
  \sum_j c_j g_{(j)}^{(1)} (\hat{\theta} - \hat{\theta}_j) + \pi \hat{C}_y
 \end{pmatrix}. 
\end{align}
$\mathcal{G}$ is a matrix given by 
\begin{equation}
 \mathcal{G}(\hat{\theta}) = 
  \begin{pmatrix}
   G(\hat{\theta} - i\epsilon) & -G(\hat{\theta} + i\epsilon) & G_K(\hat{\theta} - \frac{i\pi}{2} + i\epsilon) \\
   -\bar{G}(\hat{\theta} - i\epsilon) & \bar{G}(\hat{\theta} - i\epsilon) & G_K(\hat{\theta} + \frac{i\pi}{2} - i\epsilon) \\
   G_K(\hat{\theta} + \frac{i\pi}{2} - i\epsilon) & G_K(\hat{\theta} - \frac{i\pi}{2} + i\epsilon) & 0
  \end{pmatrix}. 
\end{equation}
Using $G(\hat{\theta}) = \bar{G}(-\hat{\theta})$ and $G_K(\hat{\theta}) = \bar{G}_K(-\hat{\theta}) = G_K(\hat{\theta})$, one obtains that $\mathcal{G}(\hat{\theta})$ satisfies 
\begin{equation} \label{prop-G}
 \mathcal{G}_{ij}(\hat{\theta}) = \mathcal{G}_{ji}(-\hat{\theta}), 
  \qquad 
  i \neq j. 
\end{equation}

Consider an integral of $\bm{lb}^{+'} \cdot \bm{lB}^+ - \bm{lb}^+ \cdot \bm{lB}^{+'}$, which is written in terms of dilogarithm functions: 
\begin{equation} \label{lhs}
\begin{split}
 &\frac{1}{2} \int_{-\infty}^{\infty} d\hat{\theta}\ 
  \Big(\bm{lb}^{+'}(\hat{\theta}) \cdot \bm{lB}^+(\hat{\theta}) - \bm{lb}^+(\hat{\theta}) \cdot \bm{lB}^{+'}(\hat{\theta})\Big)
  \\&=
 L_+(b^+(\infty)) - L_+(b^+(-\infty)) + L_+(\bar{b}^+(\infty)) - L_+(\bar{b}^+(-\infty)) + L_+(y^+(\infty)) - L_+(y^+(-\infty)). 
\end{split}
\end{equation}
On the other hand, by substituting (\ref{nlie-int}) into $\bm{lb}^{+'}$ and $\bm{lb}^+$ and then we observe that compensation occurs to terms concerning $\mathcal{G}$ due to (\ref{prop-G}). Remaining terms are obtained as 
\begin{equation} \label{rhs}
\begin{split}
 &\frac{1}{2} \int_{-\infty}^{\infty} d\hat{\theta}\ 
  \Big(\bm{lb}^{+'}(\hat{\theta}) \cdot \bm{lB}^+(\hat{\theta}) - \bm{lb}^+(\hat{\theta}) \cdot \bm{lB}^{+'}(\hat{\theta})\Big)
 \\
 &=
 2{\rm Im} \int_{-\infty}^{\infty} d\htheta\; e^{\htheta} \ln \bar{B}^+(\htheta) 
 + 2\pi \sum_j c_j e_{(j)}^{\htheta_j} \\
 &- \frac{i}{2} \Big[
 \Big(e^{\htheta} + \sum_j c_j g_{(j)}(\htheta - \htheta_j) + \pi \hat{C}_b\Big)
 (\ln B^+(\htheta) - \ln\bar{B}^+(\htheta))
 \Big]_{-\infty}^{\infty} \\
 &- \frac{i}{2} \Big[
 \Big(\sum_j c_j g^{(1)}_{(j)}(\htheta - \htheta_j) + \pi \hat{C}_y\Big)
 \ln Y^+(\htheta)
 \Big]_{-\infty}^{\infty} \\
 &+ 2\pi i \sum_{j;\htheta_j \neq \hat{h}^{(1)}_j} c_j \ln b^+_{(j)}(\htheta_j) 
 + 2\pi i \sum_{j = 1}^{N_1^+} \ln y^+(\hat{h}^{(1)}_j) \\
 &+ 2\pi^2 \hat{C}_b (N_H^+ - 2N_S^+ - M_C^+ - M_W^+ - M_{SC}^+) 
 + 2\pi^2 \hat{C}_y N_1^+, 
\end{split}
\end{equation}
where 
\begin{align}
 &\ln b_{(j)}^+(\htheta) = 
 \begin{cases}
  \ln b^+(\htheta - i\epsilon) + \ln b^+(\htheta + i\epsilon) & \text{for specials} \\
  \ln b^+(\htheta) & \text{otherwise}
 \end{cases}
 \\
 &e_{(j)}^{\htheta} = 
 \begin{cases}
  e^{\htheta - i\epsilon} + e^{\htheta + i\epsilon} & \text{for specials} \\
  e_{\rm II}^{\htheta} = 0 & \text{for wide roots} \\
  e^{\htheta} & \text{otherwise}. 
 \end{cases}
\end{align}
From definitions of auxiliary functions, we have the following relations: 
\begin{equation}
\begin{split}
 &\textstyle\sum_{j;\htheta_j \neq h^{(1)}_j} c_j \ln b^+_{(j)}(\htheta_j) 
 = 2 \pi i (I_{N_H^+} - 2(I_{N_S^+} + I_{N_V^+}) - I_{M_C^+} - I_{M_W^+}), 
 \\
 &\textstyle\sum _{j=1}^{N_1^+} \ln y^+_{(j)}(\hat{h}_j^{(1)}) = 2\pi i I_{N_1^+}
\end{split}
\end{equation}
by introduced integers $I_{A^+}$ ($A \in \{N_H, N_S, N_V, M_C, M_W\}$) which give summation of quantum numbers {\it i.e.} $I_{A^+} = \sum_{j =1}^{A^+} I_{A,j}^+ = \frac{1}{2\pi i} \sum_{j=1}^{A^+} \ln b^+(\htheta_j)$. 

Using (\ref{lhs}), (\ref{rhs}), and an energy formula (\ref{cft-energy}), we finally obtain (\ref{UVenergy}).

\if10
\end{document}
\fi
\ifx10
\documentclass{article}

%#############
%packages 
%#############
%\usepackage[utf8x]{inputenc}
\usepackage{amsmath, amsthm, amsfonts, amssymb}
\usepackage{bm}
\usepackage{a4wide}
\usepackage{graphicx} % for graphics
\usepackage{color}
\usepackage{euscript}
\usepackage{fancybox}

%#############
%newtheorems
%############
%\renewcommand{\thefootnote}{\fnsymbol{footnote}}
%\theoremstyle{plain}
%\newtheorem{thm}{Theorem}
%\newtheorem{claim}{Claim}
%\newtheorem{propn}{\bfseries Proposition}
%\newtheorem{propnn}{Proposition}
%\renewcommand{\thepropnn}{\arabic{propn}\alph{propnn}}
%\newtheorem{lem}{\bfseries Lemma}
%\newtheorem{cor}{Corollary}
%\newtheorem{conj}{\bfseries Conjecture}
%\newtheorem{defn}{Definition}
%\theoremstyle{remark}
%\newtheorem{rem}{Remark}
%\newtheorem{df}{Definition}
%\newtheorem{th1}[df]{Theorem}
%\newtheorem{lem}[df]{Lemma}
%\newtheorem{conj}[df]{Conjecture}
%\newtheorem{prop}[df]{Proposition}
%\newtheorem{cor}[df]{Corollary}
%\newtheorem{lem2}{Lemma}[section]
%\newtheorem{prop2}{Proposition}[section]
%\newtheorem{cor2}{Corollary}[section]
%\newtheorem{df2}{Definition}[section]

%##############
%newcommands
%#############
\newcommand{\cO}{\mathcal{O}}
\newcommand{\cA}{\mathcal{A}}
\newcommand{\cB}{\mathcal{B}}
\newcommand{\cC}{\mathcal{C}}
\newcommand{\cD}{\mathcal{D}}	
\newcommand{\cF}{\mathcal{F}}
\newcommand{\sP}{\mathsf{P}}
\newcommand{\sB}{\mathsf{B}}
\newcommand{\sY}{\mathsf{Y}}
\newcommand{\sy}{\mathsf{y}}
\newcommand{\sG}{\mathsf{G}}
\newcommand{\sm}{\mathsf{m}}
\newcommand{\sg}{\mathsf{g}}
\newcommand{\cT}{\mathcal{T}}
\newcommand{\cU}{\mathcal{U}}
\newcommand{\cH}{\mathcal{H}}
\newcommand{\cL}{\mathcal{L}}
\newcommand{\bC}{\mathbb{C}}
\newcommand{\txi}{\tilde{\xi}}
\newcommand{\tlambda}{\tilde{\lambda}}
\newcommand{\cI}{\mathcal{I}}
\newcommand{\cR}{\mathcal{R}}
\newcommand{\cS}{\mathcal{S}}
\newcommand{\cN}{\mathcal{N}}
\newcommand{\cM}{\mathcal{M}}
\newcommand{\cG}{\mathcal{G}}
\newcommand{\cK}{\mathcal{K}}
\newcommand{\bZ}{\mathbb{Z}}
\newcommand{\bR}{\mathbb{R}}
\newcommand{\ve}{\varepsilon}
\newcommand{\eH}{\EuScript{H}}
\newcommand{\cP}{\mathcal{P}}
\newcommand{\hT}{\hat{T}}
\newcommand{\htheta}{\hat{\theta}}
\newcommand{\sh}{\,\mathrm{sh}}
\newcommand{\ch}{\,\mathrm{ch}}
\newcommand{\fA}{\mathfrak{A}}
\newcommand{\fa}{\mathfrak{a}}
\newcommand{\fC}{\mathfrak{C}}
\newcommand{\fc}{\mathfrak{c}}
\newcommand{\fD}{\mathfrak{D}}
\newcommand{\fd}{\mathfrak{d}}
\newcommand{\fS}{\mathfrak{S}}
\newcommand{\ba}{\bar{a}}
\newcommand{\bb}{\bar{b}}
\newcommand{\dpsi}{\psi^\dag}
\newcommand{\pa}{a^{(+)}}
\newcommand{\dpa}{a^{(+)\dag}}
\newcommand{\ma}{a^{(-)}}
\newcommand{\dma}{a^{(-)\dag}}
\newcommand{\da}{a^\dag}
\newcommand{\ppsi}{\psi_+}
\newcommand{\dppsi}{\psi_+^\dag}
\newcommand{\mpsi}{\psi_-}
\newcommand{\dmpsi}{\psi_-^\dag}
\newcommand{\vphi}{\varphi}
\newcommand{\pS}{S^+}
\newcommand{\mS}{S^-}
\newcommand{\zS}{S^z}
%%%
%\newcommand{\bm}[1]{\mbox{\boldmath$#1$}}
%\newcommand{\be}{\begin{equation}}
%\newcommand{\ee}{\end{equation}}
%\newcommand{\bea}{\begin{eqnarray}}
%\newcommand{\eea}{\end{eqnarray}}
%\newcommand{\non}{\nonumber}
%\newcommand{\ra}{\rangle}
%\newcommand{\la}{\langle}
%\newcommand{\lam}{\lambda} 
%\newcommand{\Lam}{\Lambda} 
%\newcommand{\tht}{\theta} 
%\newcommand{\al}{\alpha} 
%\newcommand{\bt}{\beta} 
%\newcommand{\gm}{\gamma} 
%\newcommand{\dt}{\delta} 

\begin{document}
\fi

\section{Dilogarithm identities} \label{sec:dilog}
Dilogarithm functions appear widely in integrable systems and have been intesively studied. The dologarithm function defined by (\ref{dilog}) is connected to a Rogers' dilogarithm: 
\begin{equation}
 L(x) = -\frac{1}{2} \int_0^x dy\; \Big(
  \frac{\ln (1-y)}{y} + \frac{\ln y}{1-y}
  \Big)
\end{equation}
via a relation $L_+(x) = L(\frac{x}{1+x})$. Subsequently, remarkable relations among dilogarithms have been found mathematical physics problems \cite{bib:Z91, bib:RVT93, bib:K89, bib:BR90, bib:K93, bib:K95}. Here we list some of them which appear in the expression of eigenenergy (\ref{UVenergy}): 
\begin{equation}
\begin{split}
 &L(0) = 0, 
 \qquad
 L(\tfrac{1}{2}) = \tfrac{\pi^2}{12}, 
 \qquad
 L(-\infty) = -\tfrac{\pi^2}{6}, 
 \\
 &L(1) = L(x) + L(1 - x) = \tfrac{\pi^2}{6} 
 \qquad x \in [0,1], 
 \\
 &2L(1) = 2L(\tfrac{1}{n+1}) + \sum_{j=0}^{n-1} L(\tfrac{1}{(1 + j)^2})
 \qquad n \in \mathbb{Z}_{\geq 0}, 
 \\
 &L(1) \tfrac{3n}{n+2} = \sum_{j=0}^{n-1} L(\tfrac{\sin^2\frac{\pi}{n+2}}{\sin^2\frac{\pi(j+1)}{n+2}})
 \qquad n \in \mathbb{Z}_{\geq 0}. 
\end{split}
\end{equation}
Moreover, it has been obtained in \cite{bib:S04} that the following relation holds: 
\begin{equation}
 2 \sum_{p=1}^{k-1}\Big[
  L(\tfrac{p(p+2)}{(p+1)^2}) - L(\tfrac{\sin\frac{\pi p}{k+2} \sin\frac{\pi(p+2)}{k+2}} {\sin^2\frac{\pi(p+1)}{k+2}})
  \Big]
  + 4L(\tfrac{k}{k+1})
  =
  \tfrac{\pi^2 k}{k+2}. 
\end{equation}
Thus, we obtain the following relation for small $\gamma$: 
\begin{equation}
\begin{split}
 &L_+(b^+(-\infty)) + L_+(\bar{b}^+(-\infty)) + L_+(y^+(-\infty)) \\
 &- L_+(
 (\tfrac{1}{2}({\rm sgn}(1-H_+) + {\rm sgn}(1-H_-) + {\rm sgn}(1+H_+) + {\rm sgn}(1+H_+)))_{{\rm mod}\,2}
 ) 
 \\
 &= \begin{cases}
     \frac{\pi^2}{4} & (\frac{1}{2}({\rm sgn}(1-H_+) + {\rm sgn}(1-H_-) + {\rm sgn}(1+H_+) + {\rm sgn}(1+H_+)))_{{\rm mod}\,2} = 0 \\
     \tfrac{\pi^2}{2} & (\frac{1}{2}({\rm sgn}(1-H_+) + {\rm sgn}(1-H_-) + {\rm sgn}(1+H_+) + {\rm sgn}(1+H_+)))_{{\rm mod}\,2} = 1. 
    \end{cases}
\end{split}
\end{equation}

\ifx10
\end{document}
\fi

%Bibliography
\bibliographystyle{unsrt}
\bibliography{references}

\end{document}